\documentclass[acmlarge]{acmart}
\AtBeginDocument{%
  }

\setcopyright{acmlicensed}
\copyrightyear{2026}
\acmYear{2026}
\acmDOI{XXXXXXX.XXXXXXX}

\acmJournal{IMWUT}
\acmVolume{0}
\acmNumber{0}
\acmArticle{0}
\acmMonth{0}
\usepackage{tikz}       % for tikzpicture
\usepackage{pgfplots}
\usepackage{arydshln}% if using axis environment
\usepackage[table]{xcolor} % for \rowcolor
\pgfplotsset{compat=1.18}
\usepackage{algorithm}
\usepackage{algpseudocode}
\usepackage[utf8]{inputenc}
\usepackage{subcaption}
\usepackage{booktabs}          % For \toprule, \midrule, \bottomrule
\usepackage{xcolor}            % For row coloring (\rowcolor)
\usepackage{subcaption}        % For \subcaptionbox% For subfloat panels (a), (b), (c)
\usepackage{multirow}          % For multirow cells (if needed)
\usepackage{array}
\usepackage{makecell}
\usepackage{graphicx}% For better column alignment
\usepackage{caption}           % For caption formatting
\usepackage{amsmath}           % For math symbols like \Delta
\usepackage[utf8]{inputenc}
\DeclareUnicodeCharacter{2212}{-}
\usepackage{booktabs}
\usepackage[table]{xcolor}

\usepackage{enumitem}

\usepackage{etoolbox}
\begin{document}

%\title{Ethical Fairness without Demographics in Human-Centered AI}
\title{Ethical Fairness in Ubiquitous Health Sensing without Known Attributes}


\author{Shaily Roy}
\authornote{Equal Contribution.}
\email{shailyro@asu.edu}
\affiliation{%
  \institution{Arizona State University}
  \city{Tempe}
  \state{Arizona}
  \country{USA}
}

\author{Harshit Sharma}
\email{hsharm62@asu.edu}
\affiliation{%
  \institution{Arizona State University}
  \city{Tempe}
  \state{Arizona}
  \country{USA}
}

\author{Daniel A. Adler}
\email{adlerdan@umich.edu}
\affiliation{%
  \institution{University of Michigan}
  \city{Ann Arbor}
  \state{Michigan}
  \country{USA}
}

\author{Srijan Sen}
\email{srijan@umich.edu}
\affiliation{%
  \institution{University of Michigan}
  \city{Ann Arbor}
  \state{Michigan}
  \country{USA}
}

\author{Tanzeem Choudhury}
\email{tanzeem.choudhury@cornell.edu}
\affiliation{%
  \institution{Cornell University}
  \city{Ithaca}
  \state{New York}
  \country{USA}
}

\author{Asif Salekin}
\authornotemark[2]
\email{asalekin@asu.edu}
\affiliation{%
  \institution{Arizona State University}
  \city{Tempe}
  \state{Arizona}
  \country{USA}
}
\renewcommand{\shortauthors}{Trovato et al.}

\begin{abstract}
Wearable and mobile sensing models infer health and well-being in everyday life, yet strong average performance can conceal failures for particular people, contexts, or devices. Existing fairness methods often require sensitive attributes during training, although such attributes may be unavailable, privacy-sensitive, or restricted. Moreover, they may reduce disparities by degrading performance for some subgroups, violating the ethical principle of non-maleficence.

We present \textbf{\textsc{Flare}} (Fisher-guided LAtent-subgroup learning with do-no-harm REgularization), to our knowledge the first attribute-free framework to operationalize beneficence, non-maleficence, and justice in ubiquitous sensing. \textbf{\textsc{Flare}} uses Fisher information to regularize loss curvature and combines representation, loss, and curvature signals to discover latent performance strata. It then trains a separate model for each stratum using do-no-harm regularization and validation-controlled parameter sharing. We also introduce BHE (Benefit, Harm-Avoidance, Equity), a metric suite for evaluating ethical fairness beyond statistical parity.
Across four physiological, behavioral, and clinical sensing datasets, \textbf{\textsc{Flare}} achieves the highest mean subgroup F1 on every dataset. It is the only evaluated method that improves mean subgroup performance without degrading any evaluated subgroup and the only one that reduces disparity across all 15 evaluated sensitive attributes. Although some baselines achieve larger average disparity reductions, each harms at least one subgroup. Post-hoc analyses on two datasets show that the discovered strata are only weakly aligned with recorded attributes and identify a recurrent within-person state that participant-level attributes cannot capture. Label-free routing and inference take at most 8.1\,ms per sample on a smartphone and 10.5\,ms on a Raspberry Pi, supporting real-time on-device deployment.

\end{abstract}



%\keywords{Do, Not, Use, This, Code, Put, the, Correct, Terms, for,
%  Your, Paper}


\maketitle

\section{Introduction}\label{sec:intro}

%Artificial intelligence (AI) has become a cornerstone of modern healthcare, seamlessly integrating data-driven intelligence across clinical and human-sensing domains~\cite{wang2025recent}. Leveraging electronic health records, medical imaging, and physiological time series, AI systems accelerate triage, enhance diagnostic precision, enable risk stratification, and extend care through continuous remote monitoring~\cite{zhang2023leveraging}. 
%In the realm of human sensing, the convergence of advanced machine learning (ML) and ubiquitous sensing technologies~\cite{li2021deep,puccinelli2005wireless} empowers everyday devices—such as smartphones and wearables—to capture rich multimodal data for diverse applications, including gesture recognition~\cite{zhang2021WIDAR3} and stress detection~\cite{vos2023generalizable}. These advances facilitate automated pattern discovery in high-dimensional, multimodal datasets, offering scalable and cost-effective feedback for both clinicians and patients.

Artificial intelligence (AI) is increasingly embedded in ubiquitous and mobile health systems, where wearable, smartphone, physiological, and behavioral sensing data are used to infer health and well-being states in everyday life \cite{wang2025recent,zhang2023leveraging,li2021deep,puccinelli2005wireless,zhang2021WIDAR3,vos2023generalizable}. 
However, these systems face a persistent challenge: model performance is rarely uniform. In ubiquitous computing, machine learning models often estimate health and well-being from sensing data without using demographic subgroup information during training \cite{adler2024measuring,xu2023globem,wang2018tracking}. At the same time, heterogeneity in physiology, behavior, context, and device usage can create uneven performance across latent subgroups \cite{xiao2025human,adler2024beyond}. As a result, models that perform well on average may still systematically underserve specific groups, embedding inequities into sensing-based decision pipelines.% and motivating subgroup-aware, fairness-driven learning.}

For example, on the OhioT1DM glucose dataset, a standard model achieves an F1 of 67.7\% for users of the Medtronic 530G insulin pump but only 51.6\% for 630G users (Appendix~\ref{sec:subgroup-analysis}, Table~\ref{tab:ohio_eda_f1}). For hyperglycemia alerts, this gap means more errors for an entire device population, despite involving no demographic attribute. The gap is detectable only because OhioT1DM records pump generation. Comparable sources of heterogeneity—such as sensor placement, firmware, medication, daily routines, or care setting—are rarely recorded or known beforehand to matter. Equally large gaps along these unobserved factors may therefore evade subgroup audits and fairness methods that require prespecified attributes.

More broadly, in ubiquitous and mobile health technologies, uneven model performance can unintentionally amplify existing systemic inequities, including the ``digital divide,'' where access to devices is unequal \cite{wang2022digital}; ``digital redlining,'' where algorithmic systems exclude or underserve specific populations \cite{wang2024digital}; and broader ``digital determinants of health,'' where technological access and infrastructure shape health outcomes \cite{chidambaram2024introduction}. That means, in real-world ubiquitous health system deployments, indeed disparities may appear not only through demographic differences, such as sex or age, but also through non-demographic sources of variation, such as sensor band, device availability, clinical condition, behavioral context, and more.

Throughout this paper, we distinguish between \emph{demographic attributes} and broader \emph{heterogeneity factors}. Demographic attributes, such as sex, age, race, and ethnicity, describe population-level human variation and are commonly used in fairness analysis \cite{pal2023ensuring, hu2023parametric,liu2026fairness}. From a system perspective, however, fairness failures in ubiquitous computing can arise from broader heterogeneity factors: demographics are one important source, but sensing performance is also shaped by behavioral, contextual, clinical, physiological, data-quality, and device-level variations in real deployments. Some of these factors may be observed as known attributes during evaluation, while others remain latent. Therefore, our goal goes beyond conventional demographic-based fairness \cite{pessach2023algorithmic,andrus2022demographic,giguere2022fairness}: \emph{we aim to improve fairness across both known and latent heterogeneity factors.}% without relying on demographic or other sensitive attributes during training.

Moreover, in everyday sensing, heterogeneity is a property not only of \emph{who} is sensed but also of \emph{when} and \emph{how}: the same person moves between contexts, routines, and physiological states, so a model may underperform during particular within-person states rather than for particular people. Such strata cannot be captured by any participant-level attribute, demographic or otherwise, and therefore cannot be targeted by fairness methods that require the relevant attribute to be specified in advance. This motivates learning the relevant subgroups from the model's own behavior, without attribute supervision; our post-hoc analysis later confirms that such a recurrent within-person stratum does arise in practice (Section~\ref{sec:cluster_interpretation}).

Importantly, in ubiquitous health computing—fairness is deeply intertwined with clinical ethics \cite{varkey2021principles}. Ethically fair health AI must not only ensure equitable performance but also uphold the principles of \emph{beneficence} (promoting overall benefit), \emph{non-maleficence} (avoiding harm for any subgroups) and  \emph{justice} (ensuring equitable treatment across all) \cite{hurley2003fairness,andersson2010no,gabriel2022toward}.
This ethical fairness framing is critical for ubiquitous sensing systems, where poor performance for specific subgroups can translate into unreliable health feedback, missed alerts, and inequitable downstream decisions. 
Yet, conventional fairness-driven optimization often overlooks these broader ethical dimensions. Methods that focus narrowly on equalizing performance across demographic subgroups can miss other important subgroup failures arising from heterogeneity factors such as device type, context, physiology, or behavior \cite{adler2024measuring,yfantidou2023beyond}. They may also improve parity by degrading performance for some of those subgroups, thereby violating non-maleficence and eroding trust in ubiquitous health applications \cite{rajkomar2018ensuring,meegahapola2023generalization}. To address these gaps, this paper formalizes and operationalizes \emph{ethical fairness} for ubiquitous health sensing: it promotes beneficence, non-maleficence, and justice across latent subgroups discovered without sensitive attributes, and evaluates all three across known demographic and non-demographic heterogeneity factors.
%{Methods that narrowly focus on equalizing performance across demographic subgroups can inadvertently mask other types of subgroup failures, particularly in mobile sensing contexts when demographic information is unavailable to models, and erode trust in human-centered applications }

%\textcolor{teal}{Recent U.S. policy developments—such as updates to the Health and Human Services (HHS) nondiscrimination rule and new federal AI governance frameworks—emphasize equitable and privacy-oriented ML in healthcare~\cite{hhs1557final,ai2023artificial}. This vision is echoed in the 2025 Executive Order~\cite{house2025removing}, which calls for bias-resistant, innovation-driven AI that advances fairness \emph{through principled optimization rather than demographic categorization}, fostering equity without reinforcing engineered social constructs.

Recent U.S. policy developments emphasize equitable, privacy-oriented, and bias-resistant AI in healthcare, motivating approaches that advance fairness \emph{through principled optimization rather than demographic categorization}, fostering equity without reinforcing engineered social constructs~\cite{hhs1557final,ai2023artificial,house2025removing}. These priorities are particularly relevant for ubiquitous and mobile sensing systems, where collecting or using demographic and heterogeneity-relevant sensitive attributes during training may be infeasible due to privacy, regulatory, or ethical concerns \cite{baron2020where, pessach2023algorithmic,friedler2021possibility,stopczynski2014privacy,andrus2022demographic}.
Therefore, ethical fairness should be \emph{pursued through model behavior rather than sensitive-attribute supervision.}

%a central challenge in attaining the above-discussed \emph{goals} is that ethical fairness must be pursued through model behavior rather than demographic or sensitive attribute supervision. }

Guided by these imperatives, this paper's \emph{goal} is to develop ethically fair ubiquitous health sensing that:
\begin{itemize}[labelwidth=!, labelindent=0pt, leftmargin=!, align=left]
  \item[\textbf{\emph{goal} i:}] maximize subgroup-wise performance (F1);

\item[\textbf{\emph{goal} ii:}] reduce disparities across subgroups arising from broader heterogeneity factors, without relying on demographic or other sensitive attributes during training;
  
  \item[\textbf{\emph{goal} iii:}] avoid degradation, so that no subgroup performs worse than a baseline.

\end{itemize}

%A central challenge in attaining this \emph{goal}, as discussed above, is that {the information on demographics or sensitive attributes may be unavailable, or undesirable to collect in these sensing context due to privacy, regulatory, or ethical concerns \cite{baron2020where, pessach2023algorithmic,friedler2021possibility,stopczynski2014privacy,andrus2022demographic}}. Fairness, therefore, should be pursued through model behavior rather than demographic supervision. 

%We address these \emph{goals} through optimization geometry: the curvature of the loss landscape is tightly linked to robustness, i.e., stability and subgroup disparities. Loss-landscape sharp regions—captured by the high Hessian—are associated with parameters that are highly sensitive to perturbations; subgroups concentrated near such regions tend to lie closer to decision boundaries and experience less stable performance than others \cite{dauphin2024neglected,jastrzkebski2018relation,tran2022pruning}. Because computing full Hessians is prohibitive at scale, following the literature \cite{martens2020naturalgrad,yu2022combinatorial,lee2022masking,thomas_interplay_2020}, we adopt Fisher Information as a tractable surrogate that preserves key curvature properties.

We address these \emph{goals} through optimization geometry. Samples in high-curvature regions of the loss landscape, as captured by the Hessian, tend to lie near decision boundaries and receive less stable predictions. Subgroups concentrated in these regions may therefore be systematically underserved \cite{dauphin2024neglected,jastrzkebski2018relation,tran2022pruning}. Because computing the Hessian is expensive, we use the Fisher information as a tractable proxy for curvature (Section~\ref{sec:design}) \cite{yu2022combinatorial,lee2022masking,martens2020naturalgrad,thomas_interplay_2020}. This proxy allows us both to flatten the loss landscape and to identify samples the model treats differently.

Thus, \emph{to attain the above-discussed goals,} this paper introduces \textbf{\textsc{Flare}}—Fisher-guided LAtent-subgroup learning with do-no-harm REgularization. 
\textbf{\textsc{Flare}} is a demographic- and other sensitive-attribute-agnostic framework that uses Fisher-informed curvature regularization to learn smoother loss surfaces and more stable decision boundaries. By jointly analyzing embedding similarity, prediction loss, and curvature response, \textbf{\textsc{Flare}} discovers latent subgroups where model behavior diverges, exposing disparities without demographic or heterogeneous information. It then refines model behavior by training latent subgroup-specific extensions through conditional aggregation, enhancing inter- and intra- subgroup performance and fairness, while discouraging degradation of any subgroup as fairness improves. This design operationalizes \emph{beneficence} \textit{(goal-i)} and \emph{justice} \textit{(goal-ii)} by improving collective performance and equity, and \emph{non-maleficence} \textit{(goal-iii)} through a do-no-harm loss and validation-based model selection that guard against subgroup performance degradation. The result is an ethically grounded learning framework for ubiquitous human-sensing systems that balances efficacy, fairness, and stability under real-world heterogeneity.
This work’s primary contributions, embodied in \textbf{\textsc{Flare}}, are:

%{Through Fisher-informed curvature regularization, \textbf{\textsc{Flare}} learns smoother loss surfaces and more stable decision boundaries, reducing sensitivity to perturbations. By jointly analyzing embedding similarity, prediction loss, and curvature response, \textbf{\textsc{Flare}} discovers latent subgroups and pinpoints where model behavior diverges—exposing disparities even in the absence of demographic \textcolor{teal}{or sensitive information} and advancing the ethical principle of \emph{justice} \textit{(goal ii)}. \textbf{\textsc{Flare}} refines model behavior by training latent subgroup-specific extensions through a conditional aggregation, enhancing inter- and intra- subgroup performance and fairness, while ensuring that no subgroup deteriorates as fairness improves. This design operationalizes \emph{beneficence} \textit{(goal-i)} and \emph{justice} \textit{(goal-ii)} by improving collective performance and equity, and \emph{non-maleficence} \textit{(goal-iii)} by safeguarding against \textcolor{teal}{model performance degradation for any } subgroup}
%\textcolor{teal}{The result is an ethically grounded learning framework for ubiquitous human-sensing systems that balances efficacy, fairness, and stability under real-world heterogeneity.}

\begin{itemize}
    \item \textbf{Ethical fairness without demographic or other sensitive attributes for ubiquitous sensing:} It introduces \textbf{\textsc{Flare}}, to our knowledge the first framework that operationalizes the clinical-ethics principles of beneficence, non-maleficence, and justice for fairness without demographic or other sensitive attributes. By integrating \emph{model representation–behavior fusion}—combining embeddings, cross-entropy loss, and Fisher Information—\textbf{\textsc{Flare}} uncovers latent performance strata, identifies underserved subgroups, and promotes ethically constrained fairness across known and latent heterogeneity factors.
    
    %them, all while remaining agnostic to demographic or other {heterogeneity factors, i.e.,} sensitive attributes.
    
    \item \textbf{Do-no-harm adaptation:} It presents a novel do-no-harm training mechanism, combining a one-sided loss with validation-based model selection, that improves underserved or underperforming subgroups while guarding against degradation of others, operationalizing \emph{beneficence} (improving outcomes) and \emph{non-maleficence} (avoiding harm).
    
    \item \textbf{Ethics-grounded evaluation metrics:} 
    To address the lack of principled evaluation tools, it introduces the \emph{BHE} (Benefit, Harm-Avoidance, Equity), a new metric suite, which directly ties empirical model behavior to core AI ethics principles by quantifying collective improvement, subgroup protection, and reduction in performance disparity.
    
\item \textbf{Comprehensive Empirical Validation for Ubiquitous Sensing Systems} Through extensive evaluation across physiological (EDA), behavioral (IHS), and clinical (OhioT1DM , and Percept-R) datasets, \textbf{\textsc{Flare}} demonstrates that fairness, constrained by ethics, can be seamlessly integrated into real-world ubiquitous and  mobile behavioral and physiological sensing systems. These experiments confirm that \textbf{\textsc{Flare}} achieves its above-mentioned goals: \textit{(goal-i)} improved average subgroup-wise F1 for every sensitive attribute (Section~\ref{sec:baselines-vs-flare}), \textit{(goal-ii)} reduced disparities, thus enhanced equity across subgroups (Section~\ref{sec:baselines-vs-flare}), while \textit{(goal-iii)} no evaluated subgroup degraded, either relative to \textbf{\textsc{Flare}}'s intermediate pretrained model (Section~\ref{donot_harm}) or relative to the benign baseline (Section~\ref{sec:baselines-vs-flare}). 
Loss-landscape (Section~\ref{sec:loss-landscapes}), cluster-sensitivity (Section~\ref{sec:cluster-sensitivity}), and ablation (Section~\ref{sec:network-ablation}) analyses support its design, post-hoc profiling (Section~\ref{sec:cluster_interpretation}) characterizes the discovered latent subgroups, and runtime evaluations on diverse edge platforms (Section~\ref{on-device}) show that it remains feasible for on-device deployment.

%{Ablation, rule-based interpretability, and loss-landscape analyses validate its principled design (Section~\ref{sec:validation-design}), while runtime evaluations on diverse edge platforms (Section \ref{on-device}) confirm its scalability and readiness for trustworthy, equitable deployment.}
% Ablation and loss-landscape analyses validate its principled design 

\end{itemize}

% \textcolor{red}{I want to add this in intro later}Ensuring fairness across demographic subgroups has emerged as a central challenge in ethical AI \cite{singh2021ai, venkatasubbu2022ethical, chinta2024fairaied}. Although modern machine learning models often achieve strong overall accuracy, they can still produce uneven outcomes that disadvantage specific populations, raising concerns about both equity and trustworthiness \cite{barocas2023fairness,rajkomar2018ensuring}. These disparities risk reinforcing-or even amplifying - existing societal inequities, motivating a growing body of work on fairness in algorithmic decision making \cite{corbett-davies2018measure}. 

\section{Related Work}
This section reviews related work on fairness and ethical AI in ubiquitous health sensing, covering demographic-based methods, fairness without sensitive attributes, subgroup-oriented techniques, and ethics-aligned approaches.

\subsection{Fairness through Demographic Information}
\label{sec:demo_rel_work}

Early algorithmic fairness methods used demographic or heterogeneous attributes to constrain model predictions \cite{pessach2023algorithmic,andrus2022demographic,giguere2022fairness}. These include preprocessing methods such as reweighting or relabeling \cite{kamiran2012preprocessing}, in-processing methods that add fairness constraints to the optimization objective \cite{zafar2017fairness,zafar2019jmlr}, and post-processing methods that adjust decision thresholds to satisfy criteria such as demographic parity or equalized odds \cite{hardt2016equality,xian2023fair,mishler2021fairness}. While these approaches established the foundation of fairness research, they often trade off with accuracy, leave biased representations unchanged, or become unstable in deployment \cite{hardt2016equality,friedler2019comparative}.

In ubiquitous and wearable sensing, demographic biases related to sex, ethnicity, and age can persist across the ML pipeline, from data collection to model outputs, even after preprocessing \cite{uncoveringbias}. In-processing methods address such bias by adding fairness constraints directly to the optimization objective \cite{zafar2017fairness,zafar2019jmlr}, but often trade off with accuracy. In wearable federated learning, fairness-aware objectives have also been used to reduce sensitive-attribute bias and participation inequity across heterogeneous clients \cite{djebrouni2024bias,zhou2022leftout}. However, demographic-based approaches in ubiquitous sensing systems remain limited because sensitive attributes are often unavailable, privacy-sensitive, or ethically undesirable to collect in mobile and wearable sensing deployments \cite{pessach2023algorithmic,friedler2021possibility,yfantidou2023beyond}.

\subsection{Fairness without Demographics (FWD) in Mobile Behavioral and Physiological Sensing}\label{literature-FWD}

Mobile behavioral and physiological sensing studies often infer health and well-being outcomes without using demographic information during prediction \cite{wang2018tracking,xu2023globem,ni2024fairness}. However, recent work shows that ubiquitous sensing models have inherent bias and uneven reliability across users, datasets, and populations \cite{adler2024measuring,suresh2019framework,yfantidou2023beyond,meegahapola2023generalization}. Although some studies incorporate demographic variables to mitigate these issues \cite{baker2023using}, integrating demographic information alone may not resolve inequity, as subgroups can also emerge from behavioral, contextual, physiological, and device-level heterogeneity \cite{adler2024measuring}. Moreover, many mobile sensing datasets lack demographic annotations or provide limited information about population heterogeneity \cite{yfantidou2023beyond,xu2023globem}. These challenges motivate fairness without demographics, aiming to improve equity without requiring protected attributes during training.

Several FWD methods have been proposed to address this challenge. Adversarial reweighting emphasizes underperforming or misclassified samples to reduce hidden subgroup disparities \cite{lahoti2020fairness}, but can destabilize optimization and reduce overall accuracy. Fairness-aware knowledge distillation transfers subgroup performance improvements from a teacher to a student model without demographic labels \cite{chai2022fairness}, but may propagate teacher bias and underperform on underrepresented groups. \textit{Reckoner} partitions data into confidence-based subsets and trains dual models with pseudo-label exchange and learnable noise to improve fairness-accuracy trade-offs \cite{ni2024fairness}, but its performance can depend on threshold selection and initialization. Graph of Gradients (GoG) uses last-layer gradients to construct local gradient neighborhoods and perform Rawlsian adversarial reweighting \cite{luo2025fairness}, but depends on stable $k$-NN graph construction and shallow GCN expressivity, making it sensitive to noise and over-smoothing \cite{kang2021k,huang2020tackling}. 
CurvFed \cite{sharma2026curvfed} likewise uses curvature to attain FWD, but for federated learning, without latent-subgroup adaptation or ethical grounding; because its client--server setting differs from ours, we do not include it as a baseline.
In our experiments, we use representative centralized FWD baselines from these families, including adversarial reweighting, knowledge distillation, Reckoner, and GoG.

While these methods reduce reliance on demographic labels, they primarily target disparity reduction and often apply global or proxy-based fairness corrections. In heterogeneous mobile sensing settings, such corrections can miss local model-behavior differences, complicate optimization, and degrade performance for some latent subgroups. \textbf{\textsc{Flare}} addresses these limitations by unifying three complementary components: Fisher-based curvature regularization for stable optimization geometry, do-no-harm adaptation to preserve subgroup performance, and BHE-driven evaluation grounded in ethical principles. This structured and targeted approach simplifies optimization, guides fairness adjustments more effectively, and promotes stable learning, particularly in domains such as mobile behavioral and physiological sensing. Group-specific modeling itself is an established personalization strategy in mobile sensing, where separate or multitask models are trained for groups of users with similar behavior \cite{taylor2020personalized}; \textbf{\textsc{Flare}} differs in that it discovers groups at the sample level from model behavior rather than from user identity, and constrains group-specific adaptation with do-no-harm, validation-based selection.

%{Although these methods make progress toward fairness without relying on demographic information, they typically address these challenges in isolation. Flare distinguishes itself by unifying three complementary components: Fisher-based curvature regularization for improved optimization geometry, do-no-harm adaptation to preserve subgroup performance, and BHE-driven evaluation grounded in ethical considerations. This structured and targeted approach simplifies optimization, guides fairness adjustments more effectively, and promotes stable learning, particularly in domains such as mobile behavioral and physiological sensing.}

% Given the relevance and demonstrated impact, we adopt adversarial reweighting (ARL), knowledge distillation (KD), Graph of Gradients (GoG), and \textit{Reckoner} as our baselines for comparison in this study.

\subsection{Subgroup-oriented Fairness}\label{subgroup-DRO}

%Beyond methods that rely on explicit demographics or avoid them entirely, a growing line of work targets subgroup-oriented fairness by enforcing equity across diverse population partitions. However, Kearns et al. \cite{kearns2018gerrymandering} showed that optimizing fairness for a small set of protected groups can yield ``fairness gerrymandering,'' where other subgroups suffer large disparities. Hébert-Johnson et al.\ \cite{hebert-johnson2018multicalibration} addressed this via multicalibration, enforcing predictive calibration across broad subgroup families. 
Beyond demographic-based and fairness-without-demographics methods, subgroup-oriented fairness seeks to improve equity across diverse population partitions. Kearns et al.~\cite{kearns2018gerrymandering} showed that optimizing fairness only for a small set of protected groups can lead to ``fairness gerrymandering,'' where other subgroups still experience large disparities. Multicalibration addresses this by enforcing calibration across broad subgroup families \cite{hebert-johnson2018multicalibration}.

Related work uses distributionally robust optimization (DRO) to minimize worst-case subgroup error. Group DRO focuses training on groups with the highest loss \cite{sagawa2019distributionally}, while Just Train Twice (JTT) identifies high-loss or misclassified samples as latent subgroups and upweights them during retraining \cite{liu2021jtt}. 
Additionally, Sohoni et al.~\cite{sohoni2020no} studied \emph{hidden stratification}, where one class label can contain multiple unlabeled subclasses with different difficulty levels. They cluster samples within each class and apply Group-DRO to improve worst-case subclass accuracy, making the approach class-conditional and focused on within-class robustness.

\textbf{Why subgroup-focused DRO methods are insufficient:}
Despite their gains, DRO, JTT, and hidden-stratification methods have two key limitations. First, they assume that the relevant subgroup structure is either known or can be recovered using model-efficacy-based proxies. Group DRO requires explicit group annotations \cite{sagawa2019distributionally}, while JTT uses high-loss or misclassified samples as proxies for latent subgroups \cite{liu2021jtt}. Unlike \textbf{\textsc{Flare}}, these proxy-based approaches are not grounded in optimization geometry or learning dynamics. Hidden-stratification methods assume that disparities correspond to subclasses nested within each class \cite{sohoni2020no}. These assumptions can hold when datasets follow a clear superclass-subclass hierarchy. However, in ubiquitous computing and human sensing, inequities are often latent, heterogeneous, and shaped by complex physiological, behavioral, contextual, or environmental variation that may not follow traditional class boundaries \cite{adler2024measuring}, making class-dependent discovery insufficient. Second, these methods often focus on the identified worst-performing group, which may reduce extreme disparities but does not guarantee balanced performance across the remaining population. Thus, improvements in aggregate or worst-group metrics do not necessarily imply ethically grounded fairness; strongly regularized Group DRO can even reduce average accuracy \cite{sagawa2019distributionally}. \par Closest to our goal of avoiding harm, Ustun et al.~\cite{ustun2019fairness} train decoupled, group-specific classifiers with guarantees that no group is worse off than under a shared model, but they require group labels at training and deployment time. Blind Pareto Fairness~\cite{martinez2021blind}, extending Minimax Pareto Fairness~\cite{martinez2020minimax}, removes the need for group labels by training a single model to minimize the worst-case risk over all subgroups above a chosen minimum size. Its objective also accounts for performance on the remaining population, but it leaves subgroups smaller than the chosen size outside the guarantee, and Pareto efficiency only rules out alternatives that improve some subgroup without worsening any other; it does not by itself guarantee that every subgroup improves relative to a specified reference model, such as the model being replaced. \textbf{\textsc{Flare}} instead discovers latent subgroups from model behavior without any attribute, adapts a separate model to each, and penalizes any increase in loss relative to the pretrained model, so that improving one subgroup does not come at the expense of another.

\subsection{Ethical AI in healthcare and aligning with human-centered modeling}
\label{ethical_princ}
 Ethical AI in healthcare centers on four principles: \emph{beneficence} (promote patient well-being), \emph{non-maleficence} (avoid foreseeable harm), \emph{justice} (ensure equitable treatment across groups), and \emph{autonomy} (preserve individuals’ freedom of choice) \cite{hurley2003fairness,andersson2010no,gabriel2022toward}. 
These principles are echoed across international AI ethics guidelines and form a shared foundation for trustworthy AI \cite{jobin2019global,floridi2019unified}. 
However, for ubiquitous and mobile health systems, principles alone are insufficient—ethical principles must be translated into concrete design and evaluation choices \cite{mittelstadt2019principles}.
 %These principles appear across international guidelines and are proposed as a common foundation for trustworthy AI \cite{jobin2019global,floridi2019unified}. To be actionable, principles alone are not enough; they must be turned into concrete design and evaluation steps for real systems \cite{mittelstadt2019principles}.
This paper focuses on the first three principles that are directly tied to fairness in ubiquitous health sensing: First, \emph{Justice}, which requires that algorithms perform equitably, avoiding systematic disadvantages to specific subgroups \cite{jobin2019global}; Second, \emph{Beneficence}, that emphasizes the obligation to design and implement systems that actively promote overall well-being and deliver measurable improvements \cite{hurley2003fairness}; and finally, \emph{Non-maleficence}, that focuses on minimizing risks and preventing harm, supported by frameworks that identify and mitigate potential harms to any subgroups throughout the machine learning lifecycle \cite{suresh2019framework}.

These principles anchor ethical AI in ubiquitous health computing. We operationalize these principles as model-level objectives and evaluation criteria, while leaving \emph{autonomy} outside the scope because \textbf{\textsc{Flare}} focuses on predictive model behavior rather than user interaction, consent, or decision-support interfaces.

%} \emph{autonomy} is out of scope since our {framework} provide predictions without direct user interaction or consent mechanisms}.

\subsection{Ethical AI vs fairness}
Fairness is a central concern in machine learning, but it is only one dimension of ethical AI \cite{giovanola2023beyond}. It is commonly defined as equitable treatment across individuals, groups, or sensitive attributes, and is typically measured using statistical disparity metrics \cite{hardt2016equality,friedler2019comparative}. These metrics mainly reflect \emph{justice}, by aiming to prevent disproportionate disadvantage.
However, ethical AI requires more than disparity reduction \cite{singh2021ai,venkatasubbu2022ethical,chinta2024fairaied}. Principles such as \emph{beneficence} and \emph{non-maleficence} require models to improve overall outcomes while avoiding harm for any subgroup \cite{andersson2010no}. This distinction is especially important in healthcare, where fair predictions must also provide benefit and prevent subgroup-level degradation \cite{jobin2019global}.

\emph{To the best of our knowledge, no prior work has jointly operationalized fairness and core ethical principles in developing models that infer health and well-being from mobile behavioral and physiological sensing data.} This paper directly tackles the above-discussed gaps—advancing the goal of ethical fairness in human-centered AI for ubiquitous sensing systems. In the absence of established ethical fairness baselines, we consider state-of-the-art (SoTA) fairness-without-demographics (FWD) approaches as baselines. 

%\textit{To the best of our knowledge, no existing work has jointly combined these ethical principles alongside fairness to evaluate the ethical and fair behavior of AI models.}

%This distinction is particularly salient in healthcare, where fairness is necessary but not sufficient \cite{jobin2019global}: predictive models must not only perform equitably across patients but also improve health outcomes, avoid harmful decisions, and support trust between clinicians and patients.  Ensuring subgroup parity alone does not guarantee that models improve patient care or safeguard against harmful recommendations. Thus, while fairness methods make progress toward ethical AI, they fall short of fully realizing it-underscoring the need for evaluation frameworks that operationalize ethical principles alongside fairness.

\section{Fairness Metrics}
Existing AI fairness metrics do not fully capture ethical fairness \cite{beutel2019putting}. This section reviews common fairness metrics and introduces an ethics-grounded evaluation framework for ubiquitous sensing systems.
%Although fairness is a major focus in machine learning, no existing metric adequately captures the broader scope of ethical fairness \cite{beutel2019putting}. 
%This section reviews the commonly used fairness metrics and then introduces a new framework for comprehensive ethical fairness evaluation, aligning fairness assessment more closely with the core principles of ethical AI { applied to mobile and ubiquitous sensing systems \cite{sharma2022psychophysiological}}. %Finally, it outlines the problem statement for this paper, based on the new ethical fairness evaluation framework. 

%Although fairness has become a central concern in machine learning, existing metrics fall short of capturing the broader goals of ethical AI \cite{beutel2019putting}. This section outlines the evaluation metrics employed to assess ethical fairness in model performance and introduces a new framework that aligns fairness measurement with ethical principles.

%Fairness in machine learning is commonly evaluated through statistical criteria that quantify disparities in predictions across protected groups such as \emph{Demographic Parity (DP)}, \emph{Equal Opportunity Difference (EOD)}, and \emph{Average Odds Difference (AOD)} \cite{hardt2016equality, friedler2019comparative}.  

\subsection{Limitations of existing fairness metrics}

Fairness in machine learning is commonly evaluated using statistical disparity metrics across protected groups. Binary metrics such as Demographic Parity (DP) and Equal Opportunity Difference (EOD) compare positive prediction rates or true positive rates between groups \cite{hardt2016equality,friedler2019comparative,zhang2018equality}. However, these metrics mainly capture pairwise differences, such as male versus female, and do not fully reflect disparities across multiple subgroups. Multi-group extensions, including generalized EOD \cite{opoku2025unveiling} and Relative Disparity (RD) \cite{afrose2022subpopulation}, address this by aggregating gaps across subgroups. For example, RD compares best- and worst-performing subgroups, with values closer to 1 indicating greater equity. 
%Fairness in machine learning is typically assessed using statistical measures that quantify disparities in model outcomes across protected groups. Common binary metrics include Demographic Parity (DP)—which measures whether positive prediction rates are equal across subgroups—and Equal Opportunity Difference (EOD), which compares true positive rates to ensure equitable treatment \cite{hardt2016equality, friedler2019comparative, zhang2018equality}.
%However, these binary metrics capture only pairwise group differences (e.g., male vs. female) and fail to reflect disparities across multiple subgroups. To overcome this limitation, multi-group extensions such as generalized EOD \cite{opoku2025unveiling} and Relative Disparity (RD) \cite{afrose2022subpopulation} have been introduced. These metrics aggregate performance gaps across all subgroups to provide a broader view of fairness. For instance, RD measures the ratio between the best- and worst-performing subgroups (e.g., by recall or F1-score), with values closer to 1 indicating greater equity.
Despite these advances, these metrics still fall short in key ways. Metrics that rely solely on aggregated performance (e.g., overall accuracy or AUC) or extreme-value summaries (e.g., max–min ratios) can obscure systematic harm—masking whether certain subgroups consistently underperform \cite{kearns2018gerrymandering,sagawa2019distributionally,hashimoto2018fairness}. Thus, a model may satisfy statistical parity while still harming specific populations, conflicting with \emph{justice} and \emph{non-maleficence}.  
%A model may achieve parity in statistical terms yet still produce outcomes that disproportionately burden specific populations, conflicting with the ethical principles of \emph{justice} and \emph{non-maleficence}. 
Moreover, existing fairness metrics do not assess whether model improvements benefit all subgroups and the overall population. They equalize outcomes without ensuring collective efficacy, overlooking the ethical principle of \emph{beneficence}.

 % \textcolor{orange}{These are metrices... we need them for evaluation. Even BHE uses them. Should we say this about metrics? $\rightarrow$ In addition, these metrics rely on quantifying differences in model performance among protected attributes, often specifically demographic characteristics. However, demographic or heterogeneity factors information is typically not collected in mobile behavioral and physiological sensing systems that estimate health and well-being \cite{baron2020where, pessach2023algorithmic,friedler2021possibility,stopczynski2014privacy,andrus2022demographic}}

\subsection{BHE: Fairness Metrics Aligned with Bioethics}\label{BHE-metrics}

To bridge the gap between technical fairness and ethical accountability, we introduce the \textbf{BHE} metrics (\emph{Benefit}, \emph{Harm-Avoidance}, \emph{Equity}). BHE compares a fairness method $M$ (\textbf{\textsc{Flare}} or any fairness baseline) against the benign baseline (Section~\ref{benign}) as the reference point, using subgroup-wise F1-scores (in \%), since the sensing tasks are class-imbalanced. Subgroups are defined by the known sensitive attributes recorded in each dataset, both demographic (e.g., sex, age) and non-demographic (e.g., insulin-pump or sensor band, PHQ10 status). Subgroup F1 is computed on the test predictions pooled across all $K$ person-disjoint folds, so that every participant is evaluated exactly once\footnote{Except one OhioT1DM participant with no eligible test samples (Appendix~\ref{sec:data_stat}).}. We use sensitive attributes and test participants only for this final evaluation, never for training, tuning, clustering, or model selection (Section~\ref{data}).

For a sensitive attribute $A$ with subgroups $\mathcal{S}_A$, let $F1_s^{M}$ and $F1_s^{\mathrm{Base}}$ denote the F1-scores of method $M$ and the benign baseline on subgroup $s \in \mathcal{S}_A$, and let $\delta_s = F1_s^{M} - F1_s^{\mathrm{Base}}$. We define
\[
\Delta B_A = \frac{1}{|\mathcal{S}_A|}\sum_{s \in \mathcal{S}_A} \delta_s,
\qquad
\Delta H_A = \min_{s \in \mathcal{S}_A} \delta_s,
\qquad
\Delta E_A = \mathrm{SD}\big(\{F1_s^{\mathrm{Base}}\}_{s\in\mathcal{S}_A}\big) - \mathrm{SD}\big(\{F1_s^{M}\}_{s\in\mathcal{S}_A}\big),
\]
where $\mathrm{SD}$ is the sample standard deviation (with $|\mathcal{S}_A|-1$ in the denominator) across subgroups. All three are in percentage points. Per-attribute values are reported in Tables~\ref{tab:cluster_sensitivity_all} and~\ref{tab:ablation_bhe} and in Appendix~\ref{app:per_attr}. 
To summarize a dataset, we average each metric over the dataset's sensitive attributes; Table~\ref{tab:bhe_summary} and the \textit{Mean} rows of the other tables report these averages under the same symbols $\Delta B$, $\Delta H$, and $\Delta E$. Because the averaged $\Delta H$ is a mean of per-attribute minima rather than a minimum over all subgroups, a non-negative average alone does not show that no subgroup is harmed; we make that claim only when $\Delta H_A \geq 0$ for every attribute $A$. In the following, we drop the subscript $A$ when the attribute is clear from context.

\begin{itemize}
    \item \textbf{Benefit ($\Delta B$)} is the change in average subgroup performance and reflects \emph{beneficence}, the obligation to promote overall benefit \cite{hurley2003fairness}.

    \item \textbf{Harm-Avoidance ($\Delta H$)} is the change for the \emph{most-harmed} subgroup, i.e., the subgroup whose F1 changes least or decreases most, and reflects \emph{non-maleficence}, the obligation to avoid harm \cite{andersson2010no}. A negative $\Delta H_A$ means that at least one subgroup of $A$ is worse off than under the baseline; $\Delta H_A \geq 0$ means that no evaluated subgroup of $A$ is degraded. Because $\Delta H$ is computed on finite test data and only for observed attributes, it provides evidence of, rather than a guarantee of, harm avoidance.

    \item \textbf{Equity ($\Delta E$)} is the reduction in the standard deviation of subgroup F1 and reflects \emph{justice}, the obligation to avoid systematic advantage or disadvantage \cite{jobin2019global}. Unlike max--min summaries such as RD, the standard deviation uses every subgroup and is therefore less sensitive to a single extreme subgroup.
\end{itemize}

For all three metrics, larger positive values are ethically preferable: $\Delta B$, $\Delta H$, and $\Delta E$ correspond to \emph{goals-i}, \emph{-iii}, and \emph{-ii} in Section~\ref{sec:intro}, respectively. The three must be read jointly. A method can increase $\Delta E$ by lowering the better-performing subgroups (leveling down), which $\Delta H<0$ exposes, or increase $\Delta B$ while widening disparity, which $\Delta E<0$ exposes; parity metrics alone detect neither.

Importantly, evaluating on known attributes is a proxy for the broader goal: a method trained without sensitive attributes that improves BHE on attributes it never saw provides evidence, though not a guarantee, of ethical fairness for unobserved heterogeneity factors. Within-person strata that no attribute captures are examined separately through cluster-wise changes in F1 (Section~\ref{donot_harm}).

\section{The \textsc{Flare} Framework}
\label{sec:design}

\begin{figure}[h]
    \centering
    \includegraphics[width=0.9\textwidth]{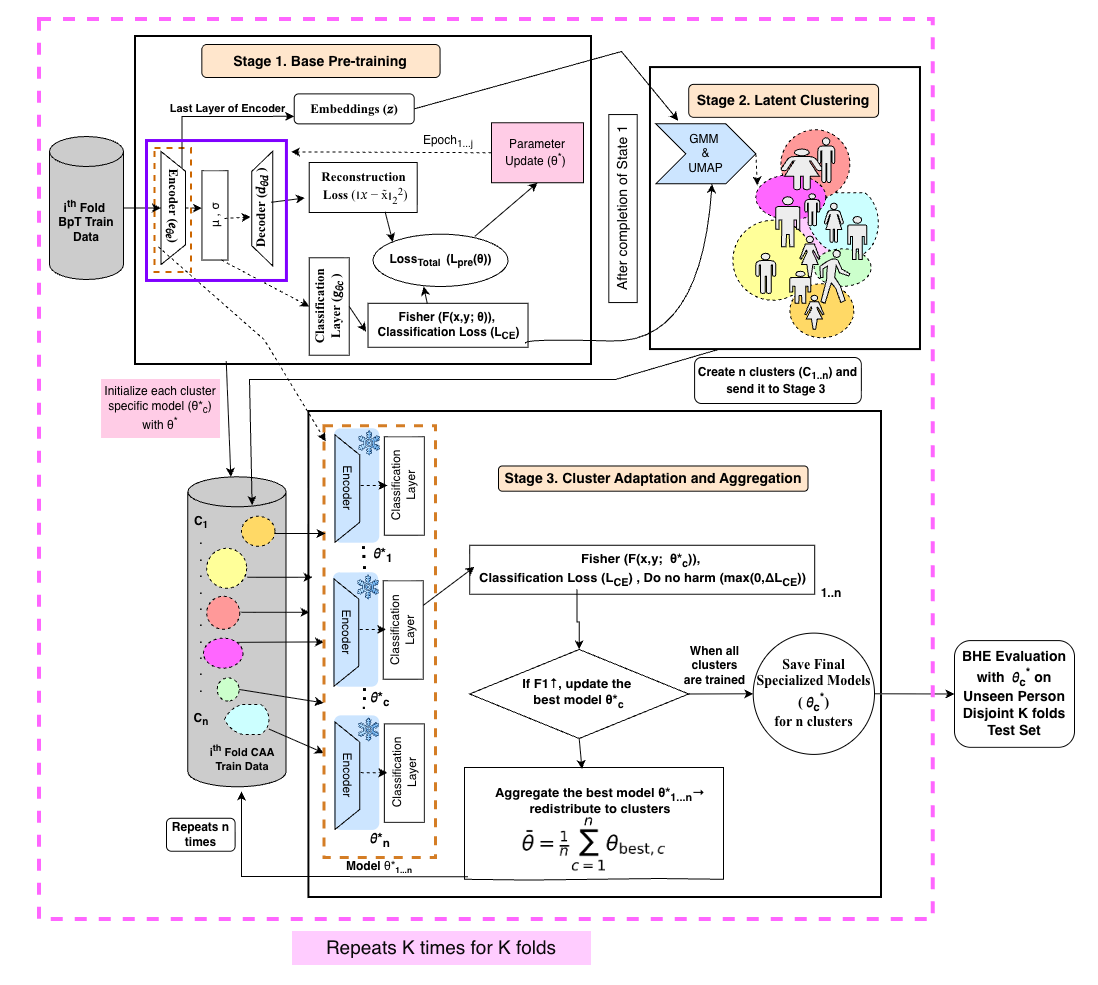}
    \vspace{-1.8em}
\caption{\textbf{Three-stage architecture of \textsc{Flare}.}
\textit{Stage 1:} Multi-task pretraining learns embeddings $z$ and model $\theta^*$.
\textit{Stage 2:} UMAP and GMM cluster descriptors $S=[z,L_{\mathrm{CE}},\mathsf{F}]$. Routing uses predicted labels $\hat{y}$, requiring no test labels.
\textit{Stage 3:} Cluster models initialized from $\theta^*$ are fine-tuned on \texttt{CAA-train}. Periodically, their best checkpoints are averaged into $\bar{\theta}$; each cluster adopts it only if its \texttt{val} F1 improves. Final models are evaluated once on held-out test participants.}
    \label{fig:approach}
\vspace{-15pt}
\end{figure}

\textbf{Rationale.} Sharp regions of the loss landscape, characterized by a large top eigenvalue of the Hessian, indicate high sensitivity to parameter perturbations and unstable generalization \cite{dauphin2024neglected, jastrzkebski2018relation}; subgroups associated with such regions tend to lie closer to the decision boundary and experience less stable, less accurate predictions, so curvature disparities can serve as a proxy for fairness gaps \cite{tran2022pruning}. Because the full Hessian is costly to compute, following prior work \cite{yu2022combinatorial,lee2022masking,thomas_interplay_2020}, we use the Fisher information matrix (FIM), which under standard regularity conditions coincides with the expected Hessian of the negative log-likelihood under the model's predictive distribution \cite{martens2020naturalgrad}, as a tractable curvature surrogate. A lower top FIM eigenvalue, which we approximate by the maximum diagonal entry of the diagonal FIM (Section~\ref{Step-1-design-choice}), reflects flatter minima and more stable decisions \cite{yu2025eigenstructure,liu2023fun}, and more uniform curvature across subgroups reflects more consistent decision boundaries across them. Because the top eigenvalue of the full FIM is itself costly to compute, \textbf{\textsc{Flare}} uses a diagonal FIM approximation, whose top eigenvalue is simply its largest diagonal entry (Section~\ref{Step-1-design-choice}). \textbf{\textsc{Flare}} uses this signal twice: as a regularizer that flattens the loss landscape (\emph{justice}), and, together with embeddings and loss, as a descriptor for discovering latent subgroups that the model treats differently, which are then improved by targeted adaptation that guards against degradation (\emph{beneficence} and \emph{non-maleficence}). Figure~\ref{fig:approach} shows its three stages; Section~\ref{sec:implementation} summarizes the procedure as algorithms.

\iffalse
% Particularly, to align design with AI ethical fairness objectives, our pipeline embodies three design principles:

% \begin{enumerate}
%     \item \textbf{Preserve generalizable representations.}
%     {Discuss why encoder-decoder structure, add citations}
%     We pretrain an encoder–decoder classifier with a composite loss that balances reconstruction fidelity, supervised accuracy, and curvature regularization via a per-sample FIM proxy $\mathsf{F}(x,y;\theta)$. This yields information-rich, stable, or generalizable embeddings $z$ suitable for downstream subgroup discovery.

%     \item \textbf{Specialize for latent heterogeneity.}
%     We form joint descriptors $s=[z,\;\ell^{\text{CE}},\;\mathsf{F}]$ that couple representation ($z$) with sample difficulty (cross-entropy) and parameter sensitivity (Fisher). UMAP reduces $s$ and a GMM partitions samples into behaviorally coherent clusters, enabling targeted adaptation without demographic attributes.

%     \item \textbf{Stabilize adaptation across clusters.}
%     Cluster-specific fine-tuning employs a \emph{do-no-harm} regularizer that prevents regressions relative to the pretrained baseline and applies Fisher only on correctly classified points to regularize reliable decision regions. Periodic, opt-in parameter aggregation transfers shared signal while guarding against harmful drift.
% \end{enumerate}

% These choices operationalize our fairness objective through optimization geometry: mitigate curvature imbalances, respect subgroup stability, and retain predictive utility.
\fi

\subsection{Stage 1: Base Pretraining (BpT)}\label{Step-1-design-choice}  

As shown in Figure~\ref{fig:approach} (Stage~1), the first stage involves training an encoder–decoder–classifier framework $(e_{\theta_e}, d_{\theta_d}, g_{\theta_g})$ to learn model-behavior-aware comprehensive representations of the data. Given an input sample $x$ with ground-truth label $y$, the encoder $e_{\theta_e}$ maps the input into a latent representation $z = e_{\theta_e}(x)$. 
The classifier $g_{\theta_g}$, with $\theta=(\theta_e,\theta_d,\theta_g)$, outputs class probabilities $p_\theta(\cdot\mid x)$ from the latent code, and the predicted label is $\hat{y}=\arg\max_k p_\theta(k\mid x)$, while the decoder $d_{\theta_d}$ reconstructs the input as $\tilde{x} = d_{\theta_d}(z)$.
%The classifier $g_{\theta_c}$ produces a label prediction $\hat{y} = g_{\theta_c}(z)$ from the latent code, while the decoder $d_{\theta_d}$ reconstructs the input as $\tilde{x} = d_{\theta_d}(z)$.  
The network is trained using a joint objective that balances reconstruction, classification, and stability:
\begin{equation}
L_{\text{pre}}(\theta) 
= (1-\alpha)\,\|x-\tilde{x}\|_2^2
+ \alpha \Big( 
   \beta \,L_{\text{CE}}(\hat{y}, y) 
   + (1-\beta)\,\mathsf{F}(x,y;\theta)\,\mathbb{I}\{\hat{y}=y\} 
   \Big).
\label{eq:lpre}
\end{equation}

Here, the first term, $\|x-\tilde{x}\|_2^2$, is the mean-squared error (MSE) reconstruction loss, encouraging the latent representation $z$ to retain as much information from $x$ as possible. $L_{\text{CE}}(\hat{y}, y)=-\log p_\theta(y\mid x)$ (with slight abuse of notation, $\hat{y}$ here stands for the predicted distribution $p_\theta(\cdot\mid x)$) is the cross-entropy loss of the prediction with respect to the true label $y$, which drives the model to make accurate classifications. 
And, $\mathsf{F}(x,y;\theta)$ denotes the Fisher penalty. Because the full FIM is too large to compute for every sample, we use its diagonal approximation, a standard choice for neural networks \cite{kirkpatrick2017overcoming,martens2020naturalgrad}, and take its maximum diagonal entry, which is the top eigenvalue of that approximation:
\begin{equation}
\mathsf{F}(x,y;\theta)=\max_{j}\,\big[\operatorname{diag}\!\big(g\,g^{\top}\big)\big]_j=\max_{j}\,g_j^{2},\qquad g=\nabla_{\theta}\log p_{\theta}(y\mid x),
\label{eq:fisher}
\end{equation}
where $j$ indexes the model parameters. It captures the sharpest parameter direction of the loss around a sample \cite{li2025fishers} and requires only one backward pass to compute.
%And, $\mathsf{F}(x,y;\theta)$ denotes Fisher penalty, i.e., promoting reduction of the top eigenvalue of the FIM, which measures the sensitivity of the model’s loss around correctly classified samples \cite{li2025fishers}. 
Minimizing this term reduces the local curvature of the loss landscape, encouraging the model to converge toward flatter minima, which correspond to more stable generalization behavior. $\mathbb{I}\{\hat{y}=y\}$ is an indicator function that equals $1$ if the sample is classified correctly and $0$ otherwise, ensuring that Fisher penalty regularization is applied only to correctly predicted samples. This training avoids penalizing the curvature of misclassified samples, since they are already penalized to encourage crossing the decision boundaries via the $L_{\text{CE}}(\hat{y}, y)$ loss. $\alpha \in [0,1]$ balances reconstruction against supervised objectives. $\beta \in [0,1]$ controls the trade-off between classification accuracy (cross-entropy) and stability (Fisher penalty) within the supervised component. %\textcolor{red}{add a line saying these two hyperparameters are found through grid search. Are these values universal across all datasets? If yes, say so, like the other paper.}

Minimizing the Fisher penalty across samples lowers its overall magnitude, which is intended to align curvature more evenly across the population and thereby promoting more consistent predictions (\emph{justice}) without access to sensitive attributes. We write $\mathsf{F}_{\mathrm{corr}}(x,y;\theta)=\mathsf{F}(x,y;\theta)\,\mathbb{I}\{\hat{y}=y\}$ for the Fisher penalty restricted to correctly classified samples. After each epoch, the model is evaluated on the validation split \texttt{val} (Section~\ref{data}), and the checkpoint with the best validation F1 is retained as $\theta^\ast$; no test data are used for this selection.

%it imprboth oves the stability and equity of the base-pretrained model $\theta^\ast$'s predictions without requiring access to sensitive demographic attributes, thereby enhancing \emph{justice} and \emph{autonomy}.
%}

However, such uniform, global regularization cannot capture how the model treats different subgroups (Section~\ref{literature-FWD}) \cite{lahoti2020fairness, chai2022fairness}. The pretrained model $\theta^\ast$ therefore also provides three per-sample behavior signals for Stage~2: the embedding $z$, indicating which samples share representational structure and are thus likely to be decided by similar rules; the cross-entropy loss, indicating how confidently a sample is predicted; and the Fisher penalty, indicating how sensitive the prediction is to perturbation, a proxy for its proximity to the decision boundary.

% encoder-decoder structure promotes identification to meaningful embeddings
% Intuitively, the reconstruction term ensures \emph{autonomy} by producing embeddings that are information-rich and interpretable, the cross-entropy term enforces \emph{beneficence} by maximizing predictive utility, and the Fisher penalty enforces \emph{non-maleficence} by discouraging brittle or overly sharp solutions. Together, these elements yield stable, generalizable embeddings that are well-suited for subsequent subgroup discovery.

% In doing so, clustering operationalizes \emph{justice} by treating subgroups defined by behavioral similarity, ensuring that disparities are surfaced and addressed.
\subsection{Stage 2: Latent Clustering}\label{design-clustering}
\begin{algorithm}[!t]
\caption{Stages 1–2: Base Pretraining and Latent Clustering}
\label{algo1}
\begin{algorithmic}[1]
%\Require Fold $k\!\in\!\{1,\dots,K\}$ with \texttt{BpT-train + CAA-train} (80\%/20\%) and person-disjoint \texttt{test}; hyperparameters $\alpha,\beta,\eta,E,C$.
\Require Fold $k\in\{1,\dots,K\}$ \textcolor{black}{with \texttt{BpT-train}, \texttt{CAA-train}, and \texttt{val} (60\%/20\%/20\%) and held-out}, person-disjoint \texttt{test}; hyperparameters $\alpha,\beta,\eta,E$.
\State Initialize encoder–decoder–classifier $M_\theta$.
\For{$\text{epoch}=1$ to $E$}
  \For{each batch $(x,y)\!\in\!\texttt{BpT-train}$}
    \State $z \leftarrow e_{\theta}(x)$; $\tilde{x} \leftarrow d_{\theta}(z)$; $\hat{y} \leftarrow g_{\theta}(z)$\Comment{Forward pass: embeddings, reconstruction, predictions}
    \State Compute Fisher penalty $\mathsf{F}_{\mathrm{corr}}$ for correctly classified samples only.
    \State $L_{\text{pre}}\!\gets\!(1-\alpha)\|x-\tilde{x}\|_2^2+\alpha[\beta L_{\text{CE}}(\hat{y},y)+(1-\beta)\mathsf{F}_{\mathrm{corr}}]$
    \State Update $\theta$ by taking a gradient step on $L_{\text{pre}}$ with learning rate $\eta$.
  \EndFor
  %\State Evaluate on \texttt{test}; save $\theta^\ast$ if validation improves.
\EndFor
%\State {Final trained weights $\theta^\ast$.} 
\State $\theta^\ast \leftarrow$ checkpoint (epoch) \textcolor{black}{with the best F1 on \texttt{val}.} 
%\State {\textbf{Feature Extraction from \texttt{BpT-train} set:} Using $\theta^\ast$, compute $(z,L_{\mathrm{CE}},\mathsf{F})$ for all samples.}
\State \textbf{Feature extraction:} using $\theta^\ast$, compute $(z, L_{\mathrm{CE}}, \mathsf{F})$ for all \texttt{BpT-train} samples.
\State Form joint descriptors $S=[z,L_{\mathrm{CE}},\mathsf{F}]$ and get $S_{\text{corr}}$ (be those of \texttt{BpT-train} samples correctly classified by $\theta^\ast$) 

%\State {\textbf{Clustering:} Fit UMAP $\varphi$ on $S_{\text{corr}}$; train GMM($n$) on $\varphi(S_{\text{corr}})$.}

\State \textbf{Clustering:} fit UMAP $\varphi$ on $S_{\text{corr}}$; fit GMM on $\varphi(S_{\text{corr}})$, with $n$ chosen by BIC on \texttt{BpT-train} (reduced until no cluster is empty).

%\State {Assign cluster IDs to all \texttt{CAA-train} samples using GMM posterior probabilities.}

%\State Assign clusters to \texttt{CAA-train}, \texttt{val}, and \texttt{test} samples using $\varphi$ and the GMM fit on \texttt{BpT-train}, unchanged.
\State Assign clusters to \texttt{CAA-train}, \texttt{val}, and \texttt{test} samples using $\varphi$ and the GMM fit on \texttt{BpT-train}, unchanged. 
\Comment{$L_{\mathrm{CE}}$, $\mathsf{F}$ computed with predicted label $\hat{y}$; no labels used}

\Ensure Save $\theta^\ast$, $\varphi$, the GMM model, and cluster assignments.
\end{algorithmic}
\end{algorithm}

%To reveal hidden stratification in model behavior, we cluster samples based not only on their learned representations or embeddings but also on how the trained encoder–decoder model $\theta^\ast$ responds to them. In addition to latent embeddings $z$, we incorporate two model behavioral indicators: prediction loss (cross-entropy) and parameter sensitivity (Fisher information). 
%The cross-entropy term identifies samples that are easier or harder for the model to classify, while the Fisher information characterizes the local curvature of the loss landscape—indicating how sensitive or stable the model is to variations around each sample \cite{lee2022masking}.

%To uncover hidden stratification with respect to model performance, we cluster samples using not only their latent embeddings but also how the model behaves on them. Alongside representation features, i.e., embeddings, we include \emph{sample difficulty} (cross-entropy loss) and \emph{parameter sensitivity} (Fisher information). Cross-entropy highlights misclassified or hard examples, while Fisher information serves as a proxy for curvature, linking to sharpness and generalization \cite{lee2022masking}.  

Building upon the baseline model $\theta^\ast$’s behavior-informed representations learned during pretraining, this Stage seeks to reveal hidden stratification—that is, differences in model treatment (i.e., performance and behavior) across latent subgroups that are not explicitly defined by demographic attributes. Even after global regularization, some samples may still reside near sharper regions of the loss landscape, corresponding to smaller decision boundary margins and higher predictive uncertainty. To uncover these disparities, we perform model behavioral clustering, grouping samples not only by how they are represented in the latent space but also by how the model $\theta^\ast$’s loss landscape behaves around them.

%For each input $x$ with label $y$: we form a vector $S = [z,\;L^{\text{CE}},\;\mathsf{F}]$, where, $z = e_{\theta_e}(x)$ is the encoder embedding, $L^{\text{CE}} = L_{\text{CE}}(\hat{y}, y)$ is the cross-entropy loss, and $\mathsf{F}$ is the Fisher penalty around this input.  
For each input $x$, we form a model-behavior descriptor $S=[z,\;L_{\mathrm{CE}},\;\mathsf{F}]$, where $z=e_{\theta_e}(x)$ is the encoder embedding, $L_{\mathrm{CE}}$ is the cross-entropy loss, and $\mathsf{F}$ is the Fisher penalty around this input. When fitting the clustering (on \texttt{BpT-train}), descriptors are computed only for correctly classified samples, using their ground-truth labels. \textcolor{black}{When assigning clusters to any other sample (\texttt{CAA-train}, \texttt{val}, and \texttt{test}),} $L_{\mathrm{CE}}$ and $\mathsf{F}$ are computed with the predicted label $\hat{y}$ instead of $y$, i.e., $L_{\mathrm{CE}}=-\log p_{\theta^\ast}(\hat{y}\mid x)$ (the negative log-confidence) and $\mathsf{F}=\mathsf{F}(x,\hat{y};\theta^\ast)$. Cluster assignment therefore requires no labels. Because the clustering is fit only on samples with $\hat{y}=y$, these label-free descriptors are computed exactly as the descriptors used to fit UMAP and the GMM.
Note that the Fisher term plays two distinct roles in \textbf{\textsc{Flare}}: as a training penalty ($\mathsf{F}_{\mathrm{corr}}$ in Eq.~\eqref{eq:lpre} and $L_{\mathrm{cl}}$), it uses ground-truth labels and is computed only on the training splits (\texttt{BpT-train} and \texttt{CAA-train}); as a routing descriptor, it uses the predicted label $\hat{y}$ and therefore requires no labels. Labels of \texttt{val} samples are used only to compute F1 for model selection, and labels of \texttt{test} samples only to compute evaluation metrics.

To form coherent behavioral clusters, we first apply Uniform Manifold Approximation and Projection (UMAP) \cite{healy2024uniform} to reduce $S$ to a lower-dimensional representation. We then fit a Gaussian Mixture Model (GMM) \cite{mclachlan2019finite} to partition samples into $n$ clusters, where the number of clusters $n$ is selected based on the Bayesian Information Criterion (BIC) \cite{schwarz1978estimating} on \texttt{BpT-train}, while ensuring that no cluster is empty in \texttt{BpT-train} or \texttt{CAA-train}. Test data are not used in this choice. Fitting on correctly classified samples only is intended to make the learned geometry reflect stable decision regions rather than transient misclassifications near the decision boundary. The fitted UMAP map and GMM are then applied unchanged to route \texttt{CAA-train}, \texttt{val}, and \texttt{test} samples; ground-truth test labels are used solely to compute evaluation metrics. The motivation for choosing BIC, UMAP, and GMM is detailed in Appendix~\ref{app:gmm}.

% \textcolor{red}{add more detail about clustering methods in appedix, reviewer asked for it right?}

Importantly, this clustering is performed at the sample level, not at the individual or demographic level. As a result, multiple samples from the same subject can belong to different clusters depending on how the model’s decision geometry interacts with each instance. This property is critical for identifying subtle, data-driven fairness gaps that demographic-based methods may overlook.

These behaviorally coherent clusters expose where the model performs inconsistently and provide the basis for targeted adaptation in Stage~3. Throughout the rest of the paper, identified clusters are used synonymously with latent subgroups.

\subsection{Stage 3: Cluster-specific Adaptation and Aggregation}  
\label{Stage3}
\begin{algorithm}[t]
\caption{Stage 3: Cluster-specific Adaptation and Conditional Aggregation}
\label{algo2}
\begin{algorithmic}[1]
\State \textbf{Inputs:} pretrained $\theta^\ast$; \textcolor{black}{clusters $\{\mathcal{D}^c_{\text{CAATrain}},\mathcal{D}^c_{\text{val}},\mathcal{D}^c_{\text{test}}\}_{c=1}^{n}$;} freeze depth $L$; weight $\gamma$; lr $\eta$; epochs $E$; aggregation interval $\tau$.
\State Initialize $\theta^*_c \leftarrow \theta^\ast$; freeze the first $L$ modules for all $c$.
\State For each cluster $c$, save $\theta^\ast$ as its initial best checkpoint. \State Set $\texttt{bestF1}^{(c)} \leftarrow \mathrm{F1}(\theta^\ast;\mathcal{D}^{c}_{\mathrm{val}})$.
\For{$\text{epoch}=1$ to $E$}
  \For{$c=1$ to $n$}
    \For{mini-batch $(x,y)\subset\mathcal{D}^c_{\text{CAATrain}}$}
      \State $\hat{y} \leftarrow$ prediction of cluster model $\theta^*_c$
      \State Compute $L_{\mathrm{CE}}$ and $\mathsf{F}_{\mathrm{corr}}$ under $\theta^*_c$, and $L^{\mathrm{pre}}_{\mathrm{CE}}$ under the frozen $\theta^\ast$.
      \State $\Delta L_{\mathrm{CE}} \leftarrow L_{\mathrm{CE}} - L^{\mathrm{pre}}_{\mathrm{CE}}$
      \State $L_{\text{cl}} \leftarrow \gamma\,L_{\mathrm{CE}} + (1-\gamma)\,\mathsf{F}_{\mathrm{corr}} + \max\{0,\Delta L_{\mathrm{CE}}\}$
      \State Update $\theta^*_c$ by a gradient step on $L_{\text{cl}}$ with learning rate $\eta$.
    \EndFor
    \State \textbf{if} $\mathrm{F1}(\theta^*_c;\mathcal{D}^c_{\text{val}}) > \texttt{bestF1}^{(c)}$ \textbf{then} save checkpoint $\theta^*_c$ and update $\texttt{bestF1}^{(c)}$ \Comment{validation F1}
  \EndFor
  \If{$\text{epoch} \bmod \tau = 0$}
    \State $\bar{\theta} \leftarrow \frac{1}{n}\sum_{c=1}^{n}\theta^{\mathrm{best}}_c$ \Comment{average of each cluster's best saved checkpoint}
    \For{$c=1$ to $n$}
      \State Update $\theta^*_c$ using Eq.~\eqref{eq:cluster_select}; if $\bar{\theta}$ is adopted and its validation F1 exceeds $\texttt{bestF1}^{(c)}$, save it and update $\texttt{bestF1}^{(c)}$ \Comment{validation F1}
    \EndFor
  \EndIf
  \State Stop when $\frac{1}{n}\sum_{c=1}^{n}\mathrm{F1}(\theta^*_c;\mathcal{D}^c_{\text{val}})$ converges. \Comment{early stopping on validation F1}
\EndFor
\For{$c=1$ to $n$}
  \State Reload best checkpoint $\theta^*_c$ and evaluate on $\mathcal{D}^c_{\text{test}}$. \Comment{test used once, after all models are frozen}
\EndFor

\end{algorithmic}

\end{algorithm}

After revealing latent subgroups through curvature- and behavior-aware clustering, this Stage of \textbf{\textsc{Flare}} focuses on ethically fair adaptation. It ensures that each cluster, i.e., latent subgroup, receives tailored optimization that respects both performance equity and ethical constraints. This Stage 3 uses \texttt{CAA-train}, which is disjoint from the \texttt{BpT-train} data used to fit $\theta^*$ and the clusters. Each cluster model is trained on its cluster's samples $\mathcal{D}^c_{\text{CAATrain}}\subset$ \texttt{CAA-train}, \textcolor{black}{and all Stage~3 selection decisions (checkpoint selection, aggregation, and early stopping) use the cluster's samples $\mathcal{D}^c_{\text{val}}$ in the separate validation split \texttt{val}, which is disjoint from both \texttt{CAA-train} and the test participants.}

\subsubsection{\textbf{Cluster Adaptation:}}
\label{Stage3-cluster-adaptation}
Once clusters are identified, we specialize models for each cluster $c$ by adapting the pretrained model $\theta^\ast$ (encoder and classifier only; cluster models have no decoder).  
The early encoder layers are frozen to preserve generic knowledge learned during pretraining, while the later encoder and classifier parameters are fine-tuned to capture cluster-specific patterns. This hierarchical adaptation enables \emph{localized fairness correction} without sacrificing generalizability.

We define the excess cross-entropy relative to the pretrained model as
\[
\Delta L_{\mathrm{CE}} 
= 
L_{\mathrm{CE}}(\hat{y},y) 
- 
L^{\mathrm{pre}}_{\mathrm{CE}},
\]
where $L^{\mathrm{pre}}_{\mathrm{CE}}$ denotes the cross-entropy loss computed on the same cluster-$c$ samples under the frozen pretrained model $\theta^\ast$. 
The term $\max\{0,\Delta L_{\mathrm{CE}}\}$ serves as a “do-no-harm” regularizer: it is positive only when the cluster-specific model's cross-entropy exceeds that of the pretrained model, thereby discouraging harmful updates; otherwise, this term vanishes.
We then define the cluster-level objective as
\[
L_{\mathrm{cl}}
=
\gamma\,L_{\mathrm{CE}}
+
(1-\gamma)\,\mathsf{F}_{\mathrm{corr}}(x,y;\theta^*_c)
+
\max\{0,\Delta L_{\mathrm{CE}}\},
\]
where \(\gamma \in [0,1]\) trades off predictive accuracy (cross-entropy) against stability (Fisher penalty).
% Additionally, $\alpha \in [0,1]$ balances accuracy (cross-entropy) against stability and fairness (Fisher regularization).  

Here: $L_{\mathrm{CE}}(\hat{y},y)$ is the cross-entropy loss between the prediction $\hat{y}$ and the ground truth $y$, measuring classification error. $\mathsf{F}_{\mathrm{corr}}(x,y;\theta^*_c)$ is the Fisher penalty under $\theta^*_c$, quantifying sensitivity of the model to parameter perturbations specific to cluster $c$ samples, encouraging flatter minima and improving decision boundary margins within each cluster. As in Section~\ref{Step-1-design-choice}, the Fisher penalty is computed \emph{only on samples correctly classified by $\theta^*_c$}, ensuring that curvature smoothing focuses on reliable decision regions, while misclassified samples are handled through the cross-entropy term that encourages boundary correction. This separation stabilizes correct predictions, promoting \emph{beneficence}, while avoiding stabilizing incorrect ones, promoting \emph{non-maleficence}. Moreover, just as in Section~\ref{Step-1-design-choice}, lowering the Fisher penalty is intended to promote more uniform curvature and thus more consistent performance across cluster $c$ samples, promoting \emph{justice}.

Through this design, each cluster-specific model evolves within a locally fairer loss geometry, improving efficacy for underperforming subgroups while discouraging degradation of others.

\subsubsection{\textbf{Aggregation:}}
\label{Stage3-aggregation}
While local adaptation refines each subgroup, independent optimization can cause clusters to drift apart, creating uneven decision geometries and potentially new disparities. To counter this, \textbf{\textsc{Flare}} introduces a periodic parameter-averaging mechanism that promotes equilibrium across clusters, ensuring that improvements remain shared and harmonized.

At scheduled intervals, the best saved checkpoints of all cluster-specific models are averaged: $\bar{\theta} = \frac{1}{n} \sum_{c=1}^n \theta^{\mathrm{best}}_c$, where $n$ denotes the number of clusters and $\theta^{\mathrm{best}}_c$ is the checkpoint of cluster $c$ with the highest validation F1 so far.

The aggregated model $\bar{\theta}$ represents a consensus parameterization, capturing shared curvature and decision geometry across subgroups \cite{izmailov2018averaging,gu2025self,zhang2023generalization}.
It is then offered to all clusters as a candidate update. Each cluster evaluates whether adopting $\bar{\theta}$ improves its F1 score on its validation (\texttt{val}) samples. If the update yields improvement, $\bar{\theta}$ replaces the prior model; otherwise, the cluster retains its previous parameters.  
%The best model for cluster $c$ upon each aggregation step is defined as $\theta_c^\ast$
After each aggregation step, the model for cluster $c$ is updated as
\begin{equation}
\theta^*_c \leftarrow \operatorname*{arg\,max}_{\theta \in \{\theta^*_c,\, \bar{\theta}\}} \mathrm{F1}\big(\theta;\, \mathcal{D}^{c}_{\text{val}}\big),
\label{eq:cluster_select}
\end{equation}
where $\mathrm{F1}(\theta;\mathcal{D}^{c}_{\text{val}})$ is the F1-score of model $\theta$ on the validation (\texttt{val}) samples of cluster $c$. That is, a cluster adopts the aggregated model $\bar{\theta}$ only if it improves that cluster's validation F1; otherwise, it retains its current $\theta^*_c$.

% Formally, the best model for cluster $c$ upon each aggregation step is defined as \theta_c^\ast:
% \[
% \theta_c^\ast = \arg\max_{\theta_c,\,\bar{\theta}} \; \text{F1}_{\text{val}}(D_c),
% \]
% where $D_c$ represents cluster $c$'s validation data distribution.
This procedure balances knowledge sharing \cite{supeksala2024private} and specialization \cite{hod2021quantifying}: clusters benefit from collective information through $\bar{\theta}$ while retaining the ability to reject it if it degrades local performance. By comparing the F1 scores between $\bar{\theta}$ and the previous cluster model $\theta^*_c$, each cluster performs a local \textit{model selection step} on its validation data, which prevents adopting an aggregated model that lowers validation F1 and thus promotes \emph{non-maleficence}.

%each cluster effectively performs a local \textit{model selection step} that guards against utility drop, promoting \emph{non-maleficence}.

Training stops early when the mean validation F1 across clusters converges. The final system is given by $\{\theta_c^\ast\}_{c=1}^{n}$, in which each cluster keeps the parameter configuration, either its adapted version or the aggregated one, that maximizes its local validation performance; each frozen cluster model is then evaluated once on its held-out test samples $\mathcal{D}^c_{\text{test}}$.

\paragraph{How non-degradation is promoted.}
\textbf{\textsc{Flare}} addresses \emph{non-maleficence} at three levels, which we distinguish explicitly. (i) During training, the do-no-harm regularizer $\max\{0,\Delta L_{\mathrm{CE}}\}$ (Section~\ref{Stage3-cluster-adaptation}) discourages updates that increase the cross-entropy relative to the pretrained model $\theta^\ast$; as a loss term, it does not by itself guarantee non-degradation. (ii) During model selection, each cluster saves a checkpoint, or adopts $\bar{\theta}$, only when doing so improves its validation (\texttt{val}) F1 (Algorithm~\ref{algo2}, Eq.~\eqref{eq:cluster_select}), which favors models that do not degrade on validation data. (iii) Whether non-degradation carries over to unseen participants, and holds relative to the benign baseline used in the BHE metrics, is an empirical question, which we test on held-out participants in Sections~\ref{sec:baselines-vs-flare} and~\ref{donot_harm}.

\subsection{Implementation}
\label{sec:implementation}
Algorithms~\ref{algo1} and~\ref{algo2} summarize the procedure. Stage~1 (Algorithm~\ref{algo1}, Lines~1--10) trains the encoder--decoder--classifier on \texttt{BpT-train} with Eq.~\eqref{eq:lpre} and selects $\theta^\ast$ by validation F1 on \texttt{val}. Stage~2 (Lines~11--14) fits UMAP and the GMM on the descriptors of correctly classified \texttt{BpT-train} samples and routes \texttt{CAA-train}, \texttt{val}, and \texttt{test} samples to clusters without labels. Stage~3 (Algorithm~\ref{algo2}) fine-tunes one model per cluster with $L_{\mathrm{cl}}$ (Lines~7--12), selects checkpoints (Line~14) and aggregation decisions (Lines~16--21, Eq.~\eqref{eq:cluster_select}) on validation F1, stops early when the mean validation F1 converges (Line~22), and evaluates each frozen cluster model once on its held-out test samples (Lines~24--26).

\section{{Experimental Results}}
\label{res-sec}
This section evaluates \textbf{\textsc{Flare}} to assess its ability to enhance ethical fairness without relying on explicit demographic or heterogeneous attribute information, while maintaining high predictive accuracy across diverse behavioral and physiological sensing tasks.
We first describe the datasets and models (Section~\ref{data}), then evaluate \textbf{\textsc{Flare}}'s ethical fairness, do-no-harm behavior and optimization geometry (Section~\ref{sec:eval}), interpret the discovered clusters (Section~\ref{sec:cluster_interpretation}), and finally validate the design choices through cluster-sensitivity and ablation analyses (Sections~\ref{sec:cluster-sensitivity}--\ref{sec:network-ablation}).

\subsection{Datasets and Models}
\label{data}
\textbf{\textsc{Flare}} is tested on  four multimodal mobile behavioral and physiological sensing datasets:
\textbf{OhioT1DM}~\cite{marling2020ohiot1dm} —continuous glucose and insulin pump data from 12 participants (normal vs.\ hyperglycemia, processed per~\cite{arefeen2023designing});
\textbf{Intern Health Study (IHS)}~\cite{adler2021identifying}—a 14-month study of 85 medical interns combining PHQ-9 scores, daily mood, and Fitbit-derived features, labels are derived from mood thresholds (>8 positive, <3 negative) in line with~\cite{manjunath2023can}; \textbf{Electrodermal Activity (EDA)}~\cite{xiao2025human}—multimodal physiological signals (EDA, HR, ACC, TEMP, HRV) from 76 participants with 340 handcrafted features for stress detection and  \textbf{Percept-R}~\cite{benway2022percept}—an open-access clinical speech corpus specialized for American English rhotic /\rotatebox[origin=c]{180}{r}/ production in children and adolescents. 
% The full corpus contains over 32 hours of citation speech collected from 281 participants aged 6 to 24 years, including both typically developing speakers and individuals with Residual Speech Sound Disorders (RSSD).
 For our evaluation, we use a publicly available subset of 76 participants of this dataset for whom the demographic information required for our BHE ethical fairness analysis is available. 
% Each sample is represented as a fixed-length multivariate sequence with 5 acoustic feature channels over 60 time steps, and labels are derived from expert and listener judgments of rhoticity, yielding a binary speech-correctness task (correct rhotic /\rotatebox[origin=c]{180}{r}/ vs.\ incorrect/derhotic /\rotatebox[origin=c]{180}{r}/). 
 Further dataset, preprocessing, fold composition, and model details are provided in Appendix~\ref{appendix:data_models}

\begin{table}[h!]
\centering

\color{black}

\arrayrulecolor{black}
\caption{Sensitive attributes used during BHE evaluation.}
\label{tab:bhe_subgroup_attributes}
\vspace{-10pt}
\resizebox{0.80\linewidth}{!}{

\begin{tabular}{p{0.22\linewidth}|p{0.34\linewidth}|p{0.34\linewidth}}
\hline
\textbf{Dataset} & \textbf{Demographic attributes} & \textbf{Heterogeneous attributes} \\
\hline

\textbf{OhioT1DM}~\cite{marling2020ohiot1dm}
&
Age; Sex
&
Pump; Sensor Band (coincides with collection cohort: 2018 = Basis, 2020 = Empatica) \\
\hline

\textbf{EDA}~\cite{xiao2025human}
&
Sex
&
Group\_label \\
\hline

\textbf{IHS}~\cite{adler2021identifying}
&
Sex; Age;  Ethnicity
&
Specialty; PHQ10$>$0 \\
\hline

\textbf{Percept-R}~\cite{benway2022percept}
&
Sex; Age (4 bins,in months); Race;  Ethnicity
&
Not available \\
\hline

\end{tabular}
}
\vspace{-12pt}
\end{table}

\paragraph{\textbf{Known Sensitive Attributes for Evaluating Subgroup-wise Ethical Fairness through BHE Metrics}}
In this work, a \emph{sensitive attribute} refers to any attribute along which differences in model performance may indicate a potential ethical fairness concern. A \emph{subgroup} refers to one specific category within such an attribute. For example, \emph{sex} is a sensitive attribute, whereas \emph{male} and \emph{female} are subgroups of that attribute. Bias in human sensing models should be evaluated beyond demographic attributes because performance disparities can also arise from how the data are collected, what devices are used, and what clinical or contextual conditions shape the sensed signals. Table~\ref{tab:bhe_subgroup_attributes} summarizes the sensitive attributes used for BHE evaluation and distinguishes between demographic attributes and heterogeneity factor attributes.
% Demographic attributes, such as age, sex, race, and ethnicity, capture identity-related sources of disparity. In contrast, heterogeneous attributes capture domain-specific sources of bias that may affect model behavior even when demographic groups appear balanced. For example, in EDA dataset, \emph{Group\_label} can introduce clinical-condition bias because physiological signals may differ across control, pre-dose, and post-dose groups; In OhioT1DM dataset, pump model and sensor band can introduce device bias because different hardware may produce signals with different noise levels, sampling properties, or measurement characteristics; In IHS dataset, cohort can introduce temporal or protocol bias because data collected in different years may reflect changes in participants, devices, or collection procedures; and specialty and PHQ10$>$0 can introduce clinical-context or baseline-condition bias because participants may differ in professional environment, stress, or mental-health status. Notably, Percept-R is evaluated only using reported demographic attributes, whereas OhioT1DM, EDA, and IHS include both demographic and heterogeneous subgroup attributes.
Additional subgroup of each sensitive attribute and model details are provided in Appendix~\ref{appendix:data_models}.

% {
% \paragraph{\textbf{Known attributes for subgroup-wise BHE evaluation.}}
% Table~\ref{tab:bhe_subgroup_attributes} summarizes the attributes used to evaluate subgroup-wise ethical fairness through BHE metrics. We distinguish \emph{demographic attributes}, such as sex, age, race, and ethnicity, from \emph{heterogeneous attributes}, which capture non-demographic sources of variation including device type, cohort year, medication status, clinical condition, residency specialty, and baseline mental-health indicators. In our evaluation, Percept-R is assessed only along reported demographic attributes, while OhioT1DM, EDA, and IHS additionally include domain-specific heterogeneous attributes. 
% }
% For all three dataset, We used a subject-disjoint cross-validation setup where all test folds together covered the full dataset, ensuring each user appeared in the test set exactly once. After selecting test users, the remaining participants were split into training and holdout training (\textit{train2}) sets using an 80:20 random split.

\begin{figure*}[h]
    \centering
    \includegraphics[width=\textwidth]{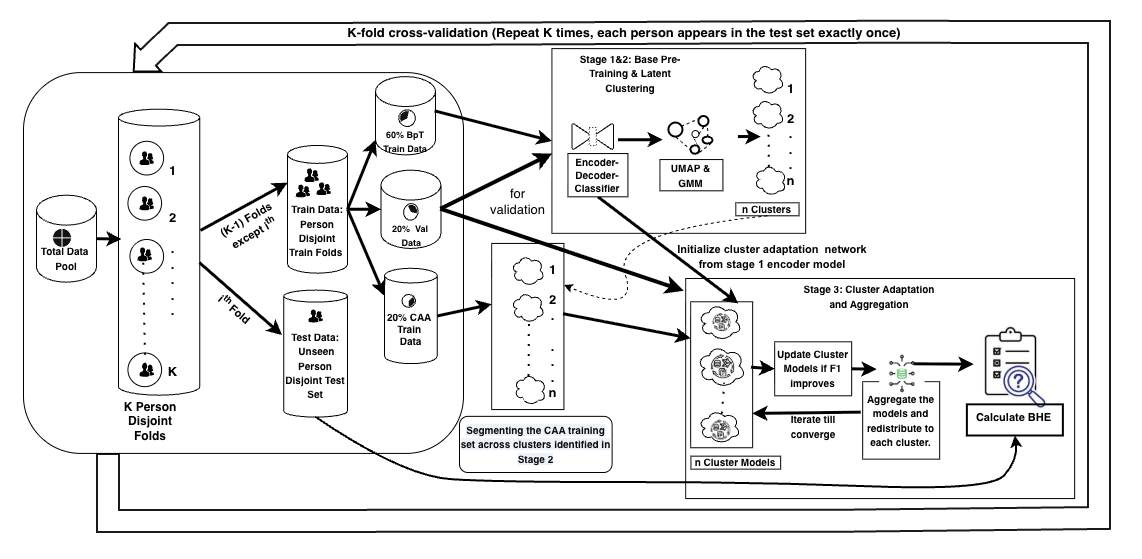}
    \hspace{-25pt}
    \vspace{-1em}
\caption{\textbf{K-fold person-disjoint evaluation.}
Each fold holds out unseen test participants. Samples from the remaining participants are randomly split 60:20:20 into \texttt{BpT-train} (pretraining and clustering), \texttt{CAA-train} (adaptation), and \texttt{val} (checkpoint selection, early stopping, tuning, and aggregation decisions). Inner splits are sample-disjoint and may share participants. Each participant is tested once after model selection, except one OhioT1DM participant (Appendix~\ref{sec:data_stat}); BHE uses pooled out-of-fold test predictions.}
    \label{fig:eval_overview}
\vspace{-12pt}
\end{figure*}

\paragraph{\textbf{Evaluation Setup}} 
Figure~\ref{fig:eval_overview} summarizes the $K$-fold person-disjoint protocol used throughout our experiments. In each fold, participants are partitioned into a held-out \texttt{test} set and a training pool, with no participant shared between them. As noted in Section~\ref{BHE-metrics}, test sets do not overlap across folds, and each participant appears in a test set exactly once across the full cross-validation procedure (with one OhioT1DM exception; Appendix~\ref{sec:data_stat}). This ensures evaluation over the entire population while avoiding repeated testing of the same participant.

Within each outer fold, the samples contributed by the training participants are randomly divided 60:20:20 into \texttt{BpT-train}, \texttt{CAA-train}, and \texttt{val}. These inner partitions are sample-disjoint but may contain samples from the same participants. Person-disjointness applies between the entire outer training pool and the held-out test participants. The \texttt{BpT-train} split is used for base pretraining and latent clustering (Sections~\ref{Step-1-design-choice} and~\ref{design-clustering}), and the \texttt{CAA-train} split is used for Cluster-specific Adaptation and Aggregation (Section~\ref{Stage3}). \texttt{val} is never used for training; it is used only for model selection in all stages: checkpoint selection, early stopping, hyperparameter tuning, and aggregation decisions. Test participants are never used for training, tuning, clustering, or selection, and are evaluated once after all models are frozen. 
At test time, samples are routed to cluster models without labels (Section~\ref{design-clustering}); test labels are used only to compute F1 and BHE.
Per-fold participant counts are given in Appendix~\ref{sec:data_stat}. Notably, all models (including all baselines) use the same $K$-fold person-disjoint train-test split.

\paragraph{\textbf{Models}} 
Each dataset uses an autoencoder–classifier backbone tailored to its modality, with dataset-specific architectures for OhioT1DM, IHS, EDA, and Percept-R. 
All models follow a shared design: an encoder to learn latent representations, a classifier head for prediction, and a symmetric decoder for reconstruction learning, enabling joint representation regularization. These `pretrained autoencoder–classifier' networks (from Stage~1 in Section \ref{Step-1-design-choice}) serve as the initialization stage for \textbf{\textsc{Flare}}, on which it performs cluster-specific adaptation and cross-cluster aggregation. Further architectural details are provided in Appendix~\ref{appendix:data_models}.
%All training hyperparameters and optimization settings follow prior literature~\cite{marling2020ohiot1dm, xiao2025human}, with IHS hyperparameters tuned empirically due to dataset size constraints. 
\paragraph{\textbf{Benign Baseline Models}}
\label{benign}

For each dataset, we select the best-performing benign baseline \textcolor{black}{by F1-score on the validation (\texttt{val}) split from standard} fully connected (MLP), convolutional (CNN), and recurrent (RNN/LSTM) models (for OhioT1DM, convolutional and recurrent models only, following prior work). Baselines follow prior settings reported in literature for OhioT1DM~\cite{marling2020ohiot1dm,arefeen2023designing}, EDA~\cite{xiao2024reading,xiao2025human}, and Percept-R~\cite{benway2022percept}; For IHS, baseline is tuned with Optuna and manual refinement. Full details are provided in Appendix~\ref{benign_eval}.
% {We evaluate multiple standard deep learning architectures (MLP, CNN, and LSTM) for each dataset and select the best-performing model based on F1-score as the \emph{benign baseline}. Details of this selection process are provided in Appendix~\ref{benign_eval}. These baselines are trained without any fairness or ethical regularization and serve as reference points for evaluating the BHE metrics of \textsc{Flare} and other Fairness with and without demographics approaches.}

% These baselines adopt established architectures and hyperparameter configurations from prior work. For the \textbf{OhioT1DM} dataset, we followed the models and parameter settings of Marling and Bunescu~\cite{marling2020ohiot1dm} and Arefeen and Ghasemzadeh~\cite{arefeen2023designing}. For the \textbf{EDA} dataset, we implemented the configurations proposed by Xiao et~al.~\cite{xiao2024reading, xiao2025human}. For the \textbf{IHS} dataset, since model configuration was not provided in the literature \cite{adler2021identifying}, the baseline was optimized through a combination of \emph{Optuna}~\cite{shekhar2021comparative} hyperparameter tuning and iterative manual refinement to obtain the highest benign performance.  {For the \textbf{Percept-R} dataset, we use a compact 1D CNN benign baseline motivated by the clinical speech classification setting described in the Percept-R corpus paper~\cite{benway2022percept}. }
% \textcolor{red}{(outperforming SoTA evaluation performance on this dataset)}. 

\paragraph{\textbf{Hyperparameter Optimization}} To ensure optimal and reproducible performance, 
all hyperparameters, including learning rate,  weight decay, and the loss weights $\alpha$, $\beta$ (Stage~1), and $\gamma$ (Stage~3) are tuned using the \emph{Optuna} \cite{akiba2019optuna} hyperparameter optimization framework. Optuna maximizes F1 on the validation (\texttt{val}) split; test data are never used. Full configurations are provided in Appendix~\ref{app:hype}. 

% Comprehensive architectural, training, and preprocessing specifications are detailed in Appendix~\ref{appendix:data_models}.

%{To ensure optimal and reproducible performance, all hyperparameters—including learning rate, batch size, weight decay, and the regularization coefficients $\alpha$, $\beta$, and $\gamma$—are tuned using the \emph{Optuna} hyperparameter optimization framework.  
%Optuna employs Bayesian optimization with early stopping and dynamic pruning to efficiently explore the search space, automatically selecting the configuration that maximizes validation performance while maintaining fairness and stability objectives \cite{akiba2019optuna}.}

\subsection{Evaluation of \textbf{\textsc{Flare}}’s Effectiveness}
\label{sec:eval}

We evaluate the ethical fairness of \textbf{\textsc{Flare}} across all available demographic and heterogeneous attributes outlined in Table~\ref{tab:bhe_subgroup_attributes}. We first compare \textbf{\textsc{Flare}} with attribute-aware and attribute-free baselines (Section~\ref{sec:baselines-vs-flare}), then test the proposed \emph{do-no-harm} strategy within \textbf{\textsc{Flare}} (Section~\ref{donot_harm}), and finally examine its optimization geometry (Section~\ref{sec:loss-landscapes}).

\subsubsection{\textbf{\textsc{Flare} vs.\ Attribute-aware and Fairness-without-Demographics Baselines}}
\label{sec:baselines-vs-flare}
We compare \textbf{\textsc{Flare}} with two groups of baselines. (i) \emph{Attribute-aware} baselines that explicitly use sex during training, with demographic-parity (Fair-DP) and equalized-odds (Fair-EO) fairness losses \cite{pal2023ensuring,hu2023parametric,liu2026fairness}; with direct access to a sensitive attribute, these methods would be expected to perform best on sex-based fairness. (ii) \emph{Fairness-without-demographics} baselines, namely knowledge distillation (KD) \cite{chai2022fairness}, adversarially reweighted learning (ARL) \cite{lahoti2020fairness}, Graph of Gradients (GoG) \cite{luo2025fairness}, and Reckoner \cite{ni2024fairness}, together with the subgroup-robustness methods JTT \cite{liu2021jtt} and Group DRO \cite{sagawa2019distributionally}. \textbf{\textsc{Flare}} uses no demographic or heterogeneous attributes during training. All baselines are trained on the same 80\% of each fold's training pool that \textbf{\textsc{Flare}} uses (\texttt{BpT-train} and \texttt{CAA-train} combined), select models on \texttt{val}, and are evaluated on the same unseen test participants. For each dataset, we aggregate the true and predicted labels produced once by the frozen, validation-selected models on the person-disjoint test sets across all folds, compute the F1 score of every subgroup of each sensitive attribute, and compute the BHE changes relative to the benign baseline (Section~\ref{BHE-metrics}). All BHE values are computed from unrounded subgroup F1 scores and rounded to two decimals only for reporting; subgroup F1 scores in Appendix~\ref{sec:subgroup-analysis} are shown to one decimal, so methods whose scores differ by less than the displayed precision can appear identical. Different methods can obtain equal subgroup F1 scores without making identical sample-level predictions. Consequently, repeated $\Delta H$ values can reflect equal minimum subgroup improvements or differences smaller than the displayed precision.For \textbf{\textsc{Flare}}, every per-attribute $\Delta H$
is non-negative before rounding. Thus, a reported
$\Delta H = 0.00$ represents an exactly zero or small positive
minimum subgroup change, never a negative change concealed
by rounding. Table~\ref{tab:bhe_summary} reports these changes averaged over the sensitive attributes of each dataset; per-attribute F1 scores and BHE changes are reported in Appendix~\ref{app:per_attr} (Tables~\ref{tab:f1_demographics_models_reduced}--\ref{tab:sota_delta_grouped}).

\begin{table}[t]
\centering
\small
%\caption{Mean BHE changes (percentage points, relative to the benign baseline), averaged over the sensitive attributes of each dataset; higher is better for all three. $^\dagger$Uses sex labels during training. Per-attribute results are reported in Appendix~\ref{app:per_attr}. Best value per column in bold.}
\caption{BHE changes (percentage points) relative to the benign baseline, averaged across each dataset's sensitive attributes; higher is better. Non-negative mean $\Delta H$ alone does not establish no harm; this is assessed using per-attribute values (Appendix~\ref{app:per_attr}). $^\dagger$Uses sex labels during training. Column bests are bold.}
\label{tab:bhe_summary}
\resizebox{\linewidth}{!}{
\begin{tabular}{l|ccc|ccc|ccc|ccc}
\toprule
& \multicolumn{3}{c|}{\textbf{OhioT1DM}} & \multicolumn{3}{c|}{\textbf{EDA}} & \multicolumn{3}{c|}{\textbf{IHS}} & \multicolumn{3}{c}{\textbf{Percept-R}}\\
\textbf{Method} & $\Delta B$ & $\Delta H$ & $\Delta E$ & $\Delta B$ & $\Delta H$ & $\Delta E$ & $\Delta B$ & $\Delta H$ & $\Delta E$ & $\Delta B$ & $\Delta H$ & $\Delta E$\\
\midrule
Fair-EO$^\dagger$ & 0.67 & $-$5.80 & 2.07 & 0.06 & $-$4.34 & 0.08 & $-$0.33 & $-$11.59 & 0.65 & 0.33 & $-$2.60 & 2.38\\
Fair-DP$^\dagger$ & 0.87 & $-$4.55 & 2.01 & $-$0.38 & $-$4.79 & 0.31 & $-$1.36 & $-$5.45 & 1.08 & 0.35 & $-$2.48 & \textbf{2.89}\\
\midrule
KD        & 0.16 & $-$3.66 & 5.06 & $-$17.86 & $-$19.54 & 0.25 & $-$1.20 & $-$9.01 & 0.91 & 0.58 & $-$0.98 & 0.65\\
ARL       & 4.07 & $-$0.32 & 4.93 & 0.56 & $-$0.89 & $-$0.27 & 2.79 & $-$1.20 & 0.56 & $-$0.48 & $-$5.63 & $-$0.59\\
GoG       & 0.67 & $-$4.53 & 5.26 & $-$0.02 & $-$0.22 & $-$0.04 & $-$0.77 & $-$10.80 & $-$0.49 & 0.96 & $-$1.29 & 0.29\\
Reckoner  & 0.84 & $-$3.12 & 5.23 & $-$1.37 & $-$4.41 & $-$0.13 & $-$3.14 & $-$8.40 & 1.22 & $-$1.83 & $-$6.57 & $-$0.52\\
JTT       & 1.47 & $-$6.89 & 3.82 & 0.36 & $-$0.14 & \textbf{0.44} & $-$1.76 & $-$10.09 & 0.19 & $-$3.26 & $-$13.96 & $-$1.09\\
Group DRO & $-$3.24 & $-$8.17 & \textbf{5.55} & 0.07 & $-$0.76 & $-$0.73 & 1.15 & $-$7.25 & \textbf{1.71} & $-$1.09 & $-$7.46 & $-$0.02\\
\midrule
\textbf{\textsc{Flare}} & \textbf{5.16} & \textbf{0.15} & 5.19 & \textbf{2.22} & \textbf{1.54} & 0.32 & \textbf{5.99} & \textbf{1.74} & 1.32 & \textbf{6.25} & \textbf{1.67} & 2.74\\
\bottomrule

\end{tabular}}
\vspace{-15pt}
\end{table}
\textbf{Benefit.} \textbf{\textsc{Flare}} attains the highest mean subgroup F1 of all methods on every dataset (70.38\% on OhioT1DM, 89.27\% on EDA, 65.23\% on IHS, and 84.04\% on Percept-R; Appendix Tables~\ref{tab:f1_demographics_models_reduced} and~\ref{tab:f1_demographics_models}) and the largest $\Delta B$ on every dataset (Table~\ref{tab:bhe_summary}). Even for sex, the attribute that Fair-DP and Fair-EO optimize during training, \textbf{\textsc{Flare}} obtains higher subgroup F1 than both on all four datasets, indicating that access to a sensitive attribute during training does not necessarily translate into stronger subgroup performance at evaluation time. This is consistent with prior work showing that enforcing parity constraints generally comes at a cost in predictive accuracy \cite{menon2018cost,zhao2022inherent,corbett-davies2018measure}, and that fair classifiers often reach parity by lowering the performance of better-off groups rather than raising that of worse-off groups, a phenomenon known as \emph{leveling down} \cite{zietlow2022leveling,mittelstadt2024unfairness}.
%Even for sex, the attribute that Fair-DP and Fair-EO optimize during training, \textbf{\textsc{Flare}} obtains higher subgroup F1 than both on all four datasets, indicating that access to a sensitive attribute during training does not necessarily translate into stronger subgroup performance at evaluation time.

\textbf{Harm-Avoidance.}
\textbf{\textsc{Flare}} is the only evaluated fairness method
with positive $\Delta B$ and non-negative $\Delta H$ on all
four datasets. Every competing baseline, including both
attribute-aware ones, has negative mean $\Delta H$ on every
dataset, indicating that it degrades at least one evaluated
subgroup (for Fair-EO and Fair-DP, mean $\Delta H$ ranges
from $-11.59$ to $-2.48$). At the level of individual
attributes, \textbf{\textsc{Flare}}'s $\Delta H$ is
non-negative for every sensitive attribute of every dataset
(Appendix Tables~\ref{tab:dp_eo_flare} and~\ref{tab:sota_delta_grouped}). As noted above,these values are non-negative
before rounding; therefore, a reported $\Delta H = 0.00$
represents an exactly zero or small positive minimum subgroup
change, never a negative change concealed by rounding.
Because $\Delta H$ measures the minimum subgroup change,
a small non-negative value indicates that even the
least-improved subgroup maintains or improves its baseline
F1, while other subgroups may improve substantially.
Thus, unlike every competing baseline,
\textbf{\textsc{Flare}} improves mean subgroup performance
without making any evaluated subgroup worse off in F1,
operationalizing non-maleficence as the absence of observed
subgroup F1 degradation.

\textbf{Equity.} \textbf{\textsc{Flare}} is the only method that reduces disparity for every sensitive attribute of every dataset ($\Delta E>0$ for all 15 attributes; Appendix Tables~\ref{tab:dp_eo_flare} and~\ref{tab:sota_delta_grouped}); every baseline, including the attribute-aware ones, widens disparity for at least one attribute (e.g., Fair-EO on IHS sex, $-0.62$; JTT on IHS ethnicity, $-5.95$). In mean $\Delta E$, at least one baseline achieves a larger reduction on each dataset (Group DRO on OhioT1DM, $+5.55$ vs.\ $+5.19$; JTT on EDA, $+0.44$ vs.\ $+0.32$; Group DRO on IHS, $+1.71$ vs.\ $+1.32$; Fair-DP on Percept-R, $+2.89$ vs.\ $+2.74$), but each does so while degrading at least one subgroup ($\Delta H<0$), i.e., partly by leveling down. Likewise, on sex itself, the attribute they optimize, Fair-EO (EDA) and Fair-DP (OhioT1DM, IHS) reduce disparity more than \textbf{\textsc{Flare}}, but with lower $\Delta B$ (Appendix~\ref{app:with_demo}). \textit{Thus, \textsc{Flare}'s equity gains are smaller in places but never come at the cost of benefit or of any subgroup, making it the only method that improves all three BHE dimensions together.}

\paragraph{\textbf{Why.}} The baselines share one structural limitation: they adjust a \emph{single shared model}, so gains for some samples are paid for by others. The parity-based methods (Fair-DP, Fair-EO) equalize outcomes across sex, which can be satisfied by lowering the better-performing subgroups as readily as by raising the worse ones \cite{zietlow2022leveling,mittelstadt2024unfairness}. The fairness-without-demographics and subgroup-robustness baselines redistribute attention within that shared model: ARL and GoG adversarially upweight high-loss samples or gradient neighborhoods \cite{lahoti2020fairness,luo2025fairness}, JTT upweights the samples a first model misclassifies \cite{liu2021jtt}, and Group DRO optimizes the worst-case group loss \cite{sagawa2019distributionally}. Because these adjustments act globally, shifting the shared decision boundary toward the upweighted samples moves it away from others, which is consistent with the negative $\Delta H$ these methods show on every dataset. KD and Reckoner, in turn, depend on a teacher model or a confidence-based partition, and inherit their errors \cite{chai2022fairness,ni2024fairness}. \textbf{\textsc{Flare}} is designed to discourage this trade-off. It discovers clusters of samples that the model treats similarly, adapts a \emph{separate} model to each cluster, so improving one cluster does not move the decision boundary of another, penalizes any increase in loss relative to the pretrained model (do-no-harm term), and adopts shared parameters only when they improve a cluster's validation F1. \textit{This combination of \emph{local specialization}, \emph{harm-aware adaptation}, and \emph{controlled sharing} is what allows \textbf{\textsc{Flare}} to improve subgroups without leveling others down.} Additional comparisons with the attribute-aware baselines and a subgroup-level analysis are provided in Appendices~\ref{app:with_demo} and~\ref{sec:subgroup-analysis}.

\noindent
\subsubsection{\textbf{Empirical Validation of the ``Do Not Harm'' Strategy}}
\label{donot_harm}
%This section provides an internal \emph{non-maleficence} validation of \textbf{\textsc{Flare}}, testing whether its clustering and aggregation improve local specialization without reducing any cluster's predictive performance relative to the pretrained classifier. It analyzes the change in F1-scores between the intermediate `Base Pretrained \emph{(BpT)}' classifier (Section \ref{Step-1-design-choice}), and the final \textbf{\textsc{Flare}} model for every dataset, fold, and cluster (Figure~\ref{fig:clusterwise_improvement}). 

\begin{figure}[h]
    \centering
    \vspace{-5pt}
\includegraphics[width=\textwidth]{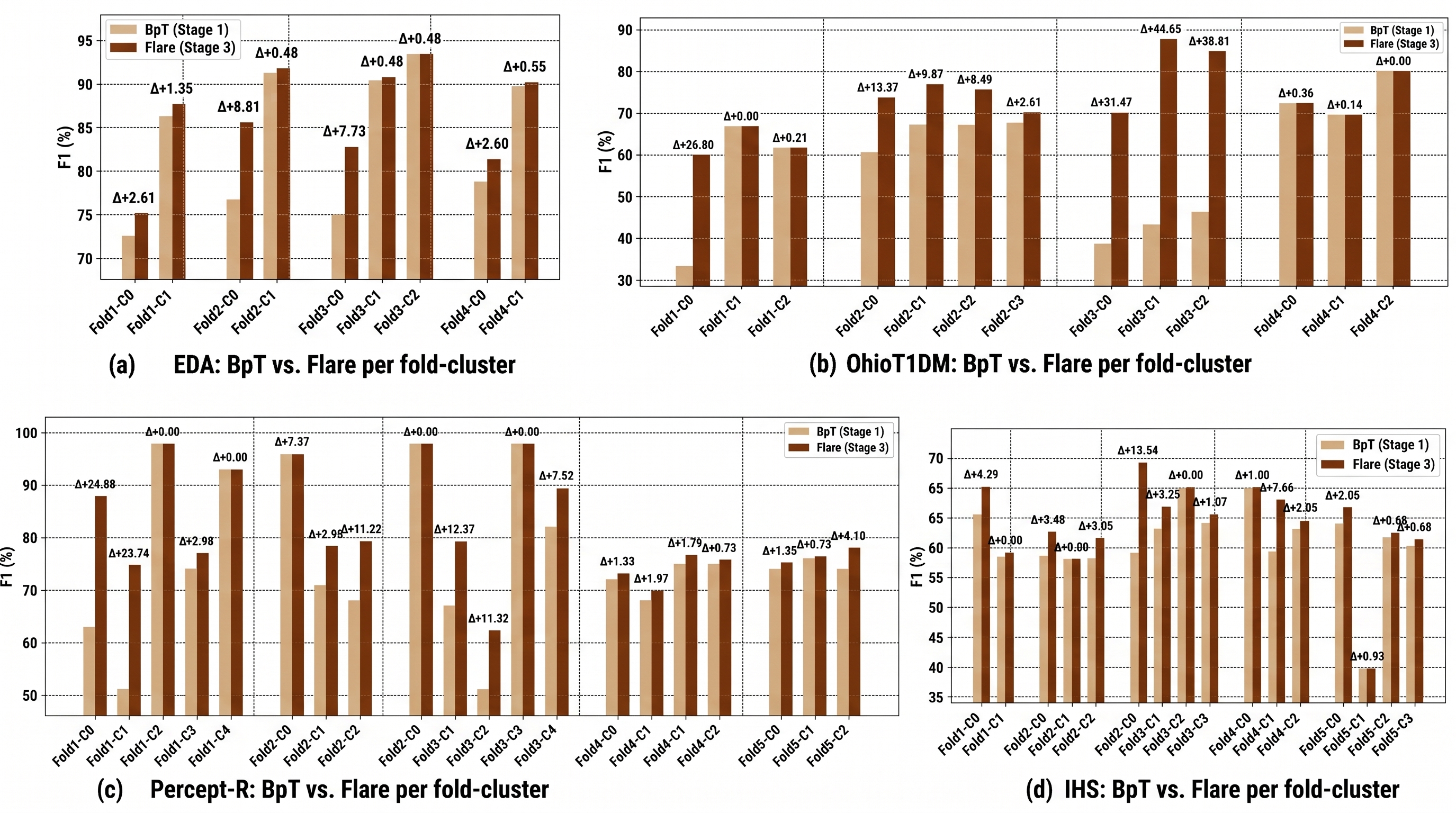}
    \caption{Fold and Cluster-wise F1 comparison across datasets within the proposed \textbf{\textsc{Flare}} framework. Bars show F1-scores for Base preTraining (BpT, light brown) as the intermediate Stage (1) and \textbf{\textsc{Flare}} (dark brown) as the final Stage (3) across all fold-cluster pairs. 
 Values above each pair ($\Delta$) indicate the change from BpT to \textbf{\textsc{Flare}}. }
    \label{fig:clusterwise_improvement}
    \vspace{-15pt}
\end{figure}

BHE measures harm only for subgroups defined by recorded attributes. The clusters that \textbf{\textsc{Flare}} adapts to, however, are formed per sample from model behavior, without any attribute, and a single participant's samples can fall into different clusters (Section~\ref{design-clustering}). As we show in Section~\ref{sec:cluster_interpretation}, these clusters are indeed not proxies for known attributes, and some are recurrent within-person states that no participant-level attribute can describe. Harm to such clusters would therefore be invisible to any attribute-based audit, including BHE. This section closes that gap by checking \emph{non-maleficence} at the level at which \textbf{\textsc{Flare}} actually operates: does adapting a model to each cluster ever make that cluster worse? For every dataset, fold, and cluster, we compare the F1-score of the intermediate Base Pretrained (\emph{BpT}) classifier $\theta^\ast$ (Section~\ref{Step-1-design-choice}) with that of the final cluster model $\theta^*_c$ (Figure~\ref{fig:clusterwise_improvement}). All model selection uses the separate validation (\texttt{val}) split, whereas $\Delta$F1 (in percentage points) is computed on held-out, person-disjoint test participants, so a non-negative $\Delta$F1 is an empirical outcome rather than a property guaranteed by construction.

\textbf{No cluster is harmed relative to the pretrained model.} Across all four datasets and every fold--cluster pair, $\Delta$F1 is non-negative: on participants never seen during training or selection, cluster-specific adaptation maintained or improved performance for every cluster. At the same time, the gains are often large, showing that avoiding harm did not come at the cost of benefit. They are largest where the pretrained model struggled most: OhioT1DM (Fold~3, Cluster~1: $+44.65$; Fold~3, Cluster~2: $+38.81$) and Percept-R (Fold~1, Cluster~0: $+24.88$; Fold~1, Cluster~1: $+23.74$; Fold~2, Cluster~2: $+11.22$). In the highly heterogeneous IHS dataset, adaptation still yields substantial gains (Fold~3, Cluster~0: $+13.54$; Fold~4, Cluster~1: $+7.66$), and even on EDA, where the pretrained model already performs well, \textbf{\textsc{Flare}} adds consistent improvements (Fold~2, Cluster~0: $+8.81$; Fold~3, Cluster~0: $+7.73$).

\textit{Because these clusters do not proxy known attributes (Section~\ref{sec:cluster_interpretation}), these are improvements that no attribute-based fairness method could have targeted. Together with the attribute-level results in Section~\ref{sec:baselines-vs-flare}, they show no observed harm both for the subgroups defined by known attributes (relative to the benign baseline) and for the latent strata \textbf{\textsc{Flare}} discovers itself (relative to the pretrained model).}

\subsubsection{\textbf{Analyzing Optimization Geometry through Loss Landscapes}}
\label{sec:loss-landscapes}
\begin{figure}[t]
    \centering
    \includegraphics[width=0.95\textwidth]{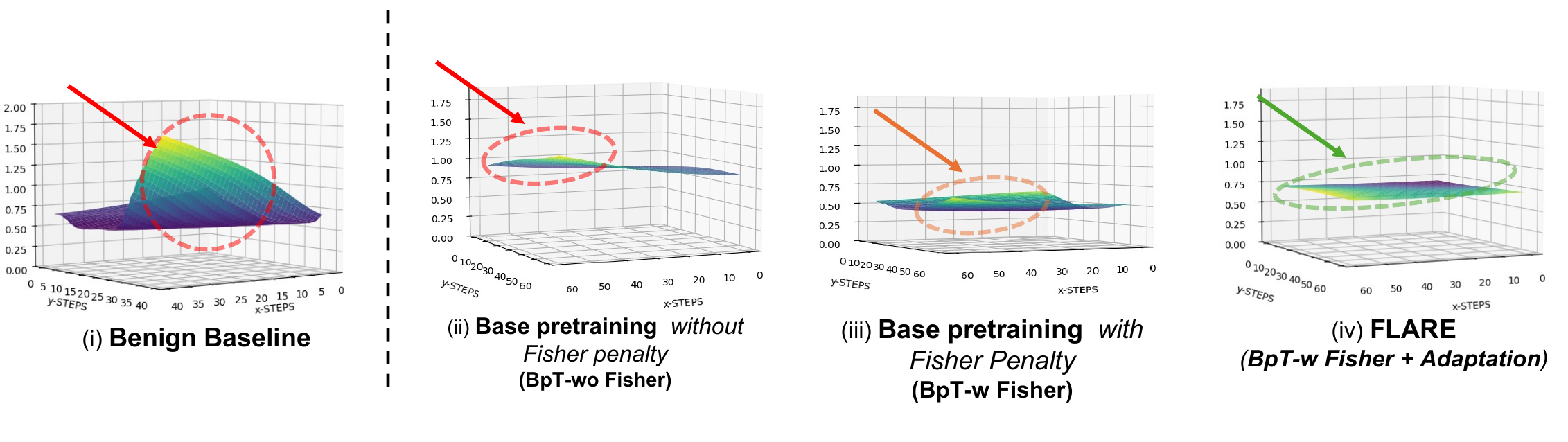}
    \vspace{-10pt}
\caption{Loss landscapes on OhioT1DM for (i) the benign baseline, (ii) base pretraining without the Fisher penalty, (iii) base pretraining with the Fisher penalty, and (iv) \textbf{\textsc{Flare}}. The \textit{red}, \textit{orange}, and \textit{green} arrows and dotted circles highlight curvature transitions, showing how Fisher regularization and cluster adaptation progressively flatten the optimization landscape. EDA and IHS are shown in Appendix~\ref{app:loss-landscape}.}
\vspace{-15pt}
\label{loss-landscape-vis}
\end{figure}

The quantitative results show \emph{that} \textbf{\textsc{Flare}} improves subgroup performance without harming any evaluated subgroup; the loss landscape helps explain \emph{why}. \textbf{\textsc{Flare}} assumes that sharp regions make predictions sensitive to small perturbations and therefore less reliable for nearby samples (Section~\ref{sec:design}), whereas flatter minima are associated with better generalization \cite{li2018visualizing,rangamani2020loss,li2025seeking}. If this premise holds, each \textbf{\textsc{Flare}} component should produce a flatter landscape than the preceding one. To test this, we visualize the loss landscapes of validation-selected models \cite{li2018visualizing}. Around the converged weights, we sweep along two random, filter-normalized, orthogonal directions ($X$- and $Y$-steps), with loss on the $Z$-axis. Using identical folds, we compare (i) the benign baseline, (ii) base pretraining without the Fisher penalty (\textit{BpT-wo Fisher}), (iii) base pretraining with the penalty (\textit{BpT-w Fisher}), and (iv) full \textbf{\textsc{Flare}} (\textit{BpT-w Fisher + Adaptation}).

\textbf{Each component further flattens the landscape.} On OhioT1DM (Figure~\ref{loss-landscape-vis}), the benign baseline occupies a sharp, irregular region with steep valleys and narrow basins (red), where small weight changes produce large loss changes. Base pretraining without the Fisher penalty smooths the surface but leaves local irregularities. Adding the Fisher penalty (orange) broadens the basin, consistent with curvature-aware regularization \cite{foret2021sam}. Full \textbf{\textsc{Flare}} (green) produces the flattest, widest basin. EDA and IHS show the same progression (Figure~\ref{fig:loss-landscape-app}, Appendix~\ref{app:loss-landscape}). \textit{This ordering is what \textbf{\textsc{Flare}}'s premise predicts: flatness increases with each added component, so the components build on rather than undo one another.}

\textbf{Adaptation specializes without sharpening.} Fitting separate models to small clusters could cause overfitting \cite{xuhong2018explicit} and convergence to sharper minima, which are associated with poorer generalization \cite{keskar2017largebatch}. \textit{Instead, \textbf{\textsc{Flare}}'s cluster-specific adaptation, do-no-harm regularization, and conditional aggregation refine the model while preserving a flat region. The ablation in Section~\ref{sec:network-ablation} provides quantitative support:} removing either the Fisher penalty or adaptation leads to harm for at least one subgroup on OhioT1DM, IHS, and Percept-R. Flatter minima are expected to make predictions less sensitive to distribution shifts \cite{cha2021swad}, such as differences in physiology or device, providing the stability needed by subgroups near decision boundaries. Because these visualizations depend on the random directions and axis scales, and sharpness itself can change under reparameterization \cite{dinh2017sharp}, we treat them as qualitative evidence that complements, rather than replaces, the quantitative results.

\subsection{Post-hoc Interpretation of Latent Clusters}
\label{sec:cluster_interpretation}
A natural question is whether the latent clusters of \textbf{\textsc{Flare}} simply recover known attributes; if they did, attribute-based fairness methods would suffice. Following post-hoc characterization in clinical clustering \cite{khan2023clustering,huang2021dice}, we learn the clusters first and only afterwards relate them to participant attributes that were never used for clustering (IHS: age, sex, specialty, ethnicity, and PHQ10 status; OhioT1DM: age group, sex, pump, and sensor band), using the training samples of a representative fold (IHS Fold~5, OhioT1DM Fold~2). Because clusters are assigned per sample, each participant is mapped to the cluster that holds most of their samples (\emph{majority assignment}). We then test single-attribute associations (permutation tests with FDR correction), joint predictability (logistic regression compared against a permutation-null AUROC), and sparse LASSO profiles \cite{tibshirani1996lasso}. Full results are reported in Appendices~\ref{sec:appendix_interpretation} and~\ref{sec:appendix_ihs_fold5}.

\begin{figure}[t]
\centering
\includegraphics[width=0.65\linewidth]{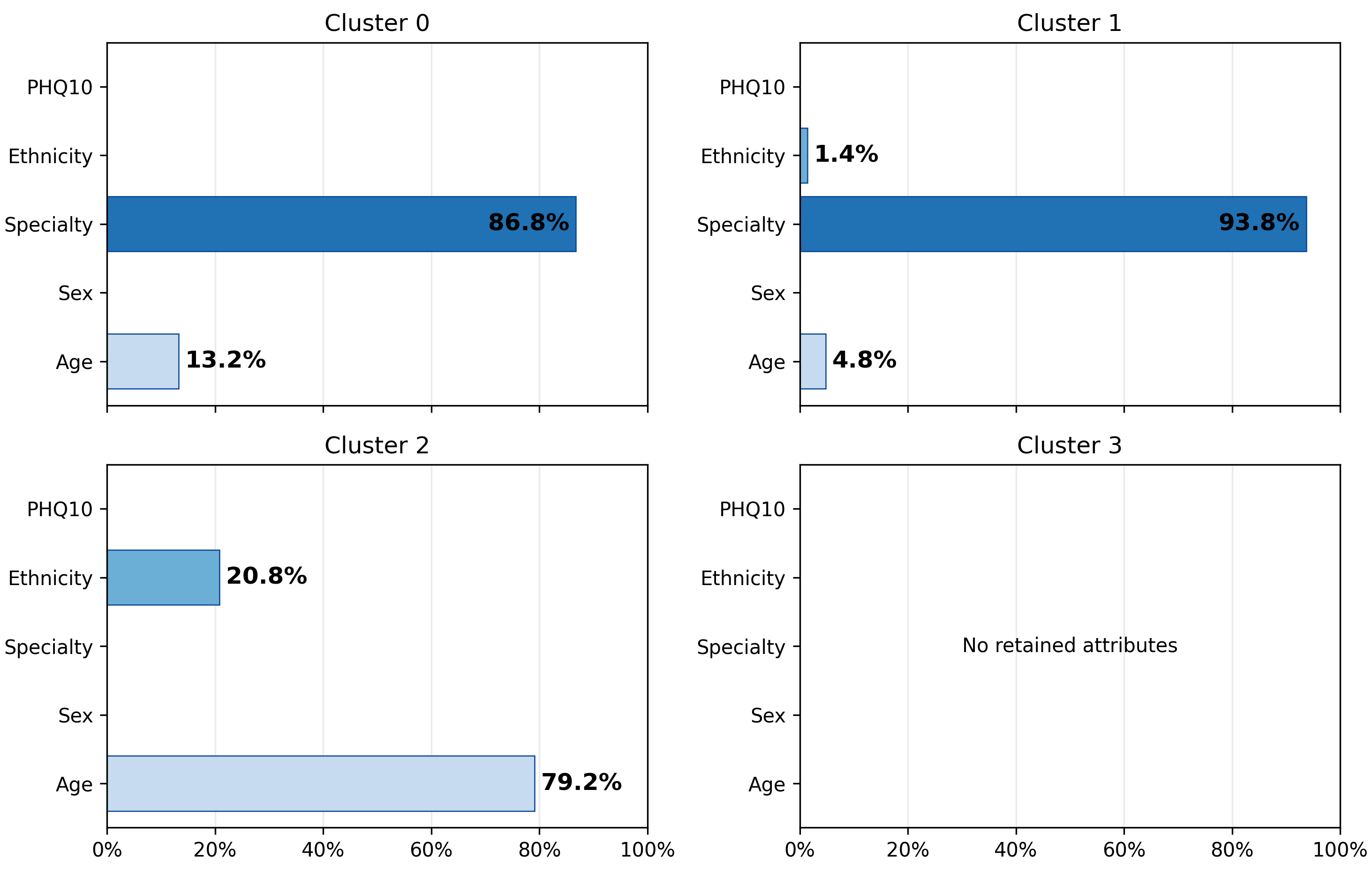}
\vspace{-2pt}
\caption{Sparse post-hoc LASSO profiles of the latent clusters in IHS Fold~5. Relative importance is the normalized absolute magnitude of the retained coefficients within each cluster; Clusters~0--2 use participant-level majority assignment. Cluster~3, which has no majority-assigned participants, retains no attribute. OhioT1DM profiles, which are at chance level, are reported in Appendix~\ref{sec:appendix_ohio_fold3} 
\vspace{-20pt}(Figure~\ref{fig:ohio_lasso}).}

\label{fig:cross_dataset_lasso}
\end{figure}

\textbf{Finding 1: No statistically reliable association with known attributes.} In IHS, the strongest association links Cluster~1 to residency specialty (Cram\'er's $V=0.538$, permutation $p=0.037$; Appendix Table~\ref{tab:ihs_cluster1_assoc}), and specialty dominates the LASSO profiles of Clusters~0 and~1 (Figure~\ref{fig:cross_dataset_lasso}). However, this association does not survive FDR correction ($q=0.554$), and all five attributes together predict Cluster~1 membership only modestly (AUROC $0.606$ vs.\ a null of $0.487$). In OhioT1DM, no attribute, including pump and sensor band, predicts cluster membership above chance   (observed AUROC $\approx$ null for every cluster; no test survives FDR correction). These analyses do not establish independence from the recorded attributes, particularly given the small participant counts, but they show that the clusters are not reducible to them.

\textbf{Finding 2: Some clusters are states, not types of people.} IHS Cluster~3 is visited by 64 of the 68 participants, yet it is not the majority cluster of any participant, and LASSO retains no attribute for it. It therefore captures a recurrent within-person state, i.e., a condition that most participants pass through at some point, rather than a group of participants.

\textbf{Takeaway: \textsc{Flare} targets performance strata that attribute-based fairness methods cannot.} The strata that \textbf{\textsc{Flare}} adapts to are only weakly aligned with demographic or device labels, and at least one of them changes within the same person over time. A method that requires a predefined attribute would not target these strata, because no recorded attribute defines them; \textbf{\textsc{Flare}} discovers them directly from model behavior.

\textbf{Known attributes can describe the clusters, not define them.} Although the specialty association is too weak to serve as a definition of any cluster, it does indicate which kinds of participants a cluster tends to contain (e.g., Cluster~1 is enriched for certain residency specialties). Thus, when sensitive attributes are available for auditing, the same post-hoc profiling can give practitioners an interpretable description of the subgroups that \textbf{\textsc{Flare}} discovers, even though these attributes are never used to form or adapt them.

%\textbf{The clusters are not proxies for known attributes.} In IHS, the strongest association links Cluster~1 to residency specialty (Cram\'er's $V=0.538$, permutation $p=0.037$; Appendix Table~\ref{tab:ihs_cluster1_assoc}), and specialty dominates the LASSO profiles of Clusters~0 and~1 (Figure~\ref{fig:cross_dataset_lasso}). However, this association does not survive FDR correction ($q=0.554$), and all five attributes together predict Cluster~1 membership only modestly (AUROC $0.606$ vs.\ a null of $0.487$). A further IHS cluster (Cluster~3) is visited by 64 of the 68 participants yet is not the majority cluster of any participant, and LASSO retains no attribute for it: it behaves as a recurrent within-person state rather than a type of person. In OhioT1DM, no attribute, including pump and sensor band, predicts cluster membership above chance (observed AUROC $\approx$ null for every cluster; no test survives FDR correction).

%These results support the premise of \textbf{\textsc{Flare}}. The performance strata it adapts to are only weakly aligned with demographic or device labels, and at least one (IHS Cluster~3) is a within-person state that no participant-level attribute can capture. A fairness method that relies on a predefined attribute would therefore not target these strata, whereas \textbf{\textsc{Flare}} discovers them from model behavior. At the same time, the partial alignment with specialty shows that, when attributes are available for auditing, post-hoc profiling can help practitioners describe the subgroups that \textbf{\textsc{Flare}} discovers.

\subsection{{Cluster sensitivity}}\label{sec:cluster-sensitivity}

\textbf{\textsc{Flare}} selects the number of clusters automatically rather than treating it as a manually tuned hyperparameter (Section~\ref{design-clustering}). Within each fold, it first estimates the count using the Bayesian Information Criterion (BIC) \cite{schwarz1978estimating}, a standard criterion for selecting mixture components \cite{fraley2002model,mclachlan2019finite}. It then lowers this estimate until every cluster contains samples in both \texttt{BpT-train} and \texttt{CAA-train}. The final count is therefore the largest feasible value at or below the BIC estimate, selected without test data or manual tuning. Counts vary across folds and datasets: Percept-R selects 5, 3, 5, 4, and 3 clusters; IHS selects 2, 3, 4, 3, and 4; EDA selects 2, 2, 3, and 2; and OhioT1DM selects 3, 4, 3, and 3. This variation suggests that no single fixed count can suit every fold.

To assess sensitivity to cluster count, we compare \textbf{\textsc{Flare}}'s fold-specific selection with configurations that use the same fixed count across all folds. Each column reports BHE computed from that configuration's pooled out-of-fold predictions. We do not report BHE per fold, because a single fold's test participants need not cover every subgroup of an attribute (e.g., a fold may contain participants of only one sex); pooling the folds evaluates each participant exactly once and covers the full population. Because different counts can perform differently across folds, the fold-specific configuration can outperform every fixed-count configuration. Its pooled BHE values are not averages of, and need not lie between, the fixed-count columns. Table~\ref{tab:cluster_sensitivity_all} reports OhioT1DM; Appendix~\ref{app:cluster_Sen} reports EDA, IHS, and Percept-R (Table~\ref{tab:appen_cluster_sensitivity_all}).

\textbf{No fixed count works across datasets.} On OhioT1DM, $n=4$ achieves the highest mean Benefit and Harm-Avoidance ($\Delta B=8.94$, $\Delta H=2.99$) but less Equity than \textbf{\textsc{Flare}} ($\Delta E=2.92$ vs.\ $5.19$). On EDA, the same count harms at least one subgroup and increases disparity ($\Delta H=-1.56$, $\Delta E=-0.91$). On the more heterogeneous IHS and Percept-R datasets, every fixed count harms at least one subgroup: mean $\Delta H$ ranges from $-3.45$ to $-19.68$ on IHS and from $-4.85$ to $-10.22$ on Percept-R. In contrast, \textbf{\textsc{Flare}} is positive on all three metrics ($+5.99$, $+1.74$, $+1.32$ on IHS; $+6.25$, $+1.67$, $+2.74$ on Percept-R). Although $n=3$ and $n=4$ reduce disparity more on Percept-R ($\Delta E=3.58$ and $4.22$), both harm at least one subgroup ($\Delta H=-6.50$ and $-4.85$), reproducing the leveling-down pattern of the baselines (Section~\ref{sec:baselines-vs-flare}).

\textbf{Both underclustering and overclustering fail.} Too few clusters merge behaviorally distinct samples, hiding poorly served subgroups within larger groups \cite{sohoni2020no}. With $n=1$, which assumes no subgroup structure, the model harms at least one subgroup on OhioT1DM, IHS, and Percept-R ($\Delta H=-6.20$, $-3.45$, and $-9.74$). Too many clusters fragment the data into small subsets on which cluster-specific fine-tuning can overfit \cite{xuhong2018explicit}. On IHS, for example, $\Delta H$ falls to $-19.68$ at $n=3$ and $-14.11$ at $n=4$.

\textbf{\textit{These results show that automated, fold-specific selection avoids both failures: \textsc{Flare} is the only setting with positive mean $\Delta B$, $\Delta H$, and $\Delta E$ on every dataset.}}

\begin{table*}[h]
\centering
\small
\captionsetup{font={color=black}}
\arrayrulecolor{black}
\color{black}
\vspace{-5pt}
%\caption{Cluster sensitivity analysis showing $(\Delta B,\Delta H,\Delta E)$ for different numbers of clusters per fold ($n=1,2,3,4$; )}
\caption{Cluster sensitivity analysis showing $(\Delta B,\Delta H,\Delta E)$ for different numbers of clusters per fold ($n=1,2,3,4$; for $n=1$, the selected model remains the Fisher-regularized pretrained model because no adapted checkpoint improved validation F1, so results equal BpT-w Fisher in Table~\ref{tab:ablation_bhe}) and \textbf{\textsc{Flare}} on OhioT1DM. Values are reported in percentage points; results for EDA, IHS, and Percept-R are in Appendix Table~\ref{tab:appen_cluster_sensitivity_all}.}
\label{tab:cluster_sensitivity_all}
\resizebox{0.94\textwidth}{!}{
\begin{tabular}{l|
ccc|ccc|ccc|ccc|ccc}
\toprule
\multirow{2}{*}{\textbf{\makecell{Sensitive \\ Attribute}}} &
\multicolumn{3}{c|}{\textbf{n=1 (\%)}} &
\multicolumn{3}{c|}{\textbf{n=2 (\%)}} &
\multicolumn{3}{c|}{\textbf{n=3 (\%)}} &
\multicolumn{3}{c|}{\textbf{n=4 (\%)}} &
\multicolumn{3}{c}{\textbf{\textsc{Flare} (\%)}} \\
&
$\Delta B$ & $\Delta H$ & $\Delta E$
& $\Delta B$ & $\Delta H$ & $\Delta E$
& $\Delta B$ & $\Delta H$ & $\Delta E$
& $\Delta B$ & $\Delta H$ & $\Delta E$
& $\Delta B$ & $\Delta H$ & $\Delta E$\\

\midrule
Age         & -4.04 & -7.87 & -3.82 &  0.38 & -2.70 & -1.33 & 3.73 & \textbf{3.40} & -0.34 &  \textbf{5.56} &  1.11 &  0.71 &  5.07 & 0.00 & \textbf{7.91} \\
Sex         & -4.02 & -4.38 &  0.50 &  4.49 &  \textbf{2.54} & \textbf{2.74} &  2.98 &  0.25 &  2.50 &  \textbf{5.57} & -0.20 & -1.80 &  1.96 & 0.48 & 2.09 \\
Pump        & -4.48 & -8.22 & -5.30 & 10.25 &  6.78 &  4.90 &  4.75 &  3.05 & -2.40 & \textbf{17.17} &  \textbf{6.87} &  \textbf{8.17} & 11.06 & 0.08 & 7.22 \\
Sensor Band & -3.87 & -4.31 & -0.63 &  5.22 &  \textbf{4.63} &  0.83 &  3.72 &  2.31 & 1.99 &  \textbf{7.44} &  4.19 & \textbf{4.59} &  2.54 & 0.03 & 3.55 \\
\rowcolor{gray!15} \textbf{Mean} 
& -4.10 & -6.20 & -2.31
&  5.09 &  2.81 &  1.79
&  3.80 &  2.25 & 0.44
&  \textbf{8.94} &  \textbf{2.99} &  2.92
& 5.16 & 0.15 & \textbf{5.19} \\

\bottomrule
\end{tabular}}
\arrayrulecolor{black}
\vspace{-15pt}
\end{table*}

\subsection{Assessing Structural Contributions via Detailed Ablation}
\label{sec:network-ablation}
\begin{table*}[!t]
\centering

\caption{Ablation study showing percentage changes $(\Delta B,\Delta H,\Delta E)$ for Base-Training with Fisher (BpT-w Fisher), Base-Training without Fisher (BpT-wo Fisher), Cluster Adaptation and Aggregation without base-trained Fisher (CAA-wo Fisher), and \textbf{\textsc{Flare} (CAA-w-Fisher)} across datasets. Values are percentage points.}
\label{tab:ablation_bhe}
\resizebox{0.8\textwidth}{!}{
\begin{tabular}{l|c|ccc|ccc|ccc|ccc}
\toprule
\textbf{Dataset} & 
\multirow{2}{*}{\textbf{\makecell{Sensitive \\ Attribute}}} &
\multicolumn{3}{c|}{\textbf{BpT-w Fisher (\%)}} &
\multicolumn{3}{c|}{\textbf{BpT-wo Fisher (\%)}} &
\multicolumn{3}{c|}{\textbf{CAA-wo Fisher (\%)}} &
\multicolumn{3}{c}{\textbf{\textsc{Flare} (\%)}}\\
& & $\Delta B$ & $\Delta H$ & $\Delta E$
 & $\Delta B$ & $\Delta H$ & $\Delta E$
 & $\Delta B$ & $\Delta H$ & $\Delta E$
 & $\Delta B$ & $\Delta H$ & $\Delta E$ \\
\midrule

OhioT1DM & Age & -4.04 & -7.87 & -3.82 & -1.24 & -4.87 & -1.82 & \textbf{5.65} & -2.22 & 5.62 & 5.07 & \textbf{0.00} & \textbf{7.91} \\
& Sex & -4.02 & -4.38 & 0.50 & -4.02 & -4.38 & 0.50 & -1.12 & -5.81 & -0.29 & \textbf{1.96} & \textbf{0.48} & \textbf{2.09} \\
& Pump & -4.48 & -8.22 & -5.30 & -4.48 & -8.22 & -5.30 & 8.36 & \textbf{1.33} & \textbf{9.94} & \textbf{11.06} & 0.08 & 7.22 \\
& Sensor Band & -3.87 & -4.31 & -0.63 & -3.89 & -4.33 & -0.62 & 1.06 & -2.74 & \textbf{5.37} & \textbf{2.54} & \textbf{0.03} & 3.55 \\
\rowcolor{gray!15}& \textit{Mean}
& -4.10 & -6.20 & -2.31
& -3.41 & -5.45 & -1.81
& 3.49 & -2.36 & 5.16
& \textbf{5.16} & \textbf{0.15} & \textbf{5.19} \\

\midrule

EDA & Group\_label & 1.70 & 0.54 & \textbf{1.04} & 1.98 & 0.82 & 0.96 & \textbf{2.17} & \textbf{1.29} & 0.56 & 2.16 & 0.85 & 0.57 \\
& Sex & 1.71 & 1.38 & -0.46 & 1.55 & 0.66 & \textbf{1.00} & \textbf{2.30} & 2.20 & -0.14 & 2.27 & \textbf{2.22} & 0.07 \\
\rowcolor{gray!15}& \textit{Mean}
& 1.71 & 0.96 & 0.29
& 1.77 & 0.74 & \textbf{0.98}
& \textbf{2.24} & \textbf{1.75} & 0.21
& 2.22 & 1.54 & 0.32 \\

\midrule

IHS & Sex & 0.00 & -0.71 & \textbf{0.76} & 0.92 & 0.65 & -0.37 & 3.68 & 3.45 & -0.32 & \textbf{4.97} & \textbf{4.93} & 0.06 \\
& Age & 3.78 & -4.85 & 1.39 & 2.45 & -4.85 & 2.20 & 5.64 & -4.80 & 1.97 & \textbf{9.72} & \textbf{0.30} & \textbf{3.07} \\
& Ethnicity & 0.29 & -0.37 & -2.74 & -0.85 & -22.73 & -9.45 & 1.22 & -15.88 & -8.57 & \textbf{8.23} & \textbf{3.38} & \textbf{0.28} \\
& Specialty & -0.06 & -10.32 & -0.35 & 0.68 & -9.57 & 0.94 & 3.78 & -10.23 & \textbf{1.07} & \textbf{5.00} & \textbf{0.03} & 0.36 \\
& PHQ10>0 & 0.00 & -0.98 & -1.39 & 1.63 & -0.11 & -2.46 & \textbf{3.36} & \textbf{2.81} & -0.77 & 2.04 & 0.04 & \textbf{2.83} \\
\rowcolor{gray!15}& \textit{Mean}
& 0.80 & -3.45 & -0.47
& 0.97 & -7.32 & -1.83
& 3.54 & -4.93 & -1.32
& \textbf{5.99} & \textbf{1.74} & \textbf{1.32} \\

\midrule

Percept-R & Sex & -2.11 & -2.91 & \textbf{0.93} & -2.95 & -4.19 & 0.30 & 2.28 & 1.02 & 0.29 & \textbf{5.12} & \textbf{4.70} & 0.59 \\
& Age & -1.90 & -11.83 & \textbf{6.75} & -2.99 & -11.06 & 5.03 & 2.42 & -6.62 & 6.28 & \textbf{9.60} & \textbf{0.02} & 3.35 \\
& Race & 0.19 & -18.02 & 2.83 & -2.08 & -19.39 & -9.84 & 2.78 & -12.82 & 4.84 & \textbf{7.43} & \textbf{0.00} & \textbf{5.77} \\
& Ethnicity & -5.03 & -6.19 & -1.64 & -4.94 & -5.10 & -0.23 & 0.78 & 0.58 & 0.28 & \textbf{2.84} & \textbf{1.96} & \textbf{1.25} \\
\rowcolor{gray!15}& \textit{Mean}
& -2.21 & -9.74 & 2.22
& -3.24 & -9.94 & -1.19
& 2.07 & -4.46 & \textbf{2.92}
& \textbf{6.25} & \textbf{1.67} & 2.74 \\

\bottomrule
\end{tabular}}
\vspace{-15pt}
\end{table*}

We conduct an ablation study to assess how each structural component of \textbf{\textsc{Flare}} contributes to ethical fairness. Specifically, we examine the effects of Fisher penalty regularization and the Cluster Adaptation and Aggregation (CAA) stage. Table~\ref{tab:ablation_bhe} summarizes the results. Overall, Fisher penalty regularization enhances representational smoothness and, on IHS and Percept-R, equity, while CAA enhances performance and helps avoid harm across latent clusters. The full \textbf{\textsc{Flare}} model, which combines both components, achieves the most balanced improvements across Benefit ($\Delta B$), Harm-Avoidance ($\Delta H$), and Equity ($\Delta E$), confirming their complementary roles without sensitive-attribute supervision. All ablation variants use the same splits and validation-based selection.

We compare four configurations:
\begin{itemize}
    \item \textbf{BpT-w Fisher}, base pretraining with Fisher penalty regularization;
  \item  \textbf{BpT-wo Fisher}, the same base pretraining procedure without Fisher penalty;
  \item  \textbf{CAA-wo Fisher}, which removes Fisher regularization from base pretraining and initializes CAA from BpT-wo Fisher, while retaining the remaining CAA procedure; and
 \item \textbf{FLARE (Full Model),} The complete framework integrating Fisher-regularized pretraining (\textit{BpT-w Fisher}) with the CAA stage (i.e., CAA-w-Fisher).
\end{itemize}

Across datasets, the ablation results show that neither Fisher regularization nor CAA alone is sufficient to produce consistently strong ethical fairness. On OhioT1DM, both base-only configurations degrade performance, with BpT-w Fisher producing $(\Delta B{=}-4.10,\ \Delta H{=}-6.20,\ \Delta E{=}-2.31)$ and BpT-wo Fisher producing $(\Delta B{=}-3.41,\ \Delta H{=}-5.45,\ \Delta E{=}-1.81)$. For OhioT1DM Sex and Pump, the with- and without-Fisher configurations produce identical BHE values at the reported precision; thus, the Fisher penalty does not yield a measurable subgroup-level change for those attributes in this ablation. Applying CAA after pretraining without the Fisher penalty (CAA-wo Fisher; Fisher removed from Stage~1 only) improves Benefit and Equity but still degrades Harm-Avoidance $(\Delta B{=}3.49,\ \Delta H{=}-2.36,\ \Delta E{=}5.16)$. Only the full \textbf{\textsc{Flare}} model avoids these regressions and achieves consistent, positive gains $(+5.16,\ +0.15,\ +5.19)$.

A similar pattern appears in EDA and IHS. On EDA, where the baseline already performs well, all configurations improve all three metrics when averaged across attributes and differ by about one point or less; \textbf{\textsc{Flare}} yields a stable outcome $(+2.22,\ +1.54,\ +0.32)$. On IHS, which contains stronger population heterogeneity, base-only and CAA-wo Fisher settings remain unstable, resulting in harm to at least one subgroup ($\Delta H = -3.45,\ -7.32,\ \text{and } -4.93$, respectively). In contrast, \textbf{\textsc{Flare}} correctly leverages both structural components to produce strong, consistent positive gains $(+5.99,\ +1.74,\ +1.32)$.

On Percept-R, Fisher regularization alone yields negative Benefit and strongly negative Harm-Avoidance $(\Delta B{=}-2.21,\ \Delta H{=}-9.74,\ \Delta E{=}2.22)$, and BpT-wo Fisher remains negative across all metrics $(\Delta B{=}-3.24,\ \Delta H{=}-9.94,\ \Delta E{=}-1.19)$. CAA-wo Fisher recovers some benefit and equity, but still suffers in harm-avoidance $(+2.07,\ -4.46,\ +2.92)$, suggesting that latent-cluster adaptation is useful, but still limited if the base representation lacks geometric stability. The full \textbf{\textsc{Flare}} model achieves the strongest Benefit and Harm-Avoidance gains with positive Equity $(+6.25,\ +1.67,\ +2.74)$.

\textit{These findings collectively demonstrate that Fisher regularization establishes a stable and equitable representational foundation, while the Cluster Adaptation and Aggregation (CAA) Stage leverages this stability to refine subgroup alignment and performance.}
\vspace{-10pt}
\section{On-device Efficiency during Training and Inference}
\label{on-device}

Evaluating on-device inference and training efficiency is critical for mobile and ubiquitous computing systems. The latency and memory consumption directly determine the feasibility of real-time deployment and energy sustainability \cite{stisen2015smart, lane2015can}. Therefore, to evaluate the practical feasibility of \textbf{\textsc{Flare}} during deployment, we benchmarked its training and inference efficiency against the benign baseline and four fairness baselines (ARL, KD, Reckoner, GoG) across seven hardware platforms, from high-end GPUs (NVIDIA RTX 4090, NVIDIA Spark GB10) and desktop-class CPUs (Apple M4, AMD Ryzen 9, Spark Cortex-X925) to resource-constrained devices (Raspberry Pi, Google Pixel 6), for all four datasets.

Tables~\ref{tab:runtime_platform_eda} and~\ref{tab:flare_breakdown_eda} summarize the deployment efficiency of \textbf{\textsc{Flare}} on Fold~1 of the EDA dataset, where the clustering mechanism identifies two latent clusters. Table~\ref{tab:runtime_platform_eda} compares end-to-end training and inference latency against the baselines across deployment platforms, while Table~\ref{tab:flare_breakdown_eda} decomposes \textbf{\textsc{Flare}} into its constituent stages to identify the source of the additional latency cost. We report wall-clock time normalized per sample, compute utilization, and peak resident set size (RSS), which together capture the latency and resource footprint relevant to on-device deployment in Table~\ref{tab:runtime_platform_eda}. For brevity, Table~\ref{tab:flare_breakdown_eda} reports only runtime latency (ms/sample), while memory and compute utilization metrics are provided in Appendix~\ref{runtime_appendix} (Table~\ref{tab:flare_breakdown_eda_rss_util}). Results for the remaining datasets are also included in Appendix~\ref{runtime_appendix}.

\textbf{\textit{Deployment workflow.}} In deployment, Stages~1--2 and the initial Stage~3 models are trained offline on a cohort. When a new user starts wearing the device, each sensed window is routed to a cluster model without labels (Section~\ref{design-clustering}) and predicted on-device. Labels that become available later (e.g., confirmed glucose readings or self-reports) can be used for lightweight cluster-specific fine-tuning on the device, while the GMM and pretrained model can be refit periodically offline when sufficient new data accumulate (this workflow is not evaluated in this paper); refitting the GMM changes the cluster definitions and therefore requires re-adapting the cluster-specific models. In practice, refitting can be triggered when monitoring on newly labeled data shows a sustained drop in cluster-level F1 relative to the validation F1 recorded at deployment, or when an increasing share of incoming samples receives low GMM posterior probability (i.e., fits no existing cluster well), both of which indicate distribution drift. Routing and inference costs are incurred per sample on each device and do not grow with the number of users, whereas pretraining and cluster fitting run offline once per cohort and are repeated only when such a trigger occurs. Because the compute-intensive pretraining and cluster-fitting stages run offline on a server or workstation, consistent with common edge deployment workflows~\cite{chen2019deep}, only lightweight cluster-specific adaptation runs on the device, which takes $\le$4.1\,ms per sample on platforms such as Raspberry Pi and Google Pixel~6 (Table~\ref{tab:flare_breakdown_eda} and Appendix Table~\ref{flare_breakdown_appendix_ihs_ohio_percept}).

\textbf{\textit{Inference-phase deployment feasibility: }} As shown in Table \ref{tab:runtime_platform_eda}(b), inference latency remains within a few milliseconds per sample (\textbf{2.03}–10.47 ms). 
The majority of this overhead arises from label-free cluster assignment: computing the descriptor $[z, L_{\mathrm{CE}}, \mathsf{F}]$ with the predicted label requires one forward and one backward pass through $\theta^\ast$, followed by UMAP projection and GMM evaluation. The cluster-specific model inference itself is comparable to the baseline (Table~\ref{tab:flare_breakdown_eda}).
%The majority of this overhead arises from cluster assignment and routing prior to invoking the cluster-specific model, while the model inference itself incurs negligible cost, comparable to the baseline as reported in Table \ref{tab:flare_breakdown_eda}. 
These results indicate that \textbf{\textsc{Flare}} is feasible for real-time inference on common edge devices such as Raspberry Pi and Google Pixel 6.
For applications with stricter latency requirements, the routing overhead can be further reduced using established edge-inference strategies such as device-edge partitioning and adaptive offloading~\cite{kang2017neurosurgeon,zeng2020coedge,li2018edge}. These optimizations are complementary to \textbf{\textsc{Flare}} and can reduce or amortize the assignment cost without changing the core specialization mechanism.

%For scenarios with stricter latency constraints, routing overhead can be further reduced using established edge-inference techniques such as device–edge partitioning and adaptive offloading ~\cite{kang2017neurosurgeon,zeng2020coedge,li2018edge}.These optimizations are complementary to \textbf{\textsc{Flare}} and can be applied to amortize or reduce the inference-time assignment cost without changing the core specialization mechanism.}

\textbf{\textit{Training-phase deployment feasibility: }}
During on-device training, \textbf{\textsc{Flare}} introduces additional latency compared with single-model baselines, as summarized in Table~\ref{tab:runtime_platform_eda}(a). The runtime breakdown in Table~\ref{tab:flare_breakdown_eda} shows that most of this overhead comes from GMM-based cluster fitting, while pre-training and cluster-specific adaptation remain comparatively small. Compute utilization and RSS memory are also comparable to the baselines, suggesting that \textbf{\textsc{Flare}} does not require substantially different resources beyond the expected cost of cluster assignment. 

%{In practical deployments, the more compute-intensive pre-training and cluster-fitting stages can be performed offline on a resource-rich server or workstation, consistent with common deep-learning edge deployment workflows~\cite{chen2019deep}. After this offline preparation, only lightweight cluster-specific adaptation needs to run on the edge device to facilitate any incremental updates with evolving data. As shown in Table~\ref{tab:flare_breakdown_eda}, this adaptation step remains within a few milliseconds ($\le$3.5ms) per sample on platforms such as Raspberry Pi and Google Pixel 6, demonstrating the feasibility of edge adaptation.}

\textbf{\textit{Ethical Fairness and runtime latency trade-off:}} \textbf{\textsc{Flare}} introduces modest training- and inference-time latency due to latent subgroup assignment and routing. However, the BHE metrics in Table~\ref{tab:bhe_summary} show that \textbf{\textsc{Flare}} is the only method with positive Benefit and non-negative Harm-Avoidance on all four datasets. Because the added latency remains in the millisecond-per-sample regime, which is practical for real-world deployment \cite{gao2020edgedrnn}, \textbf{\textsc{Flare}} offers a practical trade-off: a small runtime overhead for more ethically grounded and fair model behavior.

%\textbf{\textsc{Flare}} introduces a modest increase in both training- and inference-phase latency due to latent subgroup assignment and routing overheads. However, as evidenced by the BHE metrics in Table~\ref{tab:f1_demographics_models} (Section \ref{res-sec}), The modest increase in latency yields consistent gains in ethical fairness in comparison to Baseline, KD, ARL, GoG and Reckoner approaches. Importantly, the additional latency remains within the millisecond-per-sample regime during both training- and inference-phase, making it practical for real-world deployment \cite{gao2020edgedrnn}. Overall, this represents a favorable trade-off: a modest latency overhead enables systematically improved ethically grounded fair model behavior.}

%and ethically grounded model behavior \textit{without compromising fairness or harming any subgroup}.}

\begin{table}[h!]
\color{black}
\centering
\scriptsize
\vspace{-10pt}
\caption{Training and Inference runtime latency and resource comparison across deployment platforms for the EDA dataset (wall time in ms/sample; compute utilization in \%; peak RSS memory in MB).}
\label{tab:runtime_platform_eda}
\begin{subtable}{\textwidth}
\label{tab:train_runtime_platform_eda}
\vspace{-10pt}
\caption{Training latency and resource comparison for the EDA dataset.}
\resizebox{\textwidth}{!}{%
\begin{tabular}{lrrrrrrrrrrrrrrrrrrrrr}
\toprule
Approach & \multicolumn{3}{c}{AMD CPU} & \multicolumn{3}{c}{NVIDIA GPU} & \multicolumn{3}{c}{Spark CPU} & \multicolumn{3}{c}{Spark GPU} & \multicolumn{3}{c}{Apple M4} & \multicolumn{3}{c}{Google Pixel 6} & \multicolumn{3}{c}{Raspberry Pi} \\
\cmidrule(lr){2-4} \cmidrule(lr){5-7} \cmidrule(lr){8-10} \cmidrule(lr){11-13} \cmidrule(lr){14-16} \cmidrule(lr){17-19} \cmidrule(lr){20-22}
 & Time & Util. & RSS & Time & Util. & RSS & Time & Util. & RSS & Time & Util. & RSS & Time & Util. & RSS & Time & Util. & RSS & Time & Util. & RSS \\
\midrule
Baseline & 0.237 & 3.09 & 551.7 & 0.407 & 62.35 & 986.4 & 0.272 & 4.97 & 485.3 & 0.651 & 4.07 & 1365.8 & 0.362 & 4.91 & 295.1 & 0.930 & 10.88 & 321.9 & 1.391 & 20.99 & 411.6 \\
KD & 0.389 & 3.12 & 552.7 & 0.678 & 76.98 & 1023.0 & 0.465 & 5.04 & 485.8 & 1.114 & 7.11 & 1430.8 & 0.273 & 9.74 & 297.2 & 1.845 & 11.71 & 322.2 & 2.304 & 25.05 & 413.6 \\
ARL & 0.266 & 3.11 & 552.6 & 0.421 & 64.49 & 1002.3 & 0.291 & 5.02 & 486.3 & 0.601 & 3.99 & 1390.5 & 0.170 & 9.83 & 296.9 & 1.040 & 11.21 & 322.3 & 1.224 & 24.96 & 414.0 \\
GoG & 0.427 & 3.13 & 757.6 & 0.246 & 43.03 & 1166.1 & 0.453 & 5.05 & 677.5 & 0.336 & 0.00 & 1635.4 & 0.182 & 9.70 & 525.5 & 1.200 & 11.38 & 547.6 & 1.254 & 24.84 & 630.3 \\
Reckoner & 0.648 & 3.13 & 581.1 & 0.800 & 80.43 & 1004.2 & 0.787 & 5.05 & 506.9 & 1.395 & 10.79 & 1393.9 & 0.457 & 9.91 & 330.8 & 3.328 & 12.04 & 350.8 & 3.975 & 25.09 & 433.4 \\
\textsc{Flare} & 6.369 & 2.39 & 910.5 & 7.386 & 57.49 & 1510.3 & 7.783 & 3.94 & 858.8 & 8.021 & 1.79 & 1979.8 & 12.188 & 6.42 & 619.9 & 19.205 & 9.87 & 494.6 & 25.637 & 19.81 & 692.7 \\
\bottomrule
\end{tabular}%
}
\end{subtable}

\begin{subtable}{\textwidth}
\centering
\scriptsize
\caption{Inference runtime and resource comparison for the EDA dataset.}
\label{tab:infer_runtime_platform_eda}
\resizebox{\textwidth}{!}{%
\begin{tabular}{lrrrrrrrrrrrrrrrrrrrrr}
\toprule
Approach & \multicolumn{3}{c}{AMD CPU} & \multicolumn{3}{c}{NVIDIA GPU} & \multicolumn{3}{c}{Spark CPU} & \multicolumn{3}{c}{Spark GPU} & \multicolumn{3}{c}{Apple M4} & \multicolumn{3}{c}{Google Pixel 6} & \multicolumn{3}{c}{Raspberry Pi} \\
\cmidrule(lr){2-4} \cmidrule(lr){5-7} \cmidrule(lr){8-10} \cmidrule(lr){11-13} \cmidrule(lr){14-16} \cmidrule(lr){17-19} \cmidrule(lr){20-22}
 & Time & Util. & RSS & Time & Util. & RSS & Time & Util. & RSS & Time & Util. & RSS & Time & Util. & RSS & Time & Util. & RSS & Time & Util. & RSS \\
\midrule
Baseline & 0.015 & 1.53 & 551.9 & 0.016 & 90.00 & 986.9 & 0.015 & 4.85 & 485.3 & 0.016 & 10.00 & 1365.6 & 0.009 & 5.28 & 295.1 & 0.057 & 11.44 & 322.2 & 0.061 & 24.54 & 411.6 \\
KD & 0.015 & 1.53 & 552.9 & 0.018 & 92.00 & 1023.2 & 0.015 & 2.41 & 485.8 & 0.016 & 10.00 & 1430.8 & 0.009 & 5.39 & 297.2 & 0.067 & 12.59 & 322.6 & 0.061 & 24.61 & 413.6 \\
ARL & 0.015 & 1.53 & 552.8 & 0.016 & 89.00 & 1002.6 & 0.015 & 2.43 & 486.3 & 0.017 & 10.00 & 1390.4 & 0.009 & 5.57 & 296.9 & 0.064 & 10.19 & 322.7 & 0.061 & 24.53 & 414.0 \\
GoG & 0.015 & 1.53 & 593.1 & 0.016 & 66.00 & 1166.3 & 0.015 & 4.88 & 504.1 & 0.016 & 0.00 & 1635.5 & 0.009 & 5.47 & 525.5 & 0.065 & 11.37 & 355.2 & 0.061 & 24.53 & 608.5 \\
Reckoner & 0.015 & 3.05 & 572.8 & 0.024 & 94.00 & 999.9 & 0.031 & 3.57 & 496.7 & 0.016 & 11.00 & 1393.2 & 0.019 & 6.34 & 330.8 & 0.118 & 12.65 & 343.2 & 0.121 & 24.69 & 433.4 \\
\textsc{Flare} & 2.375 & 3.09 & 1027.4 & 2.413 & 67.67 & 1655.9 & 2.893 & 4.96 & 971.8 & 2.897 & 6.50 & 2133.8 & 2.032 & 9.14 & 709.3 & 4.624 & 12.48 & 547.3 & 10.474 & 24.80 & 785.2 \\
\bottomrule
\end{tabular}%
}
\end{subtable}
\vspace{-10pt}
\end{table}

\begin{table}[h]
\color{black}
\centering
\scriptsize
\vspace{-5pt}
\caption{\textsc{FLARE} training and inference stage latency across platforms for the EDA dataset (wall time in ms/sample).}
\vspace{-5pt}
\label{tab:flare_breakdown_eda}
\resizebox{\linewidth}{!}{%
\begin{tabular}{llccccccc}
\toprule
Phase & \textbf{\textsc{Flare}} pipeline stage 
& AMD CPU 
& NVIDIA GPU 
& Spark CPU 
& Spark GPU 
& Apple M4 
& Google Pixel 6 
& Raspberry Pi \\
\midrule
Training & Pre-training stage 
& 0.211 & 0.338 & 0.308 & 0.596 & 0.157 & 1.456 & 1.714 \\
Training & Cluster fitting stage 
& 5.416 & 5.630 & 6.365 & 5.769 & 11.472 & 10.815 & 17.691 \\
Training & Training cluster assignment 
& 0.012 & 0.013 & 0.014 & 0.013 & 0.015 & 0.014 & 0.012 \\
Training & Adaptation for Cluster 1 
& 0.382 & 0.733 & 0.554 & 0.904 & 0.299 & 3.471 & 3.093 \\
Training & Adaptation for Cluster 2 
& 0.348 & 0.672 & 0.542 & 0.739 & 0.245 & 3.449 & 3.126 \\
\midrule
Inference & Inference cluster assignment 
& 2.360 & 2.387 & 2.870 & 2.881 & 2.023 & 4.476 & 10.338 \\
Inference & Cluster-specific model inference 
& 0.015 & 0.026 & 0.023 & 0.016 & 0.009 & 0.148 & 0.136 \\
\bottomrule
\end{tabular}%
}
\vspace{-10pt}
\end{table}

\section{Implications for Ubiquitous Health Sensing}
\label{sec:implications}
\textbf{Audit beyond demographics.} Device and clinical-context disparities were comparable to, or larger than, demographic disparities. For the benign baseline, the subgroup F1 standard deviation on OhioT1DM was 11.37 points across insulin-pump types, compared with 12.75 across age groups and 3.17 across sex; on EDA, it was 4.47 points across experimental groups (control, pre-dose, and post-dose) versus 1.13 across sex (Appendix Table~\ref{tab:f1_demographics_models_reduced}). Fairness audits in ubiquitous health sensing should therefore report BHE across non-demographic sources of heterogeneity, such as device type and clinical context, whenever they are recorded, not only across demographics.

\textbf{Report harm, not only parity.} Lower disparity does not guarantee that subgroups improve. Fair-DP and Fair-EO have negative mean $\Delta H$ on all four datasets, even though their mean $\Delta E$ is positive on every dataset (Table~\ref{tab:bhe_summary}). Reporting $\Delta H$ alongside parity reveals such trade-offs, including leveling down, which a disparity metric alone would count as an improvement.

\textbf{Collect attributes for evaluation, not training.} Fairness evaluation in mobile and wearable computing remains limited, partly because sensitive attributes are not always collected or reported \cite{yfantidou2023beyond}, and practitioners face legal, privacy, and organizational barriers to collecting them \cite{andrus2021we}. Because \textbf{\textsc{Flare}} requires no sensitive attributes for training, studies can collect a minimal, consented set solely for auditing and exclude it from training and deployment, a separation \textbf{\textsc{Flare}} shares with other fairness-without-demographics methods but attribute-aware methods lack. This practice aligns with emerging regulation, such as the EU AI Act's provision allowing sensitive data to be processed strictly for bias detection and correction with safeguards against reuse \cite{smuha2025regulation}, and with earlier proposals for fairness auditing without sensitive data in the modeling pipeline \cite{veale2017fairer}.

\textbf{Budget for routing.} \textbf{\textsc{Flare}} can run where ubiquitous health sensing actually happens: on the user's device. Label-free routing and inference take at most 8.1\,ms per sample on a Google Pixel~6 smartphone and 10.5\,ms on a Raspberry Pi across all four datasets (Section~\ref{on-device}; Table~\ref{tab:runtime_platform_eda} and Appendix Table~\ref{inference_runtime_appendix_ihs_ohio_percept}). This is negligible relative to typical sensing intervals: even at the 5-minute rate at which CGM readings arrive in OhioT1DM \cite{marling2020ohiot1dm}, it is less than 0.004\% of the interval. As described in Section~\ref{on-device}, the compute-intensive pretraining and cluster fitting run offline, so only label-free routing, prediction, and optional lightweight fine-tuning need to run on the device. Ethically fair prediction can therefore stay close to the data, without a server round-trip for each sample.

\section{Discussions and Limitations}
\label{limit}
% Overhead cost as we are creating a skeleton, then clustering, and then adaptation.
%Although \textbf{\textsc{Flare}} advances ethical and demographic-agnostic fairness, it introduces certain computational and methodological limitations.
\textbf{\textsc{Flare}} improves ethical fairness without requiring sensitive attributes during training. However, it also presents computational and methodological considerations that open avenues for future research and refinement.

\begin{enumerate}
    \item \textbf{Computational overhead.} \textbf{\textsc{Flare}} adds modest training and inference latency through cluster fitting, label-free assignment, and routing. Section~\ref{on-device} quantifies these costs and explains how the more expensive steps can be performed offline.

    \item \textbf{Evaluation protocol and comparison with prior work.} Our OhioT1DM results are lower than those reported in prior work~\cite{marling2020ohiot1dm}, which uses data from the same participants for training and testing. Our person-disjoint protocol instead evaluates each participant using a model trained only on other participants. This setting better reflects deployment to new users, but its absolute F1 scores are not directly comparable with results from participant-dependent evaluations.

    \item \textbf{Meaning of the latent subgroups.} \textbf{\textsc{Flare}} discovers subgroups from embeddings, loss, and curvature without using sensitive attributes. The post-hoc analysis in Section~\ref{sec:cluster_interpretation} finds only weak associations between these clusters and known attributes. The strongest association, with IHS specialty, does not remain significant after FDR correction, and one cluster behaves as a recurrent within-person state rather than a group of people. This suggests that the clusters capture patterns beyond recorded attributes. However, the meaning of some clusters remains unclear, and the analysis covers only one representative fold per dataset. Because the clusters reflect optimization behavior rather than predefined categories, we cannot link their disparities precisely to specific demographic or other heterogeneity factors. Moreover, the known attributes used during evaluation are only proxies for broader heterogeneity. Existing FWD approaches share this limitation \cite{lahoti2020fairness,chai2022fairness,ni2024fairness,luo2025fairness}. Making latent subgroups easier to interpret remains important future work.

    \item \textbf{Calibration-dependent routing.} At test time, cluster assignment uses the pretrained model's predicted label (Section~\ref{design-clustering}). If the model makes a high-confidence error, the sample receives a low-loss descriptor and may be routed as if it were correctly classified. Routing quality therefore depends on the calibration of the pretrained model.

    \item \textbf{Scope and uncertainty of BHE.} BHE evaluates subgroups defined by recorded attributes. Within-person strata are evaluated only through cluster-wise $\Delta$F1, measured relative to the pretrained model rather than the benign baseline (Section~\ref{donot_harm}). We report point estimates from fixed-seed models using test predictions pooled across all person-disjoint folds. We do not report seed-based confidence intervals because each BHE value compares two independently trained models: the evaluated method and the benign baseline. Seed variability affects both models, so the interval would depend on how their seeds are paired. In addition, $\Delta H$ is a minimum and $\Delta E$ is a difference between standard deviations, making averages across seeds difficult to interpret. Participant sampling remains the main source of uncertainty. Some subgroups contain few participants, so small values, especially of $\Delta H$, should be interpreted cautiously. Consistency across four datasets and fifteen attributes provides stronger evidence than any single estimate.

    \item \textbf{Limited large-scale evaluation.} Large human-sensing datasets with detailed demographic annotations remain scarce, preventing evaluation on broader cohorts. Our experiments span wearable sensing (EDA), mobile sensing in the wild (IHS), clinical monitoring (OhioT1DM), and clinical speech sensing (Percept-R). However, few datasets combine meaningful scale with sufficiently rich sensitive-attribute annotations \cite{stateofalgobias}. Even large initiatives such as GLOBEM \cite{xu2023globem} omit demographic or sensitive information because of privacy, consent, and ethical constraints \cite{linna2020ethical}. This gap also affects ubiquitous and mobile sensing research more broadly \cite{yfantidou2023beyond,uncoveringbias}. Future large-scale validation of \textbf{\textsc{Flare}} and fairness research in human-centered AI will require ethically curated datasets that balance participant privacy with information about demographic and other sources of heterogeneity.

    \item \textbf{Autonomy is not explicitly addressed.} \textbf{\textsc{Flare}} operationalizes ethical principles: beneficence, non-maleficence, and justice, but not \textit{autonomy}. Supporting autonomy requires interactive, explainable mechanisms that let users understand, question, and override AI decisions. Because this work focuses primarily on fairness and its ethical dimensions, autonomy-supporting interaction design remains an important direction for future work.%Because this work focuses on fairness and distribution-level ethical behavior, autonomy-supporting interaction design remains an important direction for future work.

\end{enumerate}

\section{Conclusion}
This paper introduced \textbf{\textsc{Flare}}, to our knowledge the first attribute-free framework to operationalize beneficence, non-maleficence, and justice in ubiquitous health sensing. Rather than relying on sensitive-attribute supervision, it discovers the latent subgroups that a model treats differently from its own behavior, and adapts a separate model to each with a do-no-harm objective. Across four health-sensing datasets, \textbf{\textsc{Flare}} was the only method that improved mean subgroup F1 without degrading any evaluated subgroup and the only one that reduced disparity for every sensitive attribute, although some baselines achieved larger average disparity reductions at the cost of harming a subgroup. Its millisecond-range inference on a smartphone and a Raspberry~Pi further supports its practicality for on-device deployment.
As sensing models increasingly shape health and well-being decisions in everyday life, reducing disparity alone is not enough: fairness gains should not come at the expense of those already well served. \textbf{\textsc{Flare}} demonstrates that models can discover underserved strata without knowing demographic or other sensitive attributes and improve their performance without leveling down others. Together, \textbf{\textsc{Flare}} and the BHE metrics offer a concrete foundation for building and auditing ubiquitous health-sensing systems in which benefit, harm avoidance, and equity are pursued together.

%This paper introduces \textbf{\textsc{Flare}}—the first principled framework for achieving ethical fairness without access to demographic attributes. By coupling Fisher-based curvature regularization with behavior-aware latent clustering and collaborative, do-no-harm adaptation, it unifies algorithmic fairness with ethical principles of beneficence, non-maleficence, justice, and autonomy.

%By integrating Fisher Information–based curvature regularization with a do-no-harm cluster adaptation, it embeds ethical principles directly into model optimization. 

%\textbf{\textsc{Flare}} achieves higher predictive accuracy (\emph{beneficence}), safeguards all demographic groups from degradation (\emph{non-maleficence}), reduces variance across groups (\emph{justice}), and enhances consistency across individuals (\emph{autonomy}). Empirical evidence across wearable datasets confirms that stability-oriented optimization can ensure fairness, equity, and robustness without relying on sensitive demographic data—demonstrating a strong step toward reliable, human-centered, and ethically aligned AI systems.

% \section{Rights Information}

% \section{Sectioning Commands}

% \section{Tables}

% \section{Math Equations}

% \section{Acknowledgments}


\bibliographystyle{ACM-Reference-Format}
\bibliography{sample-base}









\appendix
\section{Methodological Transparency \& Reproducibility Appendix (META)}
\subsection{Data and Model Details}
\label{appendix:data_models}

This section provides additional details on the datasets, preprocessing pipelines, feature composition, and neural architectures used in \textbf{\textsc{Flare}}.  
All datasets were standardized to support a consistent autoencoder–classifier formulation and trained under identical preprocessing and optimization protocols, ensuring comparability across sensing domains.

\subsubsection{OhioT1DM Dataset}
The OhioT1DM dataset~\cite{marling2020ohiot1dm} includes multimodal records from individuals with Type~1 Diabetes collected in 2018 and 2020. The data contain continuous glucose monitoring (CGM) readings, insulin pump records, and contextual annotations (e.g., meals, work, and exercise events). OhioT1DM contains 12 participants, whose data are released with predefined training and testing files~\cite{marling2020ohiot1dm}. After preprocessing, one participant's testing file contains no eligible samples; therefore, 11 participants contribute to the held-out evaluation (4 folds), and that participant appears only in training. Demographic and device information is reported in~\cite{marling2020ohiot1dm}.  
Preprocessing follows Appendix~2.4 of Arefeen and Ghasemzadeh~\cite{arefeen2023designing}, including temporal alignment of CGM and pump data, removal of incomplete sequences, and normalization within each participant session.

\paragraph{Feature Representation.}  
Each instance consists of seven features:
\begin{enumerate}
    \item \texttt{basal\_insulin} — continuous basal insulin delivery rate;
    \item \texttt{insulin\_bolus} — short-acting bolus insulin dosage;
    \item \texttt{time\_since\_last\_bolus} — elapsed time since previous insulin injection;
    \item \texttt{meal\_intake} — carbohydrate content of meals (grams);
    \item \texttt{work} — binary contextual indicator of work activity;
    \item \texttt{exercise} — binary contextual indicator of exercise activity;
    \item \texttt{three\_hour\_glucose} — CGM measurement three hours post-meal.
\end{enumerate}
The binary prediction label indicates normal glucose ($\leq 180$\,mg/dL) or hyperglycemia ($>180$\,mg/dL).

\paragraph{Demographic and Device Attributes.}  
The following attributes were used for subgroup evaluation:
\begin{itemize}
    \item \textbf{Sex:} male, female ;
    \item \textbf{Age:} 20–40, 40–60, and 60–80 years (participant-level age ranges as released with the dataset);
    \item \textbf{Pump Model:} Medtronic 530G or 630G (two generations of insulin pumps);
    \item \textbf{Sensor Band:} Empatica or Basis (wrist-worn sensor brands). Sensor band coincides with the data-collection cohort (2018 cohort = Basis, 2020 cohort = Empatica), so cohort is not evaluated as a separate attribute.
\end{itemize}
These subgroups capture both physiological (sex, age) and technical (pump, sensor) variation, which drive systematic performance differences in glucose prediction tasks.

\paragraph{Model Architecture }  
OhioT1DM Model is a fully connected autoencoder–classifier network. The encoder has four layers (256–128–64–32) with \texttt{Tanh} activations and dropout rates of 0.5, 0.3, and 0.3, respectively. The decoder mirrors this structure, and the classifier is a two-layer MLP (32–8–2) with \texttt{ReLU} and dropout (0.3). Training minimizes a weighted combination of reconstruction loss (MSE) and cross-entropy loss.

\subsubsection{Intern Health Study (IHS) Dataset}
The Intern Health Study (IHS)~\cite{adler2021identifying} is a 14-month longitudinal dataset tracking medical interns from two months before internship through the end of their first year. It includes demographics, PHQ-9 mental health assessments, and daily mood self-reports (1–10). Fitbit data provide daily summaries of sleep, step count, and heart rate. We analyze cohorts from 2018–2022.

\paragraph{Data Preprocessing.}  
We use 85 participants (17 per fold). Daily mood labels are binarized following Manjunath et~al.~\cite{manjunath2023can}: moods $>8$ are labeled positive (1), moods $<3$ negative (0), and intermediate values excluded. Sensor data are aggregated to daily averages; missing values are imputed with mean substitution, and binary “missing-indicator” flags are added. All numeric features are z-normalized per participant.

\paragraph{Feature Representation.}  
Eighteen input features are used: 9 Fitbit-derived measures (e.g., sleep duration, sleep phases, resting heart rate, step count) and 9 corresponding missing-indicator variables. These features capture daily physical activity and rest patterns linked to stress and mood variation.

\paragraph{Demographic and Behavioral Attributes.}  
Subgroups for model evaluation include:
\begin{itemize}
    \item \textbf{Sex:} male, female (self-identified at baseline);
    \item \textbf{Age:} age at baseline in single years (13 values: 23, 24, 25, 26, 27, 28, 29, 30, 32, 33, 34, 38, 51);
    \item \textbf{Ethnicity:} White, Black/African American, Latino/Hispanic, Asian, Multiracial, Arab/Middle Eastern;
    \item \textbf{Residency Specialty:} Internal Medicine, Surgery, Obstetrics/Gynecology (Ob/Gyn), Pediatrics, Psychiatry, Neurology, Emergency Medicine, Medicine/Pediatrics (Med/Peds), Family Practice, Transitional, Anesthesiology, Otolaryngology;
    \item \textbf{Baseline Depression (PHQ10$>$0):} yes, no (presence of depressive symptoms prior to internship).
\end{itemize}

These attributes represent both demographic diversity and occupational heterogeneity, capturing psychosocial and contextual factors associated with stress, sleep, and mood variation among medical interns.

\paragraph{Model Architecture}  
The IHS Model encoder uses three layers (128–64–32) with \texttt{ReLU} activations and dropout (0.1); the decoder mirrors this structure. The classifier has three layers (32–16–8–2) with \texttt{ReLU} and dropout (0.1). Models are optimized using Adam ($\eta=3.28\times10^{-4}$), and hyperparameters are tuned via \textit{Optuna} \cite{shekhar2021comparative} to maximize F1 performance and included in Appendix \ref{app:hype}.

\subsubsection{EDA Dataset}
The EDA dataset~\cite{xiao2025human} captures multimodal physiological signals, including electrodermal activity (EDA), heart rate (HR), temperature (TEMP), acceleration (ACC), and heart rate variability (HRV). Data were collected from 76 participants under both control and intervention conditions, producing 340 handcrafted features for binary stress classification.

\paragraph{Feature Composition.}  
Features cover tonic and phasic components of EDA and higher-order statistics such as mean, variance, skewness, kurtosis, peak count, energy, and entropy (permutation and SVD entropy). Similar features are extracted for HR, TEMP, ACC, and HRV modalities. All features are z-normalized per participant.

\paragraph{Demographic and Experimental Attributes.}  
Evaluation subgroups are defined as:
\begin{itemize}
    \item \textbf{Group:} control, pre-dose, and post-dose (experimental conditions);
    \item \textbf{Sex:} male, female.
\end{itemize}
These categories quantify fairness across both experimental interventions and intrinsic biological variation.

\paragraph{Model Architecture (EDA Model).}  
The encoder maps the 340-dimensional input into a 128-dimensional latent space using three fully connected layers with \texttt{ReLU} activations and dropout (0.1). The decoder reconstructs the input, and the classifier (128–64–64–2) outputs stress vs.\ non-stress logits. Training minimizes a combined reconstruction and cross-entropy loss with Fisher penalty-based stability regularization.

\subsubsection{Percept-R Dataset}
The Percept-R corpus~\cite{benway2022percept} is an open-access clinical speech repository specialized for the study of American English rhotic production /\rotatebox[origin=c]{180}{r}/ in children and adolescents. The full corpus comprises over 32 hours of citation speech collected as part of an ongoing longitudinal study from 281 participants aged 6 to 24 years, including both typically developing speakers and individuals with Residual Speech Sound Disorders (RSSD). For our evaluation, we use a publicly available subset of 76 participants.

\paragraph{Data Preprocessing.}
Audio samples were originally recorded as 16-bit PCM WAV files at 44.1~kHz. Ground-truth labels were derived from perceptual judgments by trained listeners, including speech-language pathologists and crowdsourced raters, identifying productions as either \emph{rhotic} (correct) or \emph{derhotic} (incorrect). For modeling, each sample is represented as a fixed-length multivariate sequence with 5 acoustic feature channels over 60 time steps. The training data are organized into five subject-wise folds, with each test split containing 15–16 unique users and the remaining 60–61 users forming the training pool.

\paragraph{Feature Representation.}
The Percept-R input representation captures acoustic cues relevant to rhoticity, including spectral structure associated with the second and third formants, where the F3-F2 distance is a key marker of /\rotatebox[origin=c]{180}{r}/ quality in American English. The prediction task is binary speech-correctness classification, distinguishing correct rhotic productions from incorrect ones.

\paragraph{Demographic Attributes.}
Subgroup evaluation includes:
\begin{itemize}
    \item \textbf{Sex:} male, female;
    \item \textbf{Age (months):} 82, 91, 92, 93, 96, 99, 100, 105, 107, 108, 109, 111, 112, 113, 115, 117, 118, 119, 120, 121, 122, 123, 124, 126, 127, 129, 132, 133, 135, 136, 137, 138, 139, 140, 149, 151, 153, 154, 155, 156, 163, 164, 167, 169, 170, 171, 179, 185, 186, 252 (grouped into four bins for BHE: $\le$100, 101--130, 131--160, and $>$160 months);
    \item \textbf{Race:} White, Asian, More than one race, Other, Black or African American;
    \item \textbf{Ethnicity:} Not Hispanic or Latino, Hispanic or Latino
\end{itemize}
These subgroup partitions are appropriate for fairness evaluation because they capture clinically and demographically meaningful variation in speech data.
\paragraph{Model Architecture.}
Percept-R Model is a compact autoencoder-classifier model built on a shared 1D convolutional encoder. The encoder processes 5-channel input sequences of length 60 using two convolutional layers (5$\rightarrow$3 and 3$\rightarrow$1 channels, kernel size 5, stride 1), producing a flattened 52-dimensional representation. The classifier branch applies a two-layer MLP (52-32-32-2) with \texttt{Hardswish} activations to predict binary speech correctness. A separate embedding head projects the 52-dimensional convolutional representation into a 32-dimensional latent space for reconstruction and downstream clustering. The decoder then reconstructs the original 5$\times$60 input from this embedding through a lightweight fully connected network (32-64-300). This architecture supports joint optimization of discriminative and generative objectives under the same \textbf{\textsc{Flare}} framework.
Table~\ref{tab:datasets_summary} summarizes the key dataset characteristics and base model configurations across all four datasets.
\subsubsection{Evaluation Splits and Training Protocol}
\label{sec:data_stat}

All datasets follow a subject-disjoint cross-validation protocol to ensure that each participant appears in the test set exactly once across folds. Specifically, \textbf{IHS} uses 5 folds with 17 participants per test set; \textbf{EDA} uses 4 folds with 19 participants per test set; \textbf{OhioT1DM} uses 4 folds with test sizes of 3, 3, 2, and 3 participants (the twelfth participant has no eligible test samples and appears only in training); and \textbf{Percept-R} uses 5 folds with test sizes of 16, 15, 15, 15, and 15 participants.

Across all folds, the total number of test samples is 8,371 for OhioT1DM, 2315 for IHS, 5497 for EDA, and 32,852 for Percept-R.

All models are trained using the Adam optimizer with early stopping based on validation (\texttt{val}) F1-score. The training objective combines reconstruction loss, cross-entropy loss, and a Fisher penalty term with dataset-specific weighting. Fisher information is computed only on correctly classified samples (Algorithm~\ref{algo1}).

\begin{table}[h]
\centering
\caption{Summary of datasets, test samples (summed over folds), and base model configurations.}
\label{tab:datasets_summary}
\small
\setlength{\tabcolsep}{3pt}\footnotesize\begin{tabular}{lccccc}
\toprule
\textbf{Dataset} & \textbf{Participants} & \textbf{Features} & \textbf{Samples} & \textbf{Task} & \textbf{Model (Encoder–Classifier)} \\
\midrule
OhioT1DM~\cite{marling2020ohiot1dm} 
& 12 & 7 & 8371 
& Glucose (binary) 
& input–256–128–64–32 / 32–8–2 \\

IHS~\cite{adler2021identifying} 
& 85 & 18 & 2315 
& Mood (binary) 
& input–128–64–32 / 32–16–8–2 \\

EDA~\cite{xiao2025human} 
& 76 & 340 & 5497 
& Stress (binary) 
& input–128 / 128–64–64–2 \\

Percept-R \cite{benway2022percept} 
& 76 & (60,5) & 32852 
& Speech correctness (binary) 
& input–52 / 52–32–32–2 \\
\bottomrule
\end{tabular}
\end{table}

\subsubsection{\textbf{Hyperparameters}}
\label{app:hype}
Hyperparameters are optimized through a combination of \emph{Optuna}~\cite{shekhar2021comparative} hyperparameter tuning and iterative manual refinement to obtain the best F1 Score. 
Random seeds were fixed for reproducibility, and evaluation uses the test predictions pooled across all folds (Section~\ref{BHE-metrics}). Optuna employs Bayesian optimization with early stopping and dynamic pruning to efficiently explore the search space, selecting the configuration that maximizes F1 on the validation (\texttt{val}) split; no sensitive attributes or test data are used \cite{akiba2019optuna}.
%Random seeds were fixed for reproducibility, and results were averaged across folds. Optuna employs Bayesian optimization with early stopping and dynamic pruning to efficiently explore the search space, automatically selecting the configuration that maximizes performance while maintaining fairness and stability objectives \cite{akiba2019optuna}.

\begin{table*}[t]
\centering
\color{black}
\caption{Hyperparameters used for Base Pretraining (Stage-1) and cluster-adaptation-aggregation (Stage-3).}
\begin{tabular}{lccc|ccc}
\hline
& \multicolumn{3}{c}{\textbf{Stage 1}} 
& \multicolumn{3}{c}{\textbf{Stage 3}} \\
\cline{2-7}

& $lr$ & $\alpha$ & $\beta$ 
%& $lr$ & $\alpha$ & Frozen Layers \\
& $lr$ & $\gamma$ & Frozen Modules \\

\hline

\textbf{IHS}
& $3.28\times10^{-4}$ & 0.497 & 0.0259
& $5.09\times10^{-4}$ & 0.568 & 4 \\

\textbf{Ohio}
& $1.0\times10^{-3}$ & 0.74 & 0.919
& $1.83\times10^{-5}$ & 0.614 & 9 \\

\textbf{EDA}
& $2.30\times10^{-3}$ & 0.76 & 0.79
& $4.38\times10^{-5}$ & 0.690 & 8 \\

\textbf{Percept-R}
& $1.0\times10^{-3}$ & 0.3 & 0.1
& $1.0\times10^{-4}$ & 0.5 & 3 \\

\hline
\end{tabular}
\label{tab:training_hyperparameters}
\end{table*}

As shown in Table~\ref{tab:training_hyperparameters}, Stage-1 corresponds to base pretraining. For all datasets, the representation model was trained for 200 epochs using a batch size of 1 with the specified learning rates and loss weighting coefficients ($\alpha$, $\beta$).
Stage-2 performs latent clustering on the model-behavior descriptors $S=[z,L_{\mathrm{CE}},\mathsf{F}]$ obtained from Stage-1, which are first projected using UMAP with dimensionality 3, 15 nearest neighbors, minimum distance 0.1, and Euclidean distance for all datasets. Clustering is then performed using Gaussian Mixture Models with a fixed random seed of 42.

In Stage-3, cluster-specific models are fine-tuned using the Adam optimizer with weight decay $10^{-5}$, with the learning rate, cluster loss weight $\gamma$, and number of frozen modules selected through hyperparameter tuning. The number of frozen modules counts consecutive modules across the encoder and classifier, including activation and dropout positions, not only trainable linear layers.

\subsubsection{\textbf{Justification behind Selection of GMM and UMAP for clustering}}
\label{app:gmm}
We select the number of clusters automatically using the Bayesian Information Criterion (BIC) applied to the Gaussian Mixture Model (GMM) fit on UMAP-reduced model-behavior descriptors. BIC is chosen over alternatives such as AIC because its stronger complexity penalty ($\log N$ per parameter) guards against overfitting the cluster structure to the training fold, which is particularly important in settings with limited participants~\cite{mclachlan2019finite, fraley2002model}. 
Manifold learning literature shows that high-dimensional data can reside on low-dimensional nonlinear manifolds \cite{belkin2003laplacian}, and linear techniques such as PCA (Principal Component Analysis) 
cannot recover intrinsic manifold geometry 
\cite{tenenbaum2000global}. Accordingly, we employ UMAP, a manifold-based dimensionality reduction method 
that preserves local neighborhood structure and supports embedding of unseen 
samples through a learned transformation 
\cite{mcinnes2018umap}.Unlike t-SNE, which is primarily designed for visualization and does not  naturally provide a parametric mapping for new data, UMAP allows consistent 
transformation at deployment \cite{mcinnes2018umap}. 
For clustering, we use a GMM because it provides a probabilistic formulation that models the data distribution as a mixture of components and yields posterior membership probabilities of that sample u belongs to cluster c ($p(c\mid u)$) \cite{bishop2006prml,murphy2012ml}. 
This enables cluster-specific covariance modeling (i.e., flexible subgroup geometry) and supports consistent assignment of unseen samples by evaluating $p(c\mid u)$ at inference time \cite{bishop2006prml}, where $u$ is the UMAP projection of the label-free descriptor (computed with $\hat{y}$).
In our framework, cluster assignments must remain stable because 
cluster-specific models $\theta^*_c$ are trained and later reused at deployment. 
GMM learns explicit cluster parameters during training, enabling consistent 
assignment of new samples by evaluating $p(c\mid u)$ under the learned mixture model.

Nearest-neighbor ($k$NN)-based grouping is not adopted because it is an instance-based method that determines assignment using local neighborhood relationships rather than learning explicit parametric cluster models \cite{murphy2012ml}. 
Since our framework trains and reuses cluster-specific models $\theta^*_c$, a parametric clustering model is more aligned with this design.

\subsubsection{\textbf{Selection of Benign Baseline}}

\label{benign_eval}

To establish a fair reference point, we evaluate each dataset using standard predictive models that do not explicitly incorporate fairness constraints. Specifically, we consider widely adopted architectures—MLP, CNN, and recurrent models (RNN/LSTM)—and select the best-performing model based on F1-score as the benign baseline for subsequent BHE computation. This ensures that the baseline reflects strong predictive performance without bias mitigation, allowing us to isolate the impact of fairness-aware interventions. The choice of MLP, CNN, and LSTM is motivated by their widespread use in physiological sensing, time-series modeling, and human-centered prediction tasks~\cite{faust2020deep,li2020deep}.

For the \textbf{OhioT1DM} dataset, prior work focuses on data-driven models for glucose prediction \cite{marling2020ohiot1dm, arefeen2023designing}. Because OhioT1DM inputs are temporally ordered glucose–insulin sequences, we evaluate recurrent (RNN, LSTM) and convolutional (CNN) architectures and select a 1D CNN as the benign baseline based on F1-score. The CNN consists of three convolutional layers (1→32→64→128 channels, kernel size 3) with batch normalization and ReLU activations, followed by global average pooling and a two-layer MLP classifier.

For the \textbf{EDA} dataset, following \cite{ xiao2025human}, we evaluate MLP, CNN, and LSTM architectures. The MLP baseline consists of a 4-layer fully connected network (340→128→128→128→1) with ReLU activations and dropout. Based on F1-score, the MLP is selected as the benign baseline.

For the \textbf{IHS} dataset, since model configurations are not specified in \cite{adler2021identifying}, we implement MLP, CNN, and LSTM baselines and optimize them using Optuna \cite{shekhar2021comparative}. The selected benign baseline is an MLP (18→16→8→1) with ReLU activations, chosen based on F1-score.

For the \textbf{Percept-R} dataset, we implement CNN, LSTM, and MLP architectures for speech-based classification \cite{benway2022percept}. The CNN is selected as the benign baseline based on F1-score. It consists of two 1D convolutional layers (5→3→1 channels, kernel size 5), producing a 52-dimensional representation, followed by a two-layer MLP classifier with Hardswish activations.

Table~\ref{tab:benign_eval} presents the performance of the candidate models (MLP, CNN, and LSTM; RNN, LSTM, and CNN for Ohio) across all datasets. Models are ordered by their reported mean subgroup weighted F1 on the held-out test participants. This ordering describes the evaluation results rather than the architecture-selection procedure. Specifically, CNN achieves the highest value on \textbf{Percept-R} (77.79\%) and \textbf{Ohio} (65.22\%), while MLP achieves the highest value on \textbf{EDA} (87.06\%) and \textbf{IHS} (59.24\%). These architectures were selected using validation F1 within the outer training data.
% This section is intentionally left empty as per request.
\begin{table}[t]
\small
\centering

\caption{Performance comparison across datasets (\%). For each dataset, models are ordered from lowest to highest score (best model on the right). The F1 row reports the mean subgroup weighted F1, i.e., the subgroup F1 scores averaged within and then across sensitive attributes, as in the \textit{Mean} rows of Table~\ref{tab:f1_demographics_models_reduced}.}

\color{black}{
\resizebox{0.6\linewidth}{!}{%
\begin{tabular}{l|ccc|ccc}

\hline

& \multicolumn{3}{c|}{\textbf{Percept-R}} 

& \multicolumn{3}{c}{\textbf{EDA}} \\

\hline

\textbf{Metric} 

& LSTM & MLP & CNN 

& LSTM & CNN & MLP  \\

\hline

Accuracy 

& 45.15 & 76.77 & 77.43 

& 64.55 & 85.98 & 86.76  \\

Mean subgroup weighted F1

& 42.65 & 76.25 & 77.79 

& 66.01 & 85.85 & 87.06  \\

Recall 

& 10.06 & 72.85 & 78.83 

& 65.94 & 91.23 & 88.81 \\

AUC 

& 53.54 & 80.67 & 81.41 

& 67.49 & 90.95 & 92.16  \\

\hline

& \multicolumn{3}{c|}{\textbf{Ohio}} 

& \multicolumn{3}{c}{\textbf{IHS}} \\

\hline

\textbf{Metric} 

& RNN & LSTM & CNN 

& LSTM & CNN & MLP  \\

\hline

Accuracy 

& 57.52 & 58.54 & 64.39 

& 56.13 & 58.88 & 59.98  \\

Mean subgroup weighted F1 

& 58.37 & 59.95 & 65.22 

& 55.43 & 58.57 & 59.24 \\

Recall 

& 67.65 & 60.44 & 72.12 

& 49.32 & 57.63 & 69.57  \\

AUC 

& 61.41 & 66.05 & 72.46 

& 58.12 & 61.85 & 59.88 \\

\hline

\end{tabular}}
}
\label{tab:benign_eval}

\end{table}

\subsection{Per-attribute Results}
\label{app:per_attr}
This section reports the per-attribute results summarized in Table~\ref{tab:bhe_summary} (Section~\ref{sec:baselines-vs-flare}). Tables~\ref{tab:f1_demographics_models_reduced} and~\ref{tab:dp_eo_flare} compare \textbf{\textsc{Flare}} with the attribute-aware baselines (Fair-EO and Fair-DP) in terms of subgroup-wise F1 and BHE changes, and Tables~\ref{tab:f1_demographics_models} and~\ref{tab:sota_delta_grouped} report the same for the fairness-without-demographics and subgroup-robustness baselines. For each sensitive attribute, F1 is computed separately for each subgroup; the reported mean and standard deviation are taken across those subgroup-level scores, and the BHE changes are computed relative to the benign baseline as defined in Section~\ref{BHE-metrics}.

\begin{table}[htbp]
\centering
\color{black}
\arrayrulecolor{black}
\caption{F1 score (\%) across sensitive attributes for different models in the OhioT1DM, EDA, IHS, and Percept-R datasets. For each sensitive attribute, F1 is computed separately for each subgroup, and the reported mean and standard deviation are calculated across the subgroup-level F1 scores of each sensitive attribute.}
\vspace{-5pt}
\begin{minipage}{0.499\textwidth}
\resizebox{\linewidth}{!}{
\begin{tabular}{l|c|c|c|c}
\hline
\textbf{\makecell{Sensitive \\ Attribute}} & \textbf{Baseline (\%)} & \textbf{Fair-EO (\%)} & \textbf{Fair-DP (\%)} & \textbf{\textsc{Flare} (\%)} \\
\hline

\multicolumn{5}{c}{\textbf{OhioT1DM Dataset}}\\
\hline
Age & 65.49 $\pm$ 12.75 & 61.08 $\pm$ 19.69 & 59.16 $\pm$ 20.77 & \textbf{70.56} $\pm$ \textbf{4.84} \\
Sex & 68.42 $\pm$ 3.17 & 67.31 $\pm$ 1.94 & 68.67 $\pm$ \textbf{0.74} & \textbf{70.38} $\pm$ 1.08 \\
Pump & 59.61 $\pm$ 11.37 & 68.20 $\pm$ \textbf{0.66} & 68.13 $\pm$ 1.77 & \textbf{70.68} $\pm$ 4.15 \\
Sensor Band & 67.36 $\pm$ 5.81 & 66.96 $\pm$ 2.52 & 68.41 $\pm$ \textbf{1.80} & \textbf{69.90} $\pm$ 2.26 \\
\rowcolor{gray!15}
\textit{Mean} & 65.22 $\pm$ 8.28 & 65.89 $\pm$ 6.20 & 66.09 $\pm$ 6.27 & \textbf{70.38} $\pm$ \textbf{3.08} \\
\hline

\multicolumn{5}{c}{\textbf{EDA Dataset}}\\
\hline
Group\_label & 86.91 $\pm$ 4.47 & 87.11 $\pm$ 4.75 & 86.46 $\pm$ \textbf{3.54} & \textbf{89.08} $\pm$ 3.91 \\
Sex & 87.20 $\pm$ 1.13 & 87.11 $\pm$ \textbf{0.69} & 86.90 $\pm$ 1.44 & \textbf{89.47} $\pm$ 1.06 \\
\rowcolor{gray!15}
\textit{Mean} & 87.06 $\pm$ 2.80 & 87.11 $\pm$ 2.72 & 86.68 $\pm$ 2.49 & \textbf{89.27} $\pm$ \textbf{2.48} \\
\hline
\end{tabular}
}
\end{minipage}
\hspace{-5pt}
% \hfill
\vline
\hspace{-3pt}
\begin{minipage}{0.499\textwidth}
\resizebox{\linewidth}{!}{
\begin{tabular}{l|c|c|c|c}
\hline
\textbf{\makecell{Sensitive \\ Attribute}} & \textbf{Baseline (\%)} & \textbf{Fair-EO (\%)} & \textbf{Fair-DP (\%)} & \textbf{\textsc{Flare} (\%)} \\
\hline
\multicolumn{5}{c}{\textbf{IHS Dataset}}\\
\hline
Sex & 59.66 $\pm$ 0.89 & 60.52 $\pm$ 1.51 & 59.10 $\pm$ \textbf{0.10} & \textbf{64.63} $\pm$ 0.83 \\
Age & 57.37 $\pm$ 14.40 & 58.50 $\pm$ 12.56 & 53.31 $\pm$ 13.59 & \textbf{67.09} $\pm$ \textbf{11.34} \\
Ethnicity & 56.37 $\pm$ 5.74 & 55.52 $\pm$ 9.64 & 56.74 $\pm$ 6.91 & \textbf{64.60} $\pm$ \textbf{5.46} \\
Specialty & 60.27 $\pm$ 5.70 & 59.80 $\pm$ \textbf{2.52} & 60.03 $\pm$ 3.90 & \textbf{65.26} $\pm$ 5.34 \\
PHQ10$>$0 & 62.52 $\pm$ 3.36 & 60.20 $\pm$ 0.62 & 60.23 $\pm$ \textbf{0.18} & \textbf{64.56} $\pm$ 0.52 \\
\rowcolor{gray!15}
\textit{Mean} & 59.24 $\pm$ 6.02 & 58.91 $\pm$ 5.37 & 57.88 $\pm$ 4.94 & \textbf{65.23} $\pm$ \textbf{4.70} \\
\hline

\multicolumn{5}{c}{\textbf{Percept-R Dataset}}\\
\hline
Sex & 76.75 $\pm$ 1.03 & 77.53 $\pm$ 0.92 & 77.65 $\pm$ 1.32 & \textbf{81.88} $\pm$ \textbf{0.44} \\
Age(months) & 76.69 $\pm$ 12.79 & 77.82 $\pm$ 5.73 & 77.76 $\pm$ \textbf{5.00} & \textbf{86.28} $\pm$ 9.45 \\
Race & 79.34 $\pm$ 10.25 & 77.97 $\pm$ 8.91 & 77.80 $\pm$ 7.48 & \textbf{86.77} $\pm$ \textbf{4.48} \\
Ethnicity & 78.38 $\pm$ 1.83 & 79.13 $\pm$ 0.81 & 79.32 $\pm$ \textbf{0.54} & \textbf{81.22} $\pm$ 0.58 \\
\rowcolor{gray!15}
\textit{Mean} & 77.79 $\pm$ 6.48 & 78.11 $\pm$ 4.09 & 78.13 $\pm$ \textbf{3.59} & \textbf{84.04} $\pm$ 3.74 \\
\hline
\end{tabular}
}
\end{minipage}
\label{tab:f1_demographics_models_reduced}
\vspace{-10pt}
\end{table}

\begin{table*}[h!]
\centering
\small
\color{black}
\caption{
Comparison of the fairness-with-demographics models, \textbf{Fair-EO} and \textbf{Fair-DP}, against \textbf{\textsc{Flare}}. Values are expressed as percentage-point changes, with higher values indicating stronger ethical performance. The reported $\Delta B$, $\Delta H$, and $\Delta E$ metrics are computed from subgroup-level performance statistics for each sensitive attribute.
}
\label{tab:dp_eo_flare}
% \resizebox{0.95\linewidth}{!}{%
\vspace{-5pt}
\begin{minipage}{0.499\textwidth}
\resizebox{\linewidth}{!}{
\begin{tabular}{l|ccc|ccc|ccc}
\toprule
\multirow{2}{*}{\textbf{\makecell{Sensitive \\ Attribute}}} 
& \multicolumn{3}{c|}{\textbf{Fair-EO (\%)}}
& \multicolumn{3}{c|}{\textbf{Fair-DP (\%)}}
& \multicolumn{3}{c}{\textbf{\textsc{Flare (\%)}}} \\
& $\Delta B$ & $\Delta H$ & $\Delta E$
& $\Delta B$ & $\Delta H$ & $\Delta E$
& $\Delta B$ & $\Delta H$ & $\Delta E$ \\
\midrule

\multicolumn{10}{c}{\textbf{OhioT1DM Dataset}} \\
\midrule
Age          
& $-4.41$ & $-12.28$ & $-6.94$
& $-6.33$ & $-15.63$  & $-8.02$
& $\mathbf{5.07}$ & $\mathbf{0.00}$ & $\mathbf{7.91}$ \\

Sex         
& $-1.11$ & $-4.73$ & $1.23$
& $0.25$  & $-2.52$  & $\mathbf{2.43}$
& $\mathbf{1.96}$ & $\mathbf{0.48}$ & $2.09$ \\

Pump        
& $8.59$ & $0.08$ & $\mathbf{10.71}$
& $8.52$ & $\mathbf{1.73}$ & $9.60$
& $\mathbf{11.06}$ & $0.08$ & $7.22$ \\

Sensor Band 
& $-0.39$ & $-6.27$ & $3.28$
& $1.05$  & $-1.79$ & $\mathbf{4.01}$
& $\mathbf{2.54}$ & $\mathbf{0.03}$ & $3.55$ \\

\rowcolor{gray!15}
\textit{Mean}
& $0.67$ & $-5.80$ & $2.07$
& $0.87$ & $-4.55$ & $2.01$
& $\mathbf{5.16}$ & $\mathbf{0.15}$ & $\mathbf{5.19}$ \\
\midrule

\multicolumn{10}{c}{\textbf{EDA Dataset}} \\
\midrule
Group label 
& $0.20$  & $-8.28$ & $-0.28$
& $-0.45$ & $-9.05$ & $\mathbf{0.93}$
& $\mathbf{2.16}$ & $\mathbf{0.85}$ & $0.57$ \\

Sex          
& $-0.09$  & $-0.40$ & $\mathbf{0.44}$
& $-0.30$ & $-0.52$ & $-0.31$
& $\mathbf{2.27}$ & $\mathbf{2.22}$ & $0.07$ \\

\rowcolor{gray!15}
\textit{Mean}
& $0.06$  & $-4.34$ & $0.08$
& $-0.38$ & $-4.79$ & $0.31$
& $\mathbf{2.22}$ & $\mathbf{1.54}$ & $\mathbf{0.32}$ \\
\midrule
\end{tabular}
}
\end{minipage}
\hspace{-5pt}
% \hfill
\vline
\hspace{-3pt}
\begin{minipage}{0.499\textwidth}
\resizebox{\linewidth}{!}{
\begin{tabular}{l|ccc|ccc|ccc}
\toprule
\multirow{2}{*}{\textbf{\makecell{Sensitive \\ Attribute}}} 
& \multicolumn{3}{c|}{\textbf{Fair-EO (\%)}}
& \multicolumn{3}{c|}{\textbf{Fair-DP (\%)}}
& \multicolumn{3}{c}{\textbf{\textsc{Flare (\%)}}} \\
& $\Delta B$ & $\Delta H$ & $\Delta E$
& $\Delta B$ & $\Delta H$ & $\Delta E$
& $\Delta B$ & $\Delta H$ & $\Delta E$ \\
\midrule
\multicolumn{10}{c}{\textbf{IHS Dataset}} \\
\midrule
Sex            
& $0.86$ & $0.43$ & $-0.62$
& $-0.56$ & $-1.26$  & $\mathbf{0.79}$
& $\mathbf{4.97}$ & $\mathbf{4.93}$ & $0.06$ \\

Age            
& $1.14$  & $-35.32$  & $1.84$
& $-4.06$ & $-4.80$  & $0.81$
& $\mathbf{9.72}$ & $\mathbf{0.30}$ & $\mathbf{3.07}$ \\

Ethnicity      
& $-0.85$ & $-8.54$ & $-3.90$
& $0.37$  & $-8.54$  & $-1.17$
& $\mathbf{8.23}$ & $\mathbf{3.38}$ & $\mathbf{0.28}$ \\

Specialty      
& $-0.46$ & $-9.36$ & $\mathbf{3.18}$
& $-0.24$ & $-8.08$ & $1.80$
& $\mathbf{5.00}$ & $\mathbf{0.03}$ & $0.36$ \\

PHQ10 $>$ 0        
& $-2.32$ & $-5.14$ & $2.74$
& $-2.29$ & $-4.55$ & $\mathbf{3.18}$
& $\mathbf{2.04}$ & $\mathbf{0.04}$ & $2.83$ \\

\rowcolor{gray!15}
\textit{Mean}
& $-0.33$ & $-11.59$ & $0.65$
& $-1.36$ & $-5.45$ & $1.08$
& $\mathbf{5.99}$ & $\mathbf{1.74}$ & $\mathbf{1.32}$ \\
\midrule

\multicolumn{10}{c}{\textbf{Percept-R Dataset}} \\
\midrule
Sex    
& $0.78$  & $0.71$ & $0.11$
& $0.90$  & $0.69$ & $-0.29$
& $\mathbf{5.12}$ & $\mathbf{4.70}$ & $\mathbf{0.59}$ \\

Age        
& $1.14$ & $-6.41$ & $7.06$
& $1.08$ & $-8.13$ & $\mathbf{7.79}$
& $\mathbf{9.60}$ & $\mathbf{0.02}$ & $3.35$ \\

Race      
& $-1.37$ & $-3.58$ & $1.34$
& $-1.54$ & $-1.74$ & $2.77$
& $\mathbf{7.43}$ & $\mathbf{0.00}$ & $\mathbf{5.77}$ \\

Ethnicity 
& $0.75$  & $-1.12$ & $1.02$
& $0.94$ & $-0.74$  & $\mathbf{1.29}$
& $\mathbf{2.84}$ & $\mathbf{1.96}$ & $1.25$ \\

\rowcolor{gray!15}
\textit{Mean}
& $0.33$ & $-2.60$ & $2.38$
 & $0.35$ & $-2.48$ & $\mathbf{2.89}$
& $\mathbf{6.25}$ & $\mathbf{1.67}$ & $2.74$ \\
\bottomrule
\end{tabular}%
}
\end{minipage}
\vspace{-15pt}
\end{table*}

\begin{table}[t!]
\centering\scriptsize
\caption{F1 scores (\%) across sensitive attributes for different SOTA models in the OhioT1DM, EDA, IHS, and Percept-R datasets. For each sensitive attribute, F1 is computed separately for each of its subgroups, and the reported values denote the mean $\pm$ standard deviation across those subgroup-level F1 scores.}
\vspace{-8pt}
\label{tab:f1_demographics_models} 
\begin{tabular}{ll|ccccccc}
\toprule
\textbf{Data} & \textbf{Attribute} & \textbf{KD} & \textbf{ARL} & \textbf{GoG} & \textbf{Reckoner} & \textbf{JTT} & \textbf{Group DRO} & \textbf{\textsc{Flare}}\\
\midrule
\multirow{5}{*}{\rotatebox{90}{OhioT1DM}}
& Age & $65.01{\pm}2.44$ & $70.16{\pm}2.57$ & $59.12{\pm}2.67$ & $66.42{\pm}3.21$ & $68.37{\pm}7.21$ & $62.15{\pm}\mathbf{0.92}$ & $\mathbf{70.56}{\pm}4.84$\\
& Sex & $68.50{\pm}2.83$ & $\mathbf{71.75}{\pm}2.20$ & $69.09{\pm}2.64$ & $68.51{\pm}2.80$ & $64.49{\pm}\mathbf{0.51}$ & $62.39{\pm}4.67$ & $70.38{\pm}1.08$\\
& Pump & $60.66{\pm}4.81$ & $65.80{\pm}2.97$ & $68.01{\pm}3.99$ & $61.74{\pm}3.23$ & $69.69{\pm}9.39$ & $62.48{\pm}\mathbf{0.26}$ & $\mathbf{70.68}{\pm}4.15$\\
& Band & $67.33{\pm}2.78$ & $69.46{\pm}5.67$ & $67.35{\pm}2.76$ & $67.57{\pm}2.96$ & $64.21{\pm}\mathbf{0.69}$ & $60.88{\pm}5.06$ & $\mathbf{69.90}{\pm}2.26$\\
\rowcolor{gray!12}& \textit{Mean} & $65.38{\pm}3.22$ & $69.29{\pm}3.35$ & $65.89{\pm}3.02$ & $66.06{\pm}3.05$ & $66.69{\pm}4.45$ & $61.98{\pm}\mathbf{2.73}$ & $\mathbf{70.38}{\pm}3.08$\\
\midrule
\multirow{3}{*}{\rotatebox{90}{EDA}}
& Group & $69.20{\pm}\mathbf{2.87}$ & $87.61{\pm}3.34$ & $87.03{\pm}4.58$ & $85.55{\pm}3.68$ & $87.39{\pm}4.45$ & $86.74{\pm}4.61$ & $\mathbf{89.08}{\pm}3.91$\\
& Sex & $69.20{\pm}2.24$ & $87.61{\pm}2.80$ & $87.04{\pm}1.09$ & $85.84{\pm}2.17$ & $87.43{\pm}\mathbf{0.28}$ & $87.52{\pm}2.46$ & $\mathbf{89.47}{\pm}1.06$\\
\rowcolor{gray!12}& \textit{Mean} & $69.20{\pm}2.56$ & $87.61{\pm}3.07$ & $87.04{\pm}2.84$ & $85.70{\pm}2.92$ & $87.41{\pm}\mathbf{2.36}$ & $87.13{\pm}3.53$ & $\mathbf{89.27}{\pm}2.48$\\
\midrule
\multirow{6}{*}{\rotatebox{90}{IHS}}
& Sex & $58.12{\pm}1.52$ & $59.65{\pm}0.92$ & $59.93{\pm}1.92$ & $57.52{\pm}1.64$ & $57.57{\pm}1.05$ & $60.52{\pm}1.28$ & $\mathbf{64.63}{\pm}\mathbf{0.83}$\\
& Age & $57.90{\pm}13.04$ & $62.54{\pm}12.08$ & $54.83{\pm}11.28$ & $54.32{\pm}9.96$ & $57.22{\pm}9.44$ & $59.73{\pm}\mathbf{8.77}$ & $\mathbf{67.09}{\pm}11.34$\\
& Ethnicity & $58.06{\pm}4.11$ & $62.66{\pm}7.13$ & $57.67{\pm}12.68$ & $52.85{\pm}6.79$ & $57.62{\pm}11.69$ & $59.17{\pm}\mathbf{2.98}$ & $\mathbf{64.60}{\pm}5.46$\\
& Specialty & $58.01{\pm}4.65$ & $62.62{\pm}4.30$ & $60.56{\pm}5.49$ & $56.94{\pm}\mathbf{3.22}$ & $58.62{\pm}4.91$ & $60.16{\pm}5.30$ & $\mathbf{65.26}{\pm}5.34$\\
& PHQ10 & $58.09{\pm}2.20$ & $62.68{\pm}2.87$ & $59.35{\pm}1.15$ & $58.85{\pm}2.41$ & $56.35{\pm}2.08$ & $62.36{\pm}3.21$ & $\mathbf{64.56}{\pm}\mathbf{0.52}$\\
\rowcolor{gray!12}& \textit{Mean} & $58.04{\pm}5.10$ & $62.03{\pm}5.45$ & $58.47{\pm}6.50$ & $56.10{\pm}4.80$ & $57.48{\pm}5.83$ & $60.39{\pm}\mathbf{4.31}$ & $\mathbf{65.23}{\pm}4.70$\\
\midrule
\multirow{5}{*}{\rotatebox{90}{Percept-R}}
& Sex & $77.43{\pm}1.34$ & $76.24{\pm}\mathbf{0.18}$ & $77.65{\pm}1.78$ & $73.96{\pm}1.30$ & $73.95{\pm}0.33$ & $76.25{\pm}2.78$ & $\mathbf{81.88}{\pm}0.44$\\
& Age & $77.52{\pm}13.36$ & $76.86{\pm}13.72$ & $77.30{\pm}14.60$ & $74.80{\pm}14.77$ & $74.61{\pm}14.08$ & $76.17{\pm}12.10$ & $\mathbf{86.28}{\pm}\mathbf{9.45}$\\
& Race & $80.01{\pm}8.06$ & $79.36{\pm}10.25$ & $80.68{\pm}7.50$ & $80.03{\pm}8.74$ & $73.49{\pm}14.45$ & $76.26{\pm}8.83$ & $\mathbf{86.77}{\pm}\mathbf{4.48}$\\
& Ethnicity & $78.54{\pm}\mathbf{0.53}$ & $76.77{\pm}4.09$ & $79.39{\pm}0.85$ & $75.03{\pm}3.17$ & $76.07{\pm}1.39$ & $78.10{\pm}2.26$ & $\mathbf{81.22}{\pm}0.58$\\
\rowcolor{gray!12}& \textit{Mean} & $78.37{\pm}5.82$ & $77.31{\pm}7.06$ & $78.75{\pm}6.18$ & $75.96{\pm}6.99$ & $74.53{\pm}7.56$ & $76.70{\pm}6.49$ & $\mathbf{84.04}{\pm}\mathbf{3.74}$\\
\bottomrule
\end{tabular}
\vspace{-8pt}
\end{table}

\begin{table}[htbp]
\scriptsize
\centering
\caption{Model-wise percentage changes $(\Delta B,\Delta H,\Delta E)$ for ARL, KD, Reckoner, GoG, JTT, GroupDRO and \textbf{\textsc{Flare}} across datasets. Values are reported as percentage-point improvements, where larger deltas indicate stronger fairness performance and greater ethical consistency across subgroups. The reported $\Delta B$, $\Delta H$, and $\Delta E$ metrics are computed from subgroup-level performance statistics for each sensitive attribute, using subgroup-wise mean performance and subgroup-wise standard deviation}
\label{tab:sota_delta_grouped}

% --- FIRST BLOCK OF MODELS (KD, ARL, GoG, Reckoner) ---
\resizebox{0.95\linewidth}{!}{
\begin{tabular}{ll|ccc|ccc|ccc|ccc}
\toprule
\multirow{2}{*}{\textbf{Data}} & \multirow{2}{*}{\textbf{Attr.}}
& \multicolumn{3}{c|}{\textbf{KD}} & \multicolumn{3}{c|}{\textbf{ARL}} & \multicolumn{3}{c|}{\textbf{GoG}} & \multicolumn{3}{c}{\textbf{Reckoner}}\\
& & $\Delta B$ & $\Delta H$ & $\Delta E$ & $\Delta B$ & $\Delta H$ & $\Delta E$ & $\Delta B$ & $\Delta H$ & $\Delta E$ & $\Delta B$ & $\Delta H$ & $\Delta E$\\
\midrule
\multirow{5}{*}{\rotatebox{90}{OhioT1DM}}
& Age & $-0.48$ & $-8.73$ & $10.31$ & $4.67$ & $-3.06$ & $10.18$ & $-6.37$ & $-13.78$ & $10.08$ & $0.93$ & $-6.89$ & $9.54$\\
& Sex & $0.08$ & $-0.16$ & $0.34$ & $\mathbf{3.33}$ & $-0.47$ & $0.97$ & $0.67$ & $0.30$ & $0.53$ & $0.08$ & $-0.18$ & $0.37$\\
& Pump & $1.05$ & $-3.59$ & $6.56$ & $6.18$ & $\mathbf{0.23}$ & $8.41$ & $8.40$ & $-2.46$ & $7.38$ & $2.13$ & $-3.62$ & $8.14$\\
& Band & $-0.03$ & $-2.16$ & $3.03$ & $2.10$ & $\mathbf{2.01}$ & $0.14$ & $-0.01$ & $-2.17$ & $3.05$ & $0.22$ & $-1.80$ & $2.85$\\
\rowcolor{gray!12}& \textit{Mean} & $0.16$ & $-3.66$ & $5.06$ & $4.07$ & $-0.32$ & $4.93$ & $0.67$ & $-4.53$ & $5.26$ & $0.84$ & $-3.12$ & $5.23$\\
\midrule
\multirow{3}{*}{\rotatebox{90}{EDA}}
& Group & $-17.72$ & $-20.29$ & $\mathbf{1.60}$ & $0.70$ & $-1.01$ & $1.13$ & $0.12$ & $-0.23$ & $-0.11$ & $-1.37$ & $-6.71$ & $0.79$\\
& Sex & $-18.00$ & $-18.79$ & $-1.11$ & $0.41$ & $-0.77$ & $-1.67$& $-0.16$ & $-0.20$ & $0.04$ & $-1.36$ & $-2.10$ & $-1.04$\\
\rowcolor{gray!12}& \textit{Mean} & $-17.86$ & $-19.54$ & $0.25$ & $0.56$ & $-0.89$ & $-0.27$ & $-0.02$ & $-0.22$ & $-0.04$ & $-1.37$ & $-4.41$ & $-0.13$\\
\midrule
\multirow{6}{*}{\rotatebox{90}{IHS}}
& Sex & $-1.54$ & $-1.98$ & $-0.64$ & $-0.01$ & $-0.03$ & $-0.03$ & $0.27$ & $-0.45$ & $-1.03$ & $-2.14$ & $-3.93$ & $-0.75$\\
& Age & $0.54$ & $-21.67$ & $1.36$ & $5.18$ & $-4.18$ & $2.32$ & $-2.54$ & $-31.82$ & $3.12$ & $-3.05$ & $-14.04$ & $4.44$\\
& Ethnicity & $1.69$ & $-5.47$ & $1.63$ & $6.29$ & $2.59$ & $-1.39$ & $1.30$ & $-7.50$ & $-6.94$ & $-3.52$ & $-7.79$ & $-1.05$\\
& Specialty & $-2.26$ & $-10.70$ & $1.05$ & $2.36$ & $-4.20$ & $1.40$ & $0.29$ & $-7.88$ & $0.21$ & $-3.32$ & $-11.89$ & $\mathbf{2.48}$\\
& PHQ10 & $-4.43$ & $-5.25$ & $1.16$ & $0.15$ & $-0.20$ & $0.49$ & $-3.17$ & $-6.36$ & $2.21$ & $-3.68$ & $-4.35$ & $0.95$\\
\rowcolor{gray!12}& \textit{Mean} & $-1.20$ & $-9.01$ & $0.91$ & $2.79$ & $-1.20$ & $0.56$ & $-0.77$ & $-10.80$ & $-0.49$ & $-3.14$ & $-8.40$ & $1.22$\\
\midrule
\multirow{5}{*}{\rotatebox{90}{Percept-R}}
& Sex & $0.67$ & $0.45$ & $-0.31$ & $-0.51$ & $-1.10$ & $\mathbf{0.84}$ & $0.89$ & $0.36$ & $-0.75$ & $-2.79$ & $-4.43$ & $-0.27$\\
& Age & $0.83$ & $-0.07$ & $-0.57$ & $0.18$ & $-6.00$ & $-0.93$ & $0.61$ & $-1.57$ & $-1.81$ & $-1.88$ & $-4.88$ & $-1.97$\\
& Race & $0.67$ & $-2.79$ & $2.19$ & $0.02$ & $-9.63$ & $0.00$ & $1.34$ & $-3.04$ & $2.75$ & $0.69$ & $-10.08$ & $1.51$\\
& Ethnicity & $0.16$ & $-1.52$ & $\mathbf{1.30}$ & $-1.61$ & $-5.79$ & $-2.25$ & $1.01$ & $-0.89$ & $0.98$ & $-3.35$ & $-6.90$ & $-1.34$\\
\rowcolor{gray!12}& \textit{Mean} & $0.58$ & $-0.98$ & $0.65$ & $-0.48$ & $-5.63$ & $-0.59$ & $0.96$ & $-1.29$ & $0.29$ & $-1.83$ & $-6.57$ & $-0.52$\\
\bottomrule
\end{tabular}
}

\vspace{0.8em}

% --- SECOND BLOCK OF MODELS (JTT, Group DRO, FLARE) ---
\resizebox{0.75\linewidth}{!}{
\begin{tabular}{ll|ccc|ccc|ccc}
\toprule
\multirow{2}{*}{\textbf{Data}} & \multirow{2}{*}{\textbf{Attr.}}
& \multicolumn{3}{c|}{\textbf{JTT}} & \multicolumn{3}{c|}{\textbf{Group DRO}} & \multicolumn{3}{c}{\textbf{\textsc{Flare}}}\\
& & $\Delta B$ & $\Delta H$ & $\Delta E$ & $\Delta B$ & $\Delta H$ & $\Delta E$ & $\Delta B$ & $\Delta H$ & $\Delta E$\\
\midrule
\multirow{5}{*}{\rotatebox{90}{OhioT1DM}}
& Age & $2.88$ & $-9.37$ & $5.53$ & $-3.34$ & $-13.21$ & $\mathbf{11.83}$ & $\mathbf{5.07}$ & $\mathbf{0.00}$ & $7.91$\\
& Sex & $-3.94$ & $-5.82$ & $\mathbf{2.66}$ & $-6.03$ & $-7.10$ & $-1.50$ & $1.96$ & $\mathbf{0.48}$ & $2.09$\\
& Pump & $10.07$ & $-4.60$ & $1.98$ & $2.87$ & $-5.35$ & $\mathbf{11.12}$ & $\mathbf{11.06}$ & $0.08$ & $7.22$\\
& Band & $-3.15$ & $-7.75$ & $\mathbf{5.12}$ & $-6.47$ & $-7.00$ & $0.75$ & $\mathbf{2.54}$ & $0.03$ & $3.55$\\
\rowcolor{gray!12}& \textit{Mean} & $1.47$ & $-6.89$ & $3.82$ & $-3.24$ & $-8.17$ & $\mathbf{5.55}$ & $\mathbf{5.16}$ & $\mathbf{0.15}$ & $5.19$\\
\midrule
\multirow{3}{*}{\rotatebox{90}{EDA}}
& Group & $0.48$ & $0.10$ & $0.02$ & $-0.18$ & $-0.90$ & $-0.13$ & $\mathbf{2.16}$ & $\mathbf{0.85}$ & $0.57$\\
& Sex & $0.23$ & $-0.37$ & $\mathbf{0.85}$ & $0.32$ & $-0.62$ & $-1.33$ & $\mathbf{2.27}$ & $\mathbf{2.22}$ & $0.07$\\
\rowcolor{gray!12}& \textit{Mean} & $0.36$ & $-0.14$ & $\mathbf{0.44}$ & $0.07$ & $-0.76$ & $-0.73$ & $\mathbf{2.22}$ & $\mathbf{1.54}$ & $0.32$\\
\midrule
\multirow{6}{*}{\rotatebox{90}{IHS}}
& Sex & $-2.09$ & $-2.19$ & $-0.16$ & $0.86$ & $-0.67$ & $-0.40$ & $\mathbf{4.97}$ & $\mathbf{4.93}$ & $\mathbf{0.06}$\\
& Age & $-0.15$ & $-18.48$ & $4.96$ & $2.36$ & $-18.48$ & $\mathbf{5.63}$ & $\mathbf{9.72}$ & $\mathbf{0.30}$ & $3.07$\\
& Ethnicity & $1.25$ & $-4.63$ & $-5.95$ & $2.80$ & $-0.84$ & $\mathbf{2.76}$ & $\mathbf{8.23}$ & $\mathbf{3.38}$ & $0.28$\\
& Specialty & $-1.65$ & $-15.13$ & $0.79$ & $-0.11$ & $-15.98$ & $0.40$ & $\mathbf{5.00}$ & $\mathbf{0.03}$ & $0.36$\\
& PHQ10 & $-6.17$ & $-10.00$ & $1.28$ & $-0.16$ & $-0.27$ & $0.15$ & $\mathbf{2.04}$ & $\mathbf{0.04}$ & $\mathbf{2.83}$\\
\rowcolor{gray!12}& \textit{Mean} & $-1.76$ & $-10.09$ & $0.19$ & $1.15$ & $-7.25$ & $\mathbf{1.71}$ & $\mathbf{5.99}$ & $\mathbf{1.74}$ & $1.32$\\
\midrule
\multirow{5}{*}{\rotatebox{90}{Percept-R}}
& Sex & $-2.80$ & $-3.30$ & $0.70$ & $-0.50$ & $-1.74$ & $-1.76$ & $\mathbf{5.12}$ & $\mathbf{4.70}$ & $0.59$\\
& Age & $-2.07$ & $-33.00$ & $-1.29$ & $-0.51$ & $-19.00$ & $0.69$ & $\mathbf{9.60}$ & $\mathbf{0.02}$ & $\mathbf{3.35}$\\
& Race & $-5.85$ & $-16.90$ & $-4.20$ & $-3.08$ & $-8.50$ & $1.42$ & $\mathbf{7.43}$ & $\mathbf{0.00}$ & $\mathbf{5.77}$\\
& Ethnicity & $-2.31$ & $-2.62$ & $0.44$ & $-0.28$ & $-0.58$ & $-0.43$ & $\mathbf{2.84}$ & $\mathbf{1.96}$ & $1.25$\\
\rowcolor{gray!12}& \textit{Mean} & $-3.26$ & $-13.96$ & $-1.09$ & $-1.09$ & $-7.46$ & $-0.02$ & $\mathbf{6.25}$ & $\mathbf{1.67}$ & $\mathbf{2.74}$\\
\bottomrule
\end{tabular}
}
\end{table}

\clearpage
\subsection{\textsc{FLARE} vs. Fairness in Demographic Models Using a Known Attribute}
\label{app:with_demo}
We evaluated demographic-aware fairness baselines based on demographic parity and equalized odds \cite{pal2023ensuring,hu2023parametric,liu2026fairness} which explicitly use \emph{sex} during training, whereas \textbf{\textsc{Flare}} does not use demographic or other sensitive attributes during training. The results are reported in Tables~\ref{tab:f1_demographics_models_reduced} and \ref{tab:dp_eo_flare}. 
Fair-DP and Fair-EO are fairness methods in the conventional sense: they use sex labels to reduce disparity, but they do not explicitly target benefit or harm avoidance. Their gains are therefore expected mainly in Equity ($\Delta E$), and they do reduce disparity more than \textbf{\textsc{Flare}} on several attributes, both on sex (Fair-EO on EDA; Fair-DP on OhioT1DM and IHS) and on attributes they were not trained on, such as OhioT1DM pump ($\Delta E=10.71$ for Fair-EO vs.\ $7.22$ for \textbf{\textsc{Flare}}).
%{Because Fair-DP and Fair-EO directly optimize with sex labels, they provide a strong demographic-aware comparison and can be viewed as an upper-bound reference for sex-specific fairness optimization. As expected, these baselines sometimes outperform \textbf{\textsc{Flare}} on the sex-based Equity metric ($\Delta E$). 
However, reducing disparity is not the same as ethical fairness. In every attribute where Fair-DP or Fair-EO reduce disparity more than \textbf{\textsc{Flare}}, they have lower $\Delta B$, and in 11 of these 13 cases they also have negative $\Delta H$; that is, they narrow the gap partly by lowering better-off subgroups (leveling down) rather than by raising every subgroup, sacrificing \emph{beneficence} and \emph{harm-avoidance}. 

For example, on EDA sex, Fair-EO achieves higher $\Delta E$ than \textbf{\textsc{Flare}} ($+0.44$ vs.\ $+0.07$), but it has much weaker $\Delta B$ and negative $\Delta H$ compared with \textbf{\textsc{Flare}} ($-0.09,-0.40$ vs.\ $+2.27,+2.22$). This indicates that demographic-aware optimization can reduce disparity for the attribute used during training, but may still sacrifice worst-subgroup protection or average subgroup performance. The limitation becomes clearer when the evaluation moves beyond the training attribute `sex.' Since Fair-DP and Fair-EO are trained using sex only, their gains do not consistently transfer to other demographic, device, clinical, or contextual sensitive attributes. For example, on OhioT1DM age, Fair-EO and Fair-DP cause severe degradation for at least one age subgroup ($\Delta H=-12.28$ and $-15.63$ percentage points), while \textbf{\textsc{Flare}} is non-negative on all three dimensions at the reported precision $(\Delta B=+5.07,\ \Delta H=0.00,\ \Delta E=+7.91)$. On IHS age, Fair-EO introduces massive harm ($\Delta H=-35.32$), while \textbf{\textsc{Flare}} protects the subgroup and boosts equity ($\Delta H=+0.30,\ \Delta E=+3.07$). On Percept-R ethnicity, Fair-EO and Fair-DP achieve equity gains comparable to \textbf{\textsc{Flare}} ($+1.02$ and $+1.29$ vs.\ $+1.25$), but with lower benefit ($+0.75$ and $+0.94$ vs.\ $+2.84$) and negative Harm-Avoidance ($-1.12$ and $-0.74$ vs.\ $+1.96$). Thus, these attribute-aware baselines strengthen our evaluation by showing that even demographic-aware methods, which have access to the sensitive attribute during training, do not consistently achieve balanced ethical fairness. Their advantage lies in Equity alone: where they reduce disparity more than \textbf{\textsc{Flare}}, they do so with lower Benefit and, in most cases, negative Harm-Avoidance; they do not provide robust improvements in Benefit or Harm-Avoidance, nor across broader heterogeneity factors. In contrast, \textbf{\textsc{Flare}} achieves a more balanced ethical fairness-performance trade-off across Benefit, Harm-Avoidance, and Equity without relying on demographic or sensitive attributes during training.
\subsection{Subgroup-Level Performance Analysis}
\label{sec:subgroup-analysis}

Tables~\ref{tab:ohio_eda_f1}, \ref{tab:ihs_f1}, and \ref{tab:perceptr_f1} summarize the subgroup-level F1 scores for the Baseline, KD, ARL, Reckoner, GoG, JTT, GroupDRO, and \textbf{\textsc{Flare}} models across the OhioT1DM, EDA, IHS, and Percept-R datasets. Each table reports performance disaggregated by demographic and heterogeneous attributes (e.g., sex, age, ethnicity, device type), allowing for the evaluation of both model utility and consistency across diverse populations.

Across all datasets, \textbf{\textsc{Flare}} attains the highest mean subgroup F1 on every dataset and the highest F1 in many individual subgroups, indicating improved predictive balance relative to all baselines. In the OhioT1DM dataset, \textbf{\textsc{Flare}} yields notable gains across device-related groups, demonstrating robustness to variation in pump generation and sensor band (the latter coinciding with the data-collection cohort). For example, it achieves \textbf{73.6\%} on the 630G Pump subgroup and \textbf{68.3\%} on the Empatica Band subgroup, outperforming most baselines. In the EDA dataset, where the baseline performance is already strong, \textbf{\textsc{Flare}} maintains stable accuracy across experimental conditions (Control at \textbf{89.6\%}, Postdose at \textbf{92.7\%}) and between sexes (Female at \textbf{88.7\%}, Male at \textbf{90.2\%}), outperforming adversarial and distillation-based approaches that exhibit inconsistent subgroup behavior.

In the highly heterogeneous IHS dataset, \textbf{\textsc{Flare}} achieves superior F1 scores across multiple sex, age, and ethnicity subgroups. Notably, it attains the highest scores for both Male (\textbf{64.0\%}) and Female (\textbf{65.2\%}) participants, reflecting its capacity to generalize across diverse clinical populations. Finally, in the Percept-R dataset, \textbf{\textsc{Flare}} demonstrates exceptional balance across high-cardinality attributes. It achieves peak performance across sex subgroups (Male at \textbf{81.6\%}, Female at \textbf{82.2\%}), underrepresented racial groups (Asian at \textbf{84.5\%}, Black/AA at \textbf{86.0\%}), and the youngest age bin (\textbf{97.6\%} for age $\le$ 100 months).

These results highlight that \textbf{\textsc{Flare}} not only improves overall utility but also reduces variability on most attributes and better protects vulnerable or underrepresented subgroups compared with fairness-oriented baselines such as ARL, GoG, and GroupDRO.

\begin{table}[t]
\centering\small
\caption{Scores (\%) across subgroups for different models in the OhioT1DM and EDA datasets.}
\vspace{-8pt}
\label{tab:ohio_eda_f1}
\begin{tabular}{ll|cccccccc}
\toprule
\textbf{Attr.} & \textbf{Subgroup} & Base & KD & ARL & Reck & GoG & JTT & GDRO & \textbf{\textsc{FLARE}}\\
\midrule
\multicolumn{10}{c}{\textbf{OhioT1DM}}\\
\midrule
Sex & Female & 70.7 & 70.5 & 70.2 & 70.5 & 71.0 & 64.9 & 65.7 & 71.1\\
& Male & 66.2 & 66.5 & 73.3 & 66.5 & 67.2 & 64.1 & 59.1 & 69.6\\
Age & 20--40 & 51.3 & 62.4 & 67.7 & 62.9 & 57.1 & 76.3 & 62.7 & 66.5\\
& 40--60 & 69.3 & 65.5 & 69.9 & 67.4 & 58.1 & 62.3 & 61.1 & 69.3\\
& 60--80 & 75.9 & 67.2 & 72.8 & 69.0 & 62.1 & 66.5 & 62.7 & 75.9\\
Pump & 530G & 67.7 & 64.1 & 67.9 & 64.0 & 65.2 & 63.1 & 62.3 & 67.7\\
& 630G & 51.6 & 57.3 & 63.7 & 59.5 & 70.8 & 76.3 & 62.7 & 73.6\\
Band & Basis & 71.5 & 69.3 & 73.5 & 69.7 & 69.3 & 63.7 & 64.5 & 71.5\\
& Empatica & 63.2 & 65.4 & 65.5 & 65.5 & 65.4 & 64.7 & 57.3 & \textbf{68.3}\\
\midrule
\multicolumn{10}{c}{\textbf{EDA}}\\
\midrule
Group & Control & 85.7 & 66.0 & 87.8 & 82.1 & 86.1 & 86.8 & 86.4 & \textbf{89.6}\\
& Predose & 83.2 & 70.0 & 84.2 & 89.4 & 83.0 & 83.3 & 82.3 & 84.9\\
& Postdose & 91.9 & 71.6 & 90.9 & 85.2 & 92.0 & 92.1 & 91.5 & \textbf{92.7}\\
Sex & Female & 86.4 & 67.6 & 85.6 & 84.3 & 86.3 & 87.2 & 85.8 & \textbf{88.7}\\
& Male & 88.0 & 70.8 & 89.6 & 87.4 & 87.9 & 87.6 & 89.3 & \textbf{90.2}\\
\bottomrule
\end{tabular}
\vspace{-8pt}
\end{table}

\begin{table}[t]
\centering\small
% \setlength{\tabcolsep}{2.4pt}
%\caption{Scores (\%) across subgroups for Baseline, KD, ARL, Reckoner, GOG, JTT, GroupDRO, \textbf{\textsc{Flare}} in the IHS dataset.}
  \caption{Scores (\%) across subgroups for Baseline, KD, ARL, Reckoner, GoG, JTT, GroupDRO, and \textbf{\textsc{Flare}} in the IHS dataset. Values are rounded to one decimal; all BHE metrics are computed from the unrounded scores, so subgroups that appear identical here may differ slightly. Residency-specialty subgroups (12 categories) are summarized at the attribute level in Tables~\ref{tab:f1_demographics_models} and~\ref{tab:sota_delta_grouped}.}
\vspace{-8pt}
\label{tab:ihs_f1}
\begin{tabular}{ll|cccccccc}
\toprule
\textbf{Attr.} & \textbf{Subgroup} & Base & KD & ARL & Reck & GoG & JTT & GDRO & \textbf{\textsc{FLARE}}\\
\midrule
Sex & Male & 59.0 & 57.0 & 59.0 & 58.7 & 58.6 & 56.8 & 61.4 & \textbf{64.0}\\
& Female & 60.3 & 59.2 & 60.3 & 56.4 & 61.3 & 58.3 & 59.6 & \textbf{65.2}\\
Age & 23 & 81.5 & 59.8 & 81.9 & 69.5 & 49.7 & 63.0 & 63.0 & \textbf{90.9}\\
& 24 & 31.9 & 35.4 & 45.7 & 36.8 & 35.3 & 39.1 & 59.1 & 52.1\\
& 25 & 58.7 & 59.7 & 59.1 & 61.0 & 57.6 & 56.3 & 61.9 & \textbf{65.3}\\
& 26 & 65.2 & 66.7 & 67.0 & 58.0 & 66.4 & 59.8 & 61.9 & \textbf{69.2}\\
& 27 & 65.2 & 64.4 & 68.6 & 60.2 & 61.0 & 60.3 & 65.2 & 67.2\\
& 28 & 56.2 & 52.7 & 62.7 & 53.8 & 61.4 & 57.3 & 56.8 & \textbf{63.0}\\
& 29 & 52.0 & 59.9 & 60.0 & 53.1 & 56.9 & 52.8 & 54.8 & 58.0\\
& 30 & 56.4 & 59.5 & 65.5 & 53.9 & 55.8 & 55.7 & 57.3 & 65.3\\
& 32 & 60.9 & 61.9 & 63.0 & 61.4 & 59.1 & 60.5 & 59.5 & \textbf{69.3}\\
& 33 & 51.3 & 56.4 & 54.5 & 42.6 & 32.9 & 61.9 & 57.6 & 59.1\\
& 34 & 34.4 & 36.1 & 49.7 & 39.3 & 55.7 & 49.8 & 45.5 & \textbf{75.5}\\
& 38 & 79.8 & 87.2 & 86.7 & 65.8 & 73.4 & 79.5 & 83.2 & 84.4\\
& 51 & 52.7 & 52.9 & 48.6 & 50.8 & 47.9 & 47.9 & 50.8 & 53.0\\
Ethn. & White & 61.0 & 61.9 & 64.9 & 58.1 & 60.6 & 58.4 & 61.7 & 64.4\\
& Black/AA & 51.2 & 58.5 & 61.3 & 51.9 & 47.8 & 48.6 & 57.9 & 58.9\\
& Latino & 62.1 & 56.6 & 73.0 & 58.1 & 76.9 & 76.8 & 61.4 & 73.8\\
& Asian & 58.6 & 60.3 & 61.2 & 55.1 & 58.7 & 54.0 & 57.8 & 63.5\\
& Multiracial & 57.7 & 60.7 & 64.6 & 54.1 & 61.9 & 63.8 & 61.9 & \textbf{67.4}\\
& Arab/ME & 47.6 & 50.6 & 51.0 & 39.8 & 40.1 & 44.2 & 54.4 & \textbf{59.8}\\
PHQ10 & Yes & 64.9 & 59.7 & 64.7 & 60.6 & 58.5 & 54.9 & 64.6 & 64.9\\
& No & 60.1 & 56.5 & 60.6 & 57.1 & 60.2 & 57.8 & 60.1 & 64.2\\
\bottomrule
\end{tabular}
\vspace{-10pt}
\end{table}
\clearpage
%\noindent(IHS Specialty rows omitted here for space; they follow the same pattern and are in the released score files.)

\begin{table}[H]
\centering\small
\caption{Scores (\%) across subgroups for different models (Baseline, KD, ARL, Reckoner, GOG, JTT, GroupDRO, \textbf{\textsc{Flare}}) in the Percept-R dataset. Age is grouped into four bins, with F1 computed from the pooled out-of-fold predictions within each bin. Values are rounded to one decimal; all BHE metrics are computed from the unrounded scores, so subgroups that appear identical here may differ slightly.}

\label{tab:perceptr_f1}
\vspace{-5pt}
\begin{tabular}{ll|cccccccc}
\toprule
\textbf{Attr.} & \textbf{Subgroup} & Base & KD & ARL & Reck & GoG & JTT & GDRO & \textbf{\textsc{FLARE}}\\
\midrule
Sex & Male & 76.0 & 76.5 & 76.1 & 74.9 & 76.4 & 73.7 & 74.3 & \textbf{81.6}\\
& Female & 77.5 & 78.4 & 76.4 & 73.0 & 78.9 & 74.2 & 78.2 & \textbf{82.2}\\
Age & $\le$100 & 59.5 & 59.5 & 56.8 & 54.6 & 59.0 & 90.7 & 92.0 & 97.6\\
& 101--130 & 75.7 & 75.8 & 87.2 & 76.3 & 75.5 & 72.9 & 63.5 & 75.8\\
& 131--160 & 81.9 & 85.1 & 79.9 & 78.2 & 80.3 & 78.3 & 78.5 & 82.1\\
& $>$160 & 89.6 & 89.7 & 83.6 & 90.1 & 94.4 & 56.6 & 70.6 & 89.7\\
Race & White & 77.4 & 78.8 & 79.4 & 76.9 & 79.9 & 75.4 & 76.9 & 83.8\\
& Asian & 66.4 & 72.0 & 71.8 & 78.7 & 74.1 & 49.5 & 62.3 & \textbf{84.5}\\
& Black/AA & 76.3 & 76.5 & 76.4 & 77.6 & 76.9 & 76.7 & 75.4 & \textbf{86.0}\\
& Multiple & 82.0 & 79.2 & 72.4 & 71.9 & 79.0 & 77.5 & 80.6 & 85.0\\
& \textit{Other} & \textit{94.6} & 93.5 & \textbf{96.8} & 95.0 & 93.5 & 88.5 & 86.1 & \textit{94.6}\\
Ethn. & Hispanic & 77.1 & 78.9 & 79.7 & 77.3 & 80.0 & 75.1 & 76.5 & \textbf{80.8}\\
& Not Hisp. & 79.7 & 78.2 & 73.9 & 72.8 & 78.8 & 77.1 & 79.7 & \textbf{81.6}\\
\bottomrule
\vspace{-15pt}
\end{tabular}
\end{table}

\subsection{Loss Landscapes for EDA and IHS}
\label{app:loss-landscape}
\begin{figure}[h!]
    \centering
    \begin{subfigure}[h!]{0.95\textwidth}
        \includegraphics[width=\linewidth]{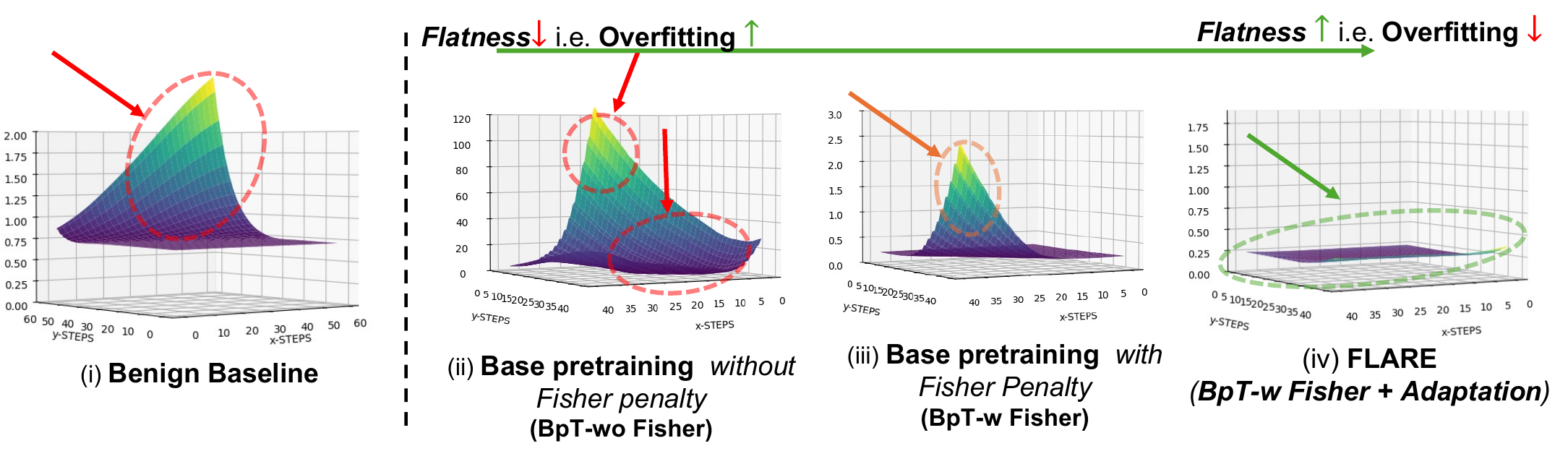}
        \caption{EDA dataset}
        \label{fig:loss-eda}
    \end{subfigure}
    \begin{subfigure}[h!]{0.95\textwidth}
        \includegraphics[width=\linewidth]{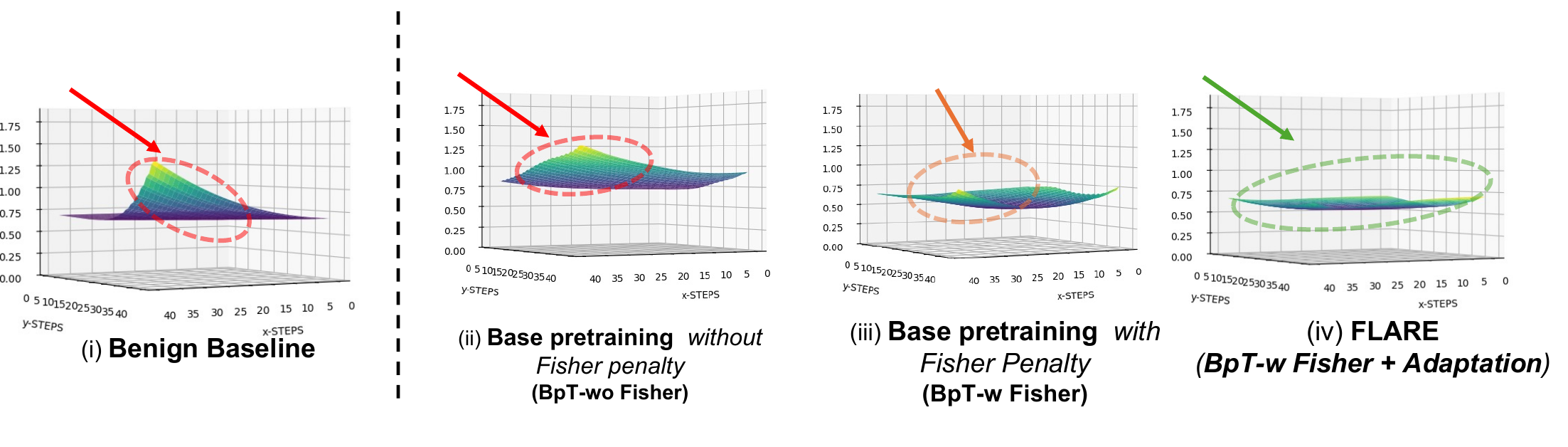}
        \caption{IHS dataset}
        \label{fig:loss-ihs}
    \end{subfigure}
\caption{Loss landscapes on EDA and IHS for (i) the benign baseline, (ii) base pretraining without the Fisher penalty, (iii) base pretraining with the Fisher penalty, and (iv) \textbf{\textsc{Flare}}. Arrows and dotted circles follow the convention of Figure~\ref{loss-landscape-vis}.}
\label{fig:loss-landscape-app}
\end{figure}

Figure~\ref{fig:loss-landscape-app} shows the loss landscapes for the EDA and IHS datasets, generated with the same procedure and variants as Figure~\ref{loss-landscape-vis} (Section~\ref{sec:loss-landscapes}). Both datasets show the same progression as OhioT1DM: the benign baseline has the sharpest surface, base pretraining smooths it, the Fisher penalty further reduces sharpness, and \textbf{\textsc{Flare}} yields the flattest and widest basin.

\subsection{Training and Inference Efficiency and Runtime Analysis of \textbf{\textsc{Flare}}}
\label{runtime_appendix}
Table~\ref{tab:flare_breakdown_eda_rss_util} reports stage wise breakdown of the CPU/GPU average compute utilization (\%) and peak process RSS memory utilization in MB for \textbf{\textsc{FLARE}} for the EDA dataset.

Tables~\ref{training_runtime_ihs_ohio_percept_appendix}, 
\ref{inference_runtime_appendix_ihs_ohio_percept}, and 
\ref{flare_breakdown_appendix_ihs_ohio_percept} extend the runtime analysis in Section~\ref{on-device} to the IHS, OhioT1DM, and Percept-R datasets. Following our setup in Section~\ref{on-device}, Fold-1 was utilized across the all the datasets to benchmark training and inference phase runtime efficiency. These results show that the deployment trends observed on EDA are consistent across datasets and hardware platforms. During both training and inference, \textbf{\textsc{Flare}} incurs higher latency than the single-model baselines, but the absolute cost remains in the millisecond-per-sample range across the evaluated devices. The  increase is expected because \textbf{\textsc{Flare}} performs sample-specific  routing and maintains cluster-specialized models rather than using a single global model.

The breakdown in Table~\ref{flare_breakdown_appendix_ihs_ohio_percept} shows that this overhead is concentrated in the same stages across datasets. During  training, the dominant additional cost is the cluster fitting step, with cluster-specific adaptation becoming more visible for datasets with more clusters, such as OhioT1DM and Percept-R. During inference, the overhead is almost entirely due to computing the label-free Fisher penalty (with the predicted label $\hat{y}$) during the cluster assignment step, while the cluster-specific model inference stage remains close to the latency of the baseline models. This pattern is consistent across desktop, server-class, and resource-constrained platforms.

Resource usage follows a similar trend. Compute utilization is largely governed by the underlying platform and backend, while \textbf{\textsc{Flare}} remains within the same order of magnitude as the baselines. Peak RSS memory is higher for \textbf{\textsc{Flare}} because it stores the GMM router and multiple cluster-specific models, but it remains comparable to other baselines like GOG. Overall, these results confirm that the conclusions drawn from the EDA dataset generalize across the remaining datasets: \textbf{\textsc{Flare}} adds predictable routing and specialization overhead, while preserving practical on-device training and inference feasibility.

\begin{table}[h]
\color{black}
\centering
% \scriptsize
\caption{\textsc{FLARE} training and inference stage resource breakdown across deployment platforms for the EDA dataset (compute utilization in \%; peak RSS memory in MB).}
\label{tab:flare_breakdown_eda_rss_util}
\resizebox{\linewidth}{!}{%
\begin{tabular}{lcrrrrrrrrrrrrrr}
\toprule
Phase & \makecell{\textbf{\textsc{Flare}}\\ Pipeline Stage} & \multicolumn{2}{c}{AMD CPU} & \multicolumn{2}{c}{NVIDIA GPU} & \multicolumn{2}{c}{Spark CPU} & \multicolumn{2}{c}{Spark GPU} & \multicolumn{2}{c}{Apple M4} & \multicolumn{2}{c}{Google Pixel 6} & \multicolumn{2}{c}{Raspberry Pi} \\
\cmidrule(lr){3-4} \cmidrule(lr){5-6} \cmidrule(lr){7-8} \cmidrule(lr){9-10} \cmidrule(lr){11-12} \cmidrule(lr){13-14} \cmidrule(lr){15-16}
 &   & Util. & RSS  & Util. & RSS  & Util. & RSS  & Util. & RSS  & Util. & RSS  & Util. & RSS  & Util. & RSS \\
\midrule
Training & Pre-training stage  & 3.17 & 555.4  & 77.09 & 1022.4  & 4.99 & 488.4  & 7.56 & 1458.1  & 9.64 & 306.5  & 12.40 & 326.4  & 25.10 & 416.9 \\
Training & Cluster fitting stage  & 3.13 & 995.0  & 42.97 & 1622.9  & 5.03 & 955.8  & 1.40 & 2069.8  & 4.64 & 716.3 & 12.41 & 605.1  & 24.44 & 762.3 \\
Training & Training cluster assignment  & 0.00 & 982.5  & 40.00 & 1608.8  & 0.00 & 936.2  & 0.00 & 2060.4  & 0.07 & 716.3  & 0.00 & 512.3  & 0.00 & 747.1 \\
Training & Adaptation for Cluster 1  & 2.56 & 1009.7  & 43.00 & 1648.6  & 4.84 & 956.7  & 0.00 & 2155.3  & 8.48 & 680.0  & 12.17 & 513.6  & 24.60 & 767.8 \\
Training & Adaptation for Cluster 2  & 3.08 & 1009.8  & 84.38 & 1648.6  & 4.85 & 956.7  & 0.00 & 2155.3  & 9.29 & 680.6  & 12.34 & 515.4  & 24.90 & 769.3 \\
\midrule
Inference & Inference cluster assignment  & 3.13 & 1045.0  & 43.00 & 1663.2  & 5.03 & 986.9  & 0.00 & 2112.3  & 9.89 & 738.0  & 12.42 & 578.7  & 25.05 & 801.1 \\
Inference & Cluster-specific model inference  & 3.06 & 1009.8  & 92.33 & 1648.6  & 4.89 & 956.7  & 13.00 & 2155.3  & 8.39 & 680.6  & 12.54 & 515.8  & 24.56 & 769.3 \\
\bottomrule
\end{tabular}%
}
\end{table}

\begin{table}[h!]
\centering
\scriptsize
\caption{Training latency and resource comparison across deployment platforms for the IHS, OhioT1DM and Percept-R datasets (wall time in ms/sample; compute utilization in \%; peak RSS memory in MB).}
\label{training_runtime_ihs_ohio_percept_appendix}
\begin{subtable}{\textwidth}
\color{black}
\caption{IHS Dataset}
\label{tab:training_runtime_ihs}
\resizebox{\textwidth}{!}{%
\begin{tabular}{lrrrrrrrrrrrrrrrrrrrrr}
\toprule
Approach & \multicolumn{3}{c}{AMD CPU} & \multicolumn{3}{c}{NVIDIA GPU} & \multicolumn{3}{c}{Spark CPU} & \multicolumn{3}{c}{Spark GPU} & \multicolumn{3}{c}{Apple M4} & \multicolumn{3}{c}{Google Pixel 6} & \multicolumn{3}{c}{Raspberry Pi} \\
\cmidrule(lr){2-4} \cmidrule(lr){5-7} \cmidrule(lr){8-10} \cmidrule(lr){11-13} \cmidrule(lr){14-16} \cmidrule(lr){17-19} \cmidrule(lr){20-22}
 & Time & Util. & RSS & Time & Util. & RSS & Time & Util. & RSS & Time & Util. & RSS & Time & Util. & RSS & Time & Util. & RSS & Time & Util. & RSS \\
\midrule
Baseline & 0.387 & 3.08 & 543.5 & 0.590 & 46.88 & 986.0 & 0.394 & 5.01 & 477.2 & 0.944 & 0.00 & 1338.4 & 0.348 & 7.71 & 287.3 & 1.272 & 9.83 & 315.5 & 1.367 & 24.90 & 404.5 \\
KD & 0.441 & 3.14 & 544.2 & 0.824 & 60.66 & 1022.8 & 0.464 & 4.92 & 477.4 & 1.328 & 4.08 & 1399.5 & 0.319 & 9.65 & 286.6 & 1.941 & 10.56 & 316.0 & 1.860 & 25.07 & 405.0 \\
ARL & 0.365 & 3.03 & 544.2 & 0.611 & 48.80 & 1002.1 & 0.381 & 5.00 & 478.0 & 0.986 & 1.25 & 1363.3 & 0.258 & 9.74 & 288.3 & 1.277 & 9.80 & 315.7 & 1.336 & 25.09 & 404.4 \\
GoG & 0.332 & 3.08 & 581.4 & 0.515 & 37.26 & 1159.4 & 0.387 & 5.02 & 515.0 & 0.684 & 0.00 & 1627.9 & 0.245 & 9.62 & 314.7 & 1.196 & 9.54 & 329.0 & 1.210 & 24.90 & 420.9 \\
Reckoner & 0.608 & 3.11 & 545.1 & 0.932 & 65.16 & 994.5 & 0.609 & 4.97 & 477.8 & 1.985 & 3.59 & 1377.0 & 0.397 & 9.92 & 287.4 & 2.679 & 11.12 & 314.7 & 2.861 & 25.05 & 405.4 \\
\textbf{\textsc{Flare}} & 8.981 & 2.34 & 889.0 & 9.973 & 46.81 & 1437.5 & 8.891 & 3.71 & 800.4 & 10.064 & 0.69 & 1887.6 & 6.997 & 7.02 & 516.5 & 22.408 & 9.77 & 467.0 & 31.186 & 19.97 & 670.2 \\
\bottomrule
\end{tabular}%
}
\end{subtable}

\begin{subtable}{\textwidth}
\centering
\scriptsize
\color{black}
\caption{OhioT1DM Dataset}
\label{tab:training_runtime_ohio}
\resizebox{\textwidth}{!}{%
\begin{tabular}{lrrrrrrrrrrrrrrrrrrrrr}
\toprule
Approach & \multicolumn{3}{c}{AMD CPU} & \multicolumn{3}{c}{NVIDIA GPU} & \multicolumn{3}{c}{Spark CPU} & \multicolumn{3}{c}{Spark GPU} & \multicolumn{3}{c}{Apple M4} & \multicolumn{3}{c}{Google Pixel 6} & \multicolumn{3}{c}{Raspberry Pi} \\
\cmidrule(lr){2-4} \cmidrule(lr){5-7} \cmidrule(lr){8-10} \cmidrule(lr){11-13} \cmidrule(lr){14-16} \cmidrule(lr){17-19} \cmidrule(lr){20-22}
 & Time & Util. & RSS & Time & Util. & RSS & Time & Util. & RSS & Time & Util. & RSS & Time & Util. & RSS & Time & Util. & RSS & Time & Util. & RSS \\
\midrule
Baseline & 0.137 & 3.13 & 547.4 & 0.230 & 85.95 & 991.1 & 0.166 & 5.04 & 480.0 & 0.275 & 10.44 & 1401.7 & 0.104 & 9.48 & 296.0 & 0.878 & 12.17 & 318.8 & 1.009 & 25.10 & 409.1 \\
KD & 0.297 & 3.14 & 548.2 & 0.500 & 89.86 & 1028.0 & 0.390 & 5.06 & 480.7 & 0.903 & 9.39 & 1462.7 & 0.208 & 9.95 & 296.9 & 2.113 & 12.36 & 317.3 & 2.391 & 25.13 & 409.7 \\
ARL & 0.142 & 3.13 & 547.7 & 0.255 & 86.74 & 977.1 & 0.171 & 5.03 & 481.1 & 0.421 & 7.17 & 1362.4 & 0.096 & 9.92 & 296.1 & 0.929 & 12.25 & 316.8 & 1.038 & 25.11 & 408.2 \\
GoG & 1.002 & 3.15 & 7400.1 & 0.052 & 56.26 & 1170.5 & 0.857 & 5.10 & 7102.0 & 0.092 & 21.60 & 1669.8 & 0.549 & 8.14 & 4698.7 & 1.012 & 12.86 & 731.8 & 1.401 & 25.10 & 766.1 \\
Reckoner & 0.594 & 3.13 & 909.4 & 0.735 & 91.20 & 1071.5 & 0.754 & 5.05 & 683.6 & 1.670 & 9.77 & 1491.5 & 0.443 & 9.95 & 532.9 & 3.908 & 12.39 & 534.2 & 4.405 & 25.12 & 561.0 \\
\textbf{\textsc{Flare}} & 2.671 & 2.60 & 1063.3 & 4.217 & 75.03 & 1715.1 & 3.227 & 4.19 & 977.3 & 5.162 & 7.13 & 2219.6 & 2.013 & 8.19 & 856.0 & 16.218 & 10.30 & 561.2 & 16.341 & 20.92 & 843.8 \\
\bottomrule
\end{tabular}%
}
\end{subtable}

\begin{subtable}{\textwidth}
\centering
\scriptsize
\color{black}
\caption{Percept-R Dataset}
\label{tab:training_runtime_percept_r}
\resizebox{\textwidth}{!}{%
\begin{tabular}{lrrrrrrrrrrrrrrrrrrrrr}
\toprule
Approach & \multicolumn{3}{c}{AMD CPU} & \multicolumn{3}{c}{NVIDIA GPU} & \multicolumn{3}{c}{Spark CPU} & \multicolumn{3}{c}{Spark GPU} & \multicolumn{3}{c}{Apple M4} & \multicolumn{3}{c}{Google Pixel 6} & \multicolumn{3}{c}{Raspberry Pi} \\
\cmidrule(lr){2-4} \cmidrule(lr){5-7} \cmidrule(lr){8-10} \cmidrule(lr){11-13} \cmidrule(lr){14-16} \cmidrule(lr){17-19} \cmidrule(lr){20-22}
 & Time & Util. & RSS & Time & Util. & RSS & Time & Util. & RSS & Time & Util. & RSS & Time & Util. & RSS & Time & Util. & RSS & Time & Util. & RSS \\
\midrule
Baseline & 0.159 & 3.13 & 585.3 & 0.233 & 85.69 & 1153.0 & 0.150 & 5.05 & 513.5 & 0.403 & 5.49 & 1535.9 & 0.101 & 9.97 & 322.6 & 0.886 & 12.18 & 351.5 & 0.892 & 25.06 & 441.1 \\
KD & 0.323 & 3.13 & 586.0 & 0.469 & 88.76 & 1189.4 & 0.304 & 5.08 & 513.8 & 0.835 & 6.96 & 1577.0 & 0.211 & 9.96 & 322.8 & 2.019 & 12.39 & 351.6 & 1.990 & 25.11 & 441.9 \\
ARL & 0.208 & 3.13 & 586.2 & 0.316 & 87.37 & 1185.3 & 0.196 & 5.01 & 514.9 & 0.711 & 5.17 & 1532.9 & 0.135 & 9.89 & 323.1 & 1.159 & 12.24 & 352.1 & 1.215 & 25.09 & 442.1 \\
GoG & 1.733 & 3.14 & 7731.5 & 0.059 & 60.53 & 1335.6 & 1.292 & 5.07 & 7651.7 & 0.103 & 22.67 & 1820.9 & 0.636 & 8.53 & 5116.1 & 1.135 & 12.90 & 1573.0 & 1.538 & 25.12 & 1471.2 \\
Reckoner & 0.704 & 3.14 & 917.6 & 0.746 & 89.25 & 1380.1 & 0.656 & 5.08 & 743.8 & 1.677 & 7.38 & 1795.9 & 0.454 & 9.94 & 622.4 & 3.990 & 12.47 & 662.6 & 4.094 & 25.10 & 572.5 \\
\textbf{\textsc{Flare}} & 3.411 & 2.72 & 1136.6 & 5.386 & 79.06 & 1892.4 & 3.607 & 4.41 & 1029.7 & 7.567 & 6.82 & 2448.3 & 2.289 & 8.44 & 934.2 & 31.838 & 10.92 & 610.2 & 20.520 & 21.93 & 910.2 \\
\bottomrule
\end{tabular}%
}
\end{subtable}
\end{table}

\begin{table}[h!]
\centering
\color{black}
\scriptsize
\caption{Inference latency and resource comparison across deployment platforms for the IHS, OhioT1DM and Percept-R datasets (wall time in ms/sample; compute utilization in \%; peak RSS memory in MB).}
\label{inference_runtime_appendix_ihs_ohio_percept}
\begin{subtable}{\textwidth}
\color{black}
\caption{IHS Dataset}
\label{tab:inference_runtime_ihs}
\resizebox{\textwidth}{!}{%
\begin{tabular}{lrrrrrrrrrrrrrrrrrrrrr}
\toprule
Approach & \multicolumn{3}{c}{AMD CPU} & \multicolumn{3}{c}{NVIDIA GPU} & \multicolumn{3}{c}{Spark CPU} & \multicolumn{3}{c}{Spark GPU} & \multicolumn{3}{c}{Apple M4} & \multicolumn{3}{c}{Google Pixel 6} & \multicolumn{3}{c}{Raspberry Pi} \\
\cmidrule(lr){2-4} \cmidrule(lr){5-7} \cmidrule(lr){8-10} \cmidrule(lr){11-13} \cmidrule(lr){14-16} \cmidrule(lr){17-19} \cmidrule(lr){20-22}
 & Time & Util. & RSS & Time & Util. & RSS & Time & Util. & RSS & Time & Util. & RSS & Time & Util. & RSS & Time & Util. & RSS & Time & Util. & RSS \\
\midrule
Baseline & 0.022 & 3.04 & 543.7 & 0.023 & 91.00 & 986.3 & 0.022 & 0.00 & 477.3 & 0.022 & 0.00 & 1338.5 & 0.026 & 1.62 & 287.3 & 0.054 & 9.75 & 315.9 & 0.064 & 16.29 & 404.5 \\
KD & 0.022 & 0.00 & 544.4 & 0.023 & 92.00 & 1023.1 & 0.022 & 0.00 & 477.4 & 0.022 & 7.00 & 1399.6 & 0.026 & 1.62 & 286.6 & 0.080 & 9.79 & 316.4 & 0.043 & 24.28 & 405.0 \\
ARL & 0.022 & 0.00 & 544.4 & 0.023 & 82.00 & 1002.3 & 0.022 & 0.00 & 478.0 & 0.023 & 3.00 & 1363.3 & 0.026 & 1.57 & 288.3 & 0.080 & 9.83 & 316.1 & 0.043 & 24.30 & 404.4 \\
GoG & 0.021 & 3.04 & 547.3 & 0.023 & 39.00 & 1159.6 & 0.022 & 0.00 & 480.8 & 0.023 & 0.00 & 1628.0 & 0.026 & 1.72 & 314.8 & 0.051 & 10.19 & 329.3 & 0.043 & 24.34 & 408.3 \\
Reckoner & 0.021 & 3.05 & 545.3 & 0.023 & 94.00 & 994.8 & 0.023 & 4.64 & 477.8 & 0.023 & 7.00 & 1377.1 & 0.026 & 2.32 & 287.4 & 0.103 & 10.17 & 315.0 & 0.064 & 24.43 & 405.4 \\
\textbf{\textsc{Flare}} & 1.148 & 3.03 & 979.6 & 1.283 & 56.85 & 1569.0 & 1.398 & 4.81 & 885.3 & 1.511 & 0.76 & 2029.3 & 1.047 & 5.91 & 545.0 & 2.495 & 12.36 & 509.7 & 5.380 & 21.72 & 740.5 \\
\bottomrule
\end{tabular}%
}
\end{subtable}

\begin{subtable}{\textwidth}
\color{black}
\centering
\scriptsize
\caption{OhioT1DM Dataset}
\label{tab:inference_runtime_ohio}
\resizebox{\textwidth}{!}{%
\begin{tabular}{lrrrrrrrrrrrrrrrrrrrrr}
\toprule
Approach & \multicolumn{3}{c}{AMD CPU} & \multicolumn{3}{c}{NVIDIA GPU} & \multicolumn{3}{c}{Spark CPU} & \multicolumn{3}{c}{Spark GPU} & \multicolumn{3}{c}{Apple M4} & \multicolumn{3}{c}{Google Pixel 6} & \multicolumn{3}{c}{Raspberry Pi} \\
\cmidrule(lr){2-4} \cmidrule(lr){5-7} \cmidrule(lr){8-10} \cmidrule(lr){11-13} \cmidrule(lr){14-16} \cmidrule(lr){17-19} \cmidrule(lr){20-22}
 & Time & Util. & RSS & Time & Util. & RSS & Time & Util. & RSS & Time & Util. & RSS & Time & Util. & RSS & Time & Util. & RSS & Time & Util. & RSS \\
\midrule
Baseline & 0.005 & 2.99 & 547.6 & 0.005 & 94.00 & 991.3 & 0.005 & 0.00 & 480.1 & 0.005 & 12.00 & 1401.2 & 0.006 & 2.68 & 296.0 & 0.022 & 10.17 & 319.1 & 0.023 & 19.45 & 409.1 \\
KD & 0.005 & 0.00 & 548.4 & 0.005 & 95.00 & 1028.2 & 0.005 & 4.81 & 480.7 & 0.005 & 10.00 & 1462.4 & 0.006 & 2.58 & 296.9 & 0.018 & 9.52 & 317.7 & 0.023 & 19.50 & 409.7 \\
ARL & 0.005 & 3.02 & 547.9 & 0.005 & 93.00 & 977.3 & 0.005 & 4.30 & 481.1 & 0.005 & 8.00 & 1361.9 & 0.006 & 2.43 & 296.1 & 0.025 & 11.39 & 317.2 & 0.019 & 18.37 & 408.2 \\
GoG & 0.005 & 0.00 & 928.3 & 0.005 & 94.00 & 1170.7 & 0.005 & 4.74 & 644.8 & 0.005 & 96.00 & 1669.8 & 0.006 & 5.28 & 295.6 & 0.023 & 12.46 & 559.0 & 0.019 & 24.39 & 659.9 \\
Reckoner & 0.005 & 2.96 & 633.3 & 0.005 & 95.00 & 1071.7 & 0.005 & 8.32 & 554.0 & 0.005 & 10.00 & 1490.0 & 0.006 & 3.90 & 527.6 & 0.036 & 11.19 & 411.0 & 0.033 & 24.47 & 532.4 \\
\textbf{\textsc{Flare}} & 0.924 & 3.09 & 1165.2 & 1.951 & 86.95 & 1825.6 & 1.268 & 3.99 & 1075.4 & 1.825 & 10.28 & 2329.9 & 0.624 & 8.29 & 966.2 & 8.060 & 12.55 & 608.7 & 7.239 & 24.87 & 929.7 \\
\bottomrule
\end{tabular}%
}
\end{subtable}

\begin{subtable}{\textwidth}
\color{black}
\centering
\scriptsize
\caption{Percept-R Dataset}
\label{tab:inference_runtime_percept_r}
\resizebox{\textwidth}{!}{%
\begin{tabular}{lrrrrrrrrrrrrrrrrrrrrr}
\toprule
Approach & \multicolumn{3}{c}{AMD CPU} & \multicolumn{3}{c}{NVIDIA GPU} & \multicolumn{3}{c}{Spark CPU} & \multicolumn{3}{c}{Spark GPU} & \multicolumn{3}{c}{Apple M4} & \multicolumn{3}{c}{ Google Pixel 6} & \multicolumn{3}{c}{Raspberry Pi} \\
\cmidrule(lr){2-4} \cmidrule(lr){5-7} \cmidrule(lr){8-10} \cmidrule(lr){11-13} \cmidrule(lr){14-16} \cmidrule(lr){17-19} \cmidrule(lr){20-22}
 & Time & Util. & RSS & Time & Util. & RSS & Time & Util. & RSS & Time & Util. & RSS & Time & Util. & RSS & Time & Util. & RSS & Time & Util. & RSS \\
\midrule
Baseline & 0.004 & 0.00 & 585.5 & 0.004 & 94.00 & 1153.2 & 0.004 & 4.78 & 513.5 & 0.004 & 6.00 & 1534.0 & 0.005 & 2.99 & 322.6 & 0.013 & 13.48 & 351.7 & 0.015 & 18.40 & 440.6 \\
KD & 0.004 & 0.00 & 586.2 & 0.004 & 86.00 & 1189.6 & 0.004 & 0.00 & 513.8 & 0.004 & 6.00 & 1575.3 & 0.005 & 7.17 & 322.9 & 0.018 & 12.44 & 351.9 & 0.019 & 19.56 & 441.4 \\
ARL & 0.004 & 0.00 & 586.4 & 0.004 & 94.00 & 1185.5 & 0.004 & 4.52 & 514.9 & 0.004 & 5.00 & 1530.9 & 0.005 & 3.16 & 323.2 & 0.019 & 11.87 & 352.5 & 0.015 & 18.37 & 441.5 \\
GoG & 0.004 & 0.00 & 656.9 & 0.004 & 96.00 & 1335.8 & 0.005 & 3.93 & 553.7 & 0.005 & 96.00 & 1821.0 & 0.005 & 3.15 & 239.1 & 0.023 & 12.02 & 588.0 & 0.015 & 18.38 & 1419.1 \\
Reckoner & 0.004 & 3.03 & 745.3 & 0.008 & 93.00 & 1380.3 & 0.008 & 2.39 & 645.4 & 0.004 & 7.00 & 1794.9 & 0.005 & 6.88 & 622.4 & 0.025 & 10.97 & 488.9 & 0.022 & 20.46 & 550.6 \\
\textbf{\textsc{Flare}} & 0.880 & 2.59 & 1213.2 & 2.102 & 89.85 & 1966.4 & 1.126 & 4.94 & 1101.2 & 1.911 & 8.87 & 2524.1 & 0.648 & 8.52 & 1019.7 & 4.953 & 12.68 & 643.2 & 6.419 & 25.58 & 973.0 \\
\bottomrule
\end{tabular}%
}
\end{subtable}

\end{table}

\begin{table}[h!]
\caption{\textbf{\textsc{Flare}} training and inference stage latency and resource breakdown across deployment platforms for IHS, OhioT1DM and Percept-R datasets (wall time in ms/sample; compute utilization in \%; peak RSS memory in MB)}
\label{flare_breakdown_appendix_ihs_ohio_percept}
\centering
\scriptsize
\begin{subtable}{\textwidth}
\color{black}
\caption{IHS Dataset}
\label{tab:flare_breakdown_ihs}
\resizebox{\textwidth}{!}{%
\begin{tabular}{llrrrrrrrrrrrrrrrrrrrrr}
\toprule
Phase & \textbf{\textsc{Flare}} pipeline stage & \multicolumn{3}{c}{AMD CPU} & \multicolumn{3}{c}{NVIDIA GPU} & \multicolumn{3}{c}{Spark CPU} & \multicolumn{3}{c}{Spark GPU} & \multicolumn{3}{c}{Apple M4} & \multicolumn{3}{c}{Google Pixel 6} & \multicolumn{3}{c}{Raspberry Pi} \\
\cmidrule(lr){3-5} \cmidrule(lr){6-8} \cmidrule(lr){9-11} \cmidrule(lr){12-14} \cmidrule(lr){15-17} \cmidrule(lr){18-20} \cmidrule(lr){21-23}
 &  & Time & Util. & RSS & Time & Util. & RSS & Time & Util. & RSS & Time & Util. & RSS & Time & Util. & RSS & Time & Util. & RSS & Time & Util. & RSS \\
\midrule
Training & Pre-training stage & 0.118 & 3.08 & 544.0 & 0.297 & 56.54 & 939.4 & 0.225 & 5.00 & 477.2 & 0.556 & 0.53 & 1337.3 & 0.084 & 9.25 & 288.0 & 0.846 & 12.38 & 315.5 & 0.758 & 24.92 & 404.7 \\
Training & Cluster fitting stage & 8.278 & 3.13 & 970.9 & 8.333 & 41.81 & 1550.5 & 7.861 & 5.04 & 876.8 & 7.997 & 0.94 & 2008.0 & 6.465 & 9.78 & 667.6 & 17.027 & 12.34 & 501.6 & 26.053 & 25.10 & 732.2 \\
Training & Training cluster assignment & 0.028 & 0.00 & 970.9 & 0.030 & 40.00 & 1550.5 & 0.032 & 0.00 & 876.8 & 0.029 & 0.00 & 2008.0 & 0.033 & 0.14 & 540.0 & 0.036 & 0.00 & 498.2 & 0.028 & 0.00 & 732.2 \\
Training & Adaptation for Cluster 1 & 0.272 & 3.04 & 979.6 & 0.600 & 39.00 & 1568.2 & 0.400 & 4.52 & 885.6 & 0.721 & 1.00 & 2042.4 & 0.231 & 7.73 & 543.5 & 2.259 & 11.86 & 509.9 & 2.354 & 24.61 & 741.1 \\
Training & Adaptation for Cluster 2 & 0.286 & 2.44 & 979.6 & 0.714 & 56.71 & 1578.9 & 0.374 & 3.98 & 885.6 & 0.761 & 1.00 & 2042.4 & 0.184 & 8.20 & 543.5 & 2.240 & 12.25 & 510.0 & 1.992 & 25.22 & 741.1 \\
\midrule
Inference & Inference cluster assignment & 1.126 & 3.08 & 979.6 & 1.260 & 43.70 & 1559.1 & 1.375 & 5.02 & 885.1 & 1.488 & 0.52 & 2016.2 & 1.021 & 9.82 & 546.6 & 2.390 & 12.26 & 509.4 & 5.295 & 25.09 & 740.0 \\
Inference & Cluster-specific model inference & 0.022 & 2.98 & 979.6 & 0.023 & 70.00 & 1578.9 & 0.023 & 4.61 & 885.6 & 0.023 & 1.00 & 2042.4 & 0.026 & 2.00 & 543.5 & 0.105 & 12.46 & 510.0 & 0.085 & 18.35 & 741.1 \\
\bottomrule
\end{tabular}%
}
\end{subtable}

\begin{subtable}{\textwidth}
\color{black}
\centering
\scriptsize
\caption{OhioT1DM Dataset}
\label{tab:flare_breakdown_ohio}
\resizebox{\textwidth}{!}{%
\begin{tabular}{llrrrrrrrrrrrrrrrrrrrrr}
\toprule
Phase & \textbf{\textsc{Flare}} pipeline stage & \multicolumn{3}{c}{AMD CPU} & \multicolumn{3}{c}{NVIDIA GPU} & \multicolumn{3}{c}{Spark CPU} & \multicolumn{3}{c}{Spark GPU} & \multicolumn{3}{c}{Apple M4} & \multicolumn{3}{c}{Google Pixel 6} & \multicolumn{3}{c}{Raspberry Pi} \\
\cmidrule(lr){3-5} \cmidrule(lr){6-8} \cmidrule(lr){9-11} \cmidrule(lr){12-14} \cmidrule(lr){15-17} \cmidrule(lr){18-20} \cmidrule(lr){21-23}
 &  & Time & Util. & RSS & Time & Util. & RSS & Time & Util. & RSS & Time & Util. & RSS & Time & Util. & RSS & Time & Util. & RSS & Time & Util. & RSS \\
\midrule
Training & Pre-training stage & 0.383 & 3.13 & 553.7 & 0.591 & 90.00 & 1158.5 & 0.496 & 5.07 & 486.7 & 1.043 & 8.90 & 1659.8 & 0.277 & 9.95 & 305.1 & 3.265 & 12.38 & 322.5 & 3.138 & 25.13 & 414.4 \\
Training & Cluster fitting stage & 1.454 & 3.13 & 1165.0 & 1.773 & 62.29 & 1820.8 & 1.590 & 5.04 & 1075.2 & 2.035 & 5.11 & 2321.3 & 1.174 & 9.79 & 965.9 & 5.502 & 12.51 & 609.3 & 6.473 & 25.09 & 929.4 \\
Training & Training cluster assignment & 0.002 & 0.00 & 1165.0 & 0.002 & 40.00 & 1820.8 & 0.002 & 0.00 & 1075.2 & 0.002 & 0.00 & 2321.3 & 0.002 & 0.12 & 965.9 & 0.002 & 0.00 & 608.6 & 0.002 & 0.00 & 929.4 \\
Training & Adaptation for Cluster 1 & 0.278 & 3.20 & 1165.3 & 0.606 & 80.23 & 1830.1 & 0.380 & 4.99 & 1075.6 & 0.697 & 6.70 & 2338.3 & 0.182 & 9.86 & 966.3 & 2.497 & 12.32 & 608.8 & 2.253 & 25.08 & 929.4 \\
Training & Adaptation for Cluster 2 & 0.277 & 3.08 & 1165.3 & 0.609 & 89.63 & 1830.1 & 0.380 & 5.06 & 1075.6 & 0.687 & 11.58 & 2338.4 & 0.195 & 9.66 & 966.3 & 2.486 & 12.37 & 608.8 & 2.236 & 25.09 & 930.0 \\
Training & Adaptation for Cluster 3 & 0.279 & 3.07 & 1165.3 & 0.637 & 88.02 & 1830.1 & 0.380 & 4.98 & 1075.6 & 0.699 & 10.47 & 2338.4 & 0.184 & 9.74 & 966.3 & 2.466 & 12.23 & 608.8 & 2.239 & 25.11 & 930.0 \\
\midrule
Inference & Inference cluster assignment & 0.915 & 3.12 & 1165.2 & 1.941 & 84.90 & 1821.0 & 1.252 & 5.05 & 1075.3 & 1.816 & 9.56 & 2321.4 & 0.618 & 9.87 & 966.1 & 7.992 & 12.43 & 608.6 & 7.179 & 25.12 & 929.4 \\
Inference & Cluster-specific model inference & 0.009 & 3.05 & 1165.3 & 0.010 & 89.00 & 1830.1 & 0.016 & 2.92 & 1075.6 & 0.010 & 11.00 & 2338.4 & 0.006 & 6.70 & 966.3 & 0.068 & 12.66 & 608.8 & 0.060 & 24.62 & 930.0 \\
\bottomrule
\end{tabular}%
}
\end{subtable}

\begin{subtable}{\textwidth}
\color{black}
\centering
\scriptsize
\caption{Percept-R Dataset}
\label{tab:flare_breakdown_percept_r}
\resizebox{\textwidth}{!}{%
\begin{tabular}{llrrrrrrrrrrrrrrrrrrrrr}
\toprule
Phase & \textbf{\textsc{Flare}} pipeline stage & \multicolumn{3}{c}{AMD CPU} & \multicolumn{3}{c}{NVIDIA GPU} & \multicolumn{3}{c}{Spark CPU} & \multicolumn{3}{c}{Spark GPU} & \multicolumn{3}{c}{Apple M4} & \multicolumn{3}{c}{Google Pixel 6} & \multicolumn{3}{c}{Raspberry Pi} \\
\cmidrule(lr){3-5} \cmidrule(lr){6-8} \cmidrule(lr){9-11} \cmidrule(lr){12-14} \cmidrule(lr){15-17} \cmidrule(lr){18-20} \cmidrule(lr){21-23}
 &  & Time & Util. & RSS & Time & Util. & RSS & Time & Util. & RSS & Time & Util. & RSS & Time & Util. & RSS & Time & Util. & RSS & Time & Util. & RSS \\
\midrule
Training & Pre-training stage & 0.382 & 3.14 & 602.3 & 0.610 & 89.83 & 1339.7 & 0.416 & 5.03 & 528.9 & 0.925 & 9.24 & 1871.8 & 0.230 & 9.94 & 336.8 & 6.445 & 12.48 & 361.7 & 2.858 & 25.11 & 465.7 \\
Training & Cluster fitting stage & 1.435 & 3.13 & 1212.1 & 1.785 & 62.57 & 1953.8 & 1.532 & 5.05 & 1100.8 & 1.897 & 5.27 & 2507.5 & 1.121 & 9.84 & 1018.7 & 5.894 & 12.57 & 656.8 & 6.064 & 25.11 & 975.7 \\
Training & Training cluster assignment & 0.002 & 0.00 & 1212.1 & 0.002 & 36.00 & 1953.8 & 0.002 & 0.00 & 1100.9 & 0.002 & 0.00 & 2507.5 & 0.002 & 0.09 & 1018.7 & 0.002 & 0.00 & 641.9 & 0.002 & 0.00 & 975.7 \\
Training & Adaptation for Cluster 1 & 0.328 & 3.08 & 1213.3 & 0.585 & 78.63 & 1977.3 & 0.330 & 4.95 & 1101.4 & 0.888 & 7.93 & 2537.7 & 0.190 & 9.41 & 1019.9 & 3.992 & 12.43 & 644.1 & 2.341 & 25.05 & 976.2 \\
Training & Adaptation for Cluster 2 & 0.313 & 3.07 & 1213.3 & 0.606 & 92.03 & 1978.5 & 0.341 & 5.07 & 1101.4 & 0.900 & 8.09 & 2540.5 & 0.196 & 9.10 & 1019.9 & 3.867 & 12.47 & 644.1 & 2.283 & 24.96 & 972.3 \\
Training & Adaptation for Cluster 3 & 0.315 & 3.20 & 1213.3 & 0.597 & 90.68 & 1978.5 & 0.328 & 5.05 & 1101.4 & 0.982 & 8.00 & 2540.5 & 0.178 & 9.88 & 1019.9 & 4.071 & 12.44 & 644.3 & 2.278 & 25.12 & 971.7 \\
Training & Adaptation for Cluster 4 & 0.320 & 3.08 & 1213.3 & 0.616 & 92.22 & 1978.8 & 0.332 & 5.05 & 1101.4 & 1.008 & 8.00 & 2540.5 & 0.178 & 9.75 & 1019.9 & 3.936 & 12.46 & 644.6 & 2.325 & 25.13 & 972.0 \\
Training & Adaptation for Cluster 5 & 0.317 & 3.08 & 1213.3 & 0.585 & 90.56 & 1978.8 & 0.326 & 5.09 & 1101.4 & 0.966 & 8.00 & 2540.5 & 0.194 & 9.55 & 1019.9 & 3.631 & 12.48 & 644.6 & 2.369 & 24.96 & 972.0 \\
\midrule
Inference & Inference cluster assignment & 0.869 & 3.14 & 1213.1 & 2.090 & 87.71 & 1954.0 & 1.115 & 5.03 & 1101.0 & 1.903 & 9.73 & 2507.7 & 0.639 & 9.95 & 1019.6 & 4.876 & 12.46 & 641.9 & 6.352 & 25.11 & 975.7 \\
Inference & Cluster-specific model inference & 0.011 & 2.04 & 1213.3 & 0.012 & 92.00 & 1978.8 & 0.011 & 4.84 & 1101.4 & 0.008 & 8.00 & 2540.5 & 0.009 & 7.09 & 1019.9 & 0.077 & 12.91 & 644.6 & 0.066 & 26.05 & 970.3 \\
\bottomrule
\end{tabular}%
}
\end{subtable}

\end{table}
\clearpage
\section{Detailed Post-hoc Cluster Interpretation}
\label{sec:appendix_interpretation}
%%%%%%%%%%%%%%%%%%%%%%%%%%%%%%%%%%%%%%%%%%%%%%%%%%%%%%%%%%%%%%%%%%%%%%%%%%%%%%%

This appendix provides the detailed analyses supporting
Section~\ref{sec:cluster_interpretation}. Following prior clinical clustering
studies, we first obtain the latent subgroups independently of the external
characterization variables and then examine whether the learned structure
corresponds to recognizable demographic, clinical, diagnostic, utilization, or
device-related characteristics
\cite{khan2023clustering,huang2021dice}.

The literature shows that both strong and weak external characterization can
be informative. Some learned classes have distinctive demographic or clinical
profiles \cite{khan2023clustering}, whereas others may cut
across conventional categories. We therefore do not require every
\textbf{\textsc{Flare}} cluster to map to a single external characteristic.

Our analysis uses three complementary post-hoc checks. We first test
individual cluster--attribute associations using effect sizes, permutation
tests, and FDR correction. We then evaluate joint prediction of cluster
membership using ordinary logistic regression and permutation-null AUROC.
Finally, we use $L_1$-regularized logistic regression to obtain sparse
attribute profiles for each cluster.

Because \textbf{\textsc{Flare}} assigns clusters at the sample level while
most external variables are participant-level, the analysis unit depends on
the observed cluster structure. We use participant-level majority assignment
when a valid participant-level comparison exists and use sample-level or
exposure-level analysis otherwise.

%%%%%%%%%%%%%%%%%%%%%%%%%%%%%%%%%%%%%%%%%%%%%%%%%%%%%%%%%%%%%%%%%%%%%%%%%%%%%%%
\section{Supplementary Post-hoc Cluster Interpretation}

\subsection{IHS Fold~5}
\label{sec:appendix_ihs_fold5}

\subsubsection{Cluster structure and analysis units}

Table~\ref{tab:appendix_ihs_fold5_structure} reports the sample counts
and participant-level majority assignments. Each participant is
majority-assigned to the cluster containing the largest fraction of
their samples. Clusters~0--2 are analyzed at the participant level,
although Cluster~2 contains only four majority-assigned participants.
Cluster~3 has no majority-assigned participants and is examined
separately through sample-level membership and participant-level
exposure.

\begin{table}[H]
\centering
\caption{\textbf{IHS Fold~5 cluster structure.}
Sample counts reflect the original sample-level clustering;
participant counts reflect post-hoc majority assignments.}
\label{tab:appendix_ihs_fold5_structure}
\begin{tabular}{lrr}
\toprule
Cluster & Samples & Majority-assigned participants \\
\midrule
0 & 313 & 23 \\
1 & 557 & 41 \\
2 & 175 & 4  \\
3 & 191 & 0  \\
\bottomrule
\end{tabular}
\end{table}

\subsubsection{Single-attribute association analysis}

For Clusters~0--2, we test each participant attribute against
one-versus-rest majority-assigned cluster membership. Effect sizes
are rank-biserial correlation for age, $\phi$ for binary attributes,
and Cram\'er's $V$ for multicategory attributes. We report
permutation $p$-values and apply Benjamini--Hochberg FDR correction
across the 15 comparisons ($3$ clusters $\times$ $5$ attributes).
Cluster~1 results are reported in Table~\ref{tab:ihs_cluster1_assoc}.

\begin{table}[h]
\centering
\caption{\textbf{Post-hoc interpretation of IHS Fold~5 Cluster~1.}
The upper panel compares participants majority-assigned to Cluster~1 with
participants majority-assigned to the remaining participant-level clusters.
FDR $q$-values are corrected across the 15 participant-level tests in this
fold ($3$ clusters $\times$ $5$ attributes). The lower panel reports
one-versus-rest ordinary logistic regression using all five participant
attributes jointly.}
\label{tab:ihs_cluster1_assoc}
\begin{tabular}{lccc}
\toprule
\multicolumn{4}{c}{\textbf{Single-attribute association}}\\
\midrule
Attribute & Effect size & Perm.\ $p$ & FDR $q$\\
\midrule
Age       & $r_{\mathrm{rb}}=0.014$ & 0.930 & 0.930\\
Sex       & $\phi=0.036$            & 0.626 & 0.783\\
Specialty & $\mathbf{V=0.538}$      & $\mathbf{0.037}$ & 0.554\\
Ethnicity & $V=0.246$               & 0.595 & 0.783\\
PHQ10     & $\phi=0.093$            & 0.397 & 0.783\\
\midrule
\multicolumn{4}{c}{\textbf{Joint prediction: ordinary logistic regression}}\\
\midrule
Accuracy & Balanced Acc. & AUROC & Null AUROC\\
\midrule
0.632 & 0.626 & 0.606 & 0.487\\
\bottomrule
\end{tabular}
\end{table}

\subsubsection{Joint prediction using ordinary logistic regression}

We fit one-versus-rest ordinary logistic regression models using
age, sex, specialty, ethnicity, and PHQ10 status to predict
participant-level majority assignment.
Table~\ref{tab:appendix_ihs_prediction_summary} reports the results
for all three participant-level clusters. Null AUROC denotes the
mean AUROC obtained after shuffling cluster labels and repeating
the prediction procedure.

\begin{table}[t]
\centering
\caption{\textbf{Participant-level prediction of IHS Fold~5 clusters
using ordinary logistic regression.}
Each model uses all five participant attributes.
$n_{+}$ denotes the number of majority-assigned participants in
the target cluster. Cluster~3 has no participant-level positive
examples and is excluded.}
\label{tab:appendix_ihs_prediction_summary}
\begin{tabular}{lrrrrr}
\toprule
Cluster & $n_{+}$ & Accuracy & Bal.\ Acc. & AUROC & Null AUROC \\
\midrule
0 & 23 & 0.559 & 0.539 & 0.569 & 0.496 \\
1 & 41 & 0.632 & 0.626 & 0.606 & 0.487 \\
2 & 4  & 0.838 & 0.563 & 0.547 & 0.488 \\
\bottomrule
\end{tabular}
\end{table}

Cluster~1 has the highest observed AUROC, although the difference
from its permutation null is not statistically significant
(permutation $p=0.293$). Cluster~0 also shows a modest numerical
difference from its null AUROC. For Cluster~2, the high raw accuracy
should be interpreted alongside the substantial class imbalance:
only four participants belong to the positive class, and balanced
accuracy and AUROC indicate limited discrimination.

\subsubsection{LASSO relative importance and prediction}

The sparse profiles in Figures~\ref{fig:cross_dataset_lasso} and~\ref{fig:ohio_lasso} are
obtained using one-versus-rest $L_1$-regularized logistic regression.
For categorical attributes represented by multiple encoded
coefficients, we sum the absolute coefficient magnitudes belonging
to the same parent attribute. Relative importance is defined as
\[
\mathrm{RI}(A)=
\frac{\sum_{j\in A}|\beta_j|}
     {\sum_k|\beta_k|}
\times 100,
\]
where $\beta_j$ denotes a fitted coefficient and $A$ identifies the
coefficients belonging to an attribute. When all coefficients are
zero, the profile is reported as empty. Otherwise, relative
importance sums to 100\% within a cluster and describes the
composition of its retained coefficients.

Table~\ref{tab:appendix_ihs_lasso_prediction_summary} reports
cross-validated LASSO prediction results separately from the
ordinary logistic regression results. Clusters~0--2 use
participant-level majority assignments; Cluster~3 uses sample-level
membership because it has no majority-assigned participants.

\begin{table}[t]
\centering
\caption{\textbf{Cross-validated LASSO prediction for IHS Fold~5.}
Clusters~0--2 are evaluated at the participant level;
Cluster~3 is evaluated at the sample level.
$n_{+}$ and $n_{-}$ denote positive and negative observations,
respectively.}
\label{tab:appendix_ihs_lasso_prediction_summary}
\begin{tabular}{lrrrrr}
\toprule
Cluster & $n_{+}$ & $n_{-}$ & Bal.\ Acc. & AUROC & Null AUROC \\
\midrule
0 & 23  & 45   & 0.583 & 0.566 & 0.493 \\
1 & 41  & 27   & 0.564 & 0.578 & 0.492 \\
2 & 4   & 64   & 0.453 & 0.406 & 0.496 \\
3 & 191 & 1045 & 0.499 & 0.498 & 0.499 \\
\bottomrule
\end{tabular}
\end{table}

The AUROC differences from the permutation null are 0.073 for
Cluster~0 (permutation $p=0.220$) and 0.086 for Cluster~1
(permutation $p=0.160$). Cluster~2 performs below its null AUROC
and has only four positive participants. Cluster~3 retains no
non-zero coefficients and performs approximately at its null level.
These results show why the relative-importance profiles should be
interpreted separately from predictive strength: an attribute can
account for most of the retained coefficient magnitude while the
model's overall discrimination remains modest.

\subsubsection{Separate characterization of Cluster~3}

Cluster~3 contains 191 samples contributed by 64 of the 68
participants, but it is not the majority cluster for any participant.
We therefore examine two comparisons: Cluster~3 samples versus all
remaining samples ($191$ versus $1{,}045$), and participants with
any Cluster~3 exposure versus those without exposure ($64$ versus
$4$).

In the sample-level comparison, effect sizes are small for all five
attributes: age ($r_{\mathrm{rb}}=-0.037$), sex ($\phi=0.018$),
specialty ($V=0.069$), ethnicity ($V=0.066$), and PHQ10 status
($\phi=0.007$). All FDR-adjusted $q$-values exceed 0.86.
Because participants contribute repeated samples, these
sample-level comparisons are interpreted descriptively.

In the participant-level exposure comparison, specialty has the
largest effect ($V=0.443$), but the association is not statistically
significant ($p=0.270$, $q=1.000$). This comparison is limited by
the four participants without exposure. The participant-specific
fraction of samples assigned to Cluster~3 is also weakly associated
with age (Spearman $\rho=0.154$, $p=0.211$).

The broad exposure, absence of majority-assigned participants,
small sample-level effects, and empty LASSO profile are consistent
with treating Cluster~3 as a recurrent within-person state without
a clear measured participant-attribute profile. These analyses do
not establish the substantive meaning of that state.

\subsection{OhioT1DM Fold~2}
\label{sec:appendix_ohio_fold3}

\subsubsection{Analysis unit and joint prediction}

All nine participants in the OhioT1DM Fold~2 \texttt{BpT-train} split are
majority-assigned to Cluster~0, although each of the four clusters
contains samples from all nine participants.
Participant-level one-versus-rest comparisons are therefore
undefined: Cluster~0 has no negative participants, while
Clusters~1--3 have no positive participants. We instead report
descriptive sample-level analyses using age group, sex, pump type,
and sensor band.

Table~\ref{tab:appendix_ohio_prediction} reports cluster sample
counts and one-versus-rest ordinary logistic regression results.
Null AUROC is obtained by shuffling cluster labels and repeating
the prediction procedure.

\begin{table}[t]
\centering
\caption{\textbf{Post-hoc characterization of OhioT1DM Fold~2 clusters
using the \texttt{BpT-train} split: 21,209 samples from nine participants.}
Each ordinary logistic
regression model uses all four external attributes.}
\label{tab:appendix_ohio_prediction}
\begin{tabular}{lrrrrr}
\toprule
Cluster & Samples & Accuracy & Bal.\ Acc. & AUROC & Null AUROC \\
\midrule
0 & 7792 & 0.536 & 0.502 & 0.499 & 0.499 \\
1 & 3462 & 0.469 & 0.499 & 0.497 & 0.497 \\
2 & 5405 & 0.485 & 0.502 & 0.501 & 0.498 \\
3 & 4550 & 0.529 & 0.499 & 0.499 & 0.497 \\
\bottomrule
\end{tabular}
\end{table}

Balanced accuracy is approximately 0.5 for every cluster, and
observed AUROC is nearly identical to the corresponding
permutation-null AUROC. Thus, the measured attributes provide
little predictive information about sample-level cluster
membership under this analysis.

\subsubsection{Single-attribute associations}

We test each of the four attributes against membership in each of
the four clusters and apply Benjamini--Hochberg correction across
the resulting 16 comparisons. The largest observed effect is the
association between Cluster~0 membership and age group
($V=0.011$, permutation $p=0.249$, FDR $q=0.751$).
None of the 16 comparisons remains significant after FDR
correction.

\begin{figure}[h]
\centering
\includegraphics[width=0.62\linewidth]{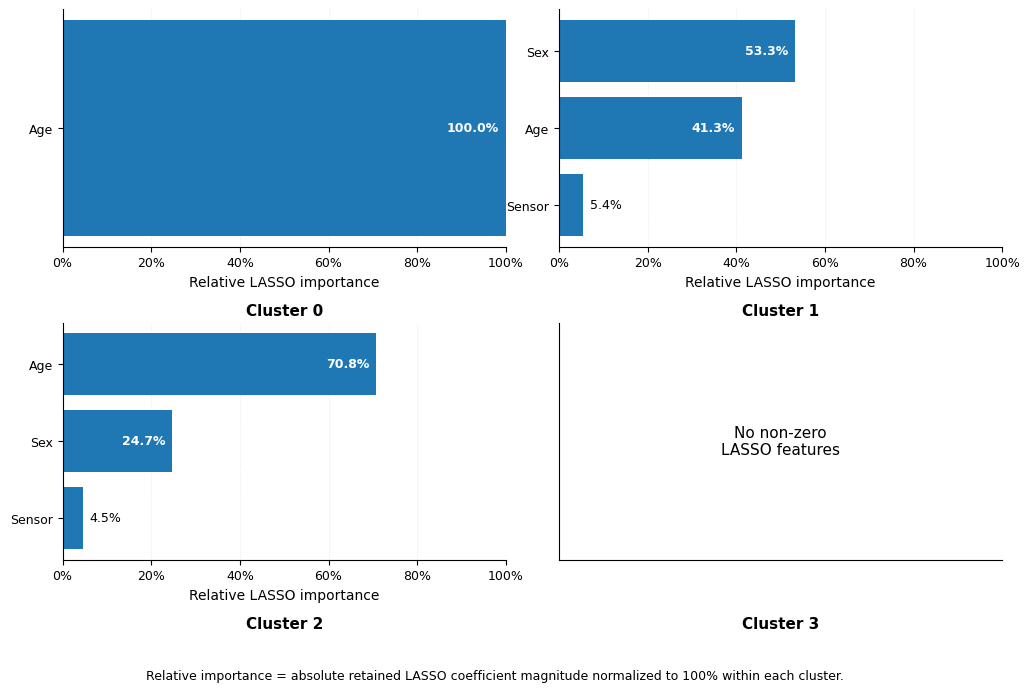}
\caption{Sparse post-hoc LASSO profiles for OhioT1DM Fold~2 at the sample level (participant-level majority assignment collapses to Cluster~0). Because cluster-membership prediction is at chance level (Table~\ref{tab:appendix_ohio_prediction}), these profiles show which coefficients survive regularization, not demographic or device predictability.}
\label{fig:ohio_lasso}
\end{figure}

These analyses contain repeated samples from only nine
participants and are therefore interpreted descriptively.
Together with the joint prediction results, they support treating
the OhioT1DM LASSO profiles in Figure~\ref{fig:ohio_lasso}
as summaries of retained coefficients rather than evidence of
strong demographic or device predictability.

\subsection{Sensitivity Analysis}
\label{app:cluster_Sen}

Table~\ref{tab:appen_cluster_sensitivity_all} presents the cluster sensitivity results for the EDA, IHS, and Percept-R datasets. 
On IHS, fixed-cluster settings show mixed behavior, especially for Sex and PHQ10$>0$ subgroups, whereas \textbf{\textsc{Flare}} provides more balanced gains. On Percept-R, larger $n$ often improves performance, but \textbf{\textsc{Flare}} matches or exceeds the strongest fixed-cluster setting for most subgroup definitions.

\begin{table*}[h]
\centering
\small
\captionsetup{font={color=black}}
\arrayrulecolor{black}
\color{black}
\caption{Cluster sensitivity analysis showing percentage changes $(\Delta B,\Delta H,\Delta E)$ for different numbers of clusters per fold ($n=1,2,3,4$; for $n=1$, the selected model remains the Fisher-regularized pretrained model because no adapted checkpoint improved validation F1, so results equal BpT-w Fisher) and \textbf{\textsc{Flare}} across the EDA, IHS, and Percept-R datasets. Values are reported in percentage points.}
\label{tab:appen_cluster_sensitivity_all}
\resizebox{0.77\textwidth}{!}{
\begin{tabular}{l|
ccc|ccc|ccc|ccc|ccc}
\toprule
\multirow{2}{*}{\textbf{\makecell{Sensitive \\ Attribute}}} &
\multicolumn{3}{c|}{\textbf{n=1 (\%)}} &
\multicolumn{3}{c|}{\textbf{n=2 (\%)}} &
\multicolumn{3}{c|}{\textbf{n=3 (\%)}} &
\multicolumn{3}{c|}{\textbf{n=4 (\%)}} &
\multicolumn{3}{c}{\textbf{\textsc{Flare} (\%)}} \\
&
$\Delta B$ & $\Delta H$ & $\Delta E$
& $\Delta B$ & $\Delta H$ & $\Delta E$
& $\Delta B$ & $\Delta H$ & $\Delta E$
& $\Delta B$ & $\Delta H$ & $\Delta E$
& $\Delta B$ & $\Delta H$ & $\Delta E$\\
\midrule
\multicolumn{16}{c}{\textbf{EDA Dataset}}\\
\midrule
Group\_label 
& 1.70 &  0.54 &  \textbf{1.04}
& 1.78 & -1.47 & -1.31
& -0.22 & -4.13 & -1.47
& 0.69 & -2.46 & -0.01
& \textbf{2.16} & \textbf{0.85} & 0.57 \\

Sex 
& 1.71 &  1.38 & -0.46
& 0.65 & -0.71 & -1.92
& 1.39 &  0.38 & -1.43
& 0.61 & -0.66 & -1.80
& \textbf{2.27} & \textbf{2.22} & \textbf{0.07} \\

\rowcolor{gray!15}\textbf{Mean} 
& 1.71 &  0.96 &  0.29
& 1.22 & -1.09 & -1.62
& 0.59 & -1.88 & -1.45
& 0.65 & -1.56 & -0.91
& \textbf{2.22} & \textbf{1.54} & \textbf{0.32} \\

\midrule
\multicolumn{16}{c}{\textbf{IHS Dataset}}\\
\midrule
Sex            &  0.00 & -0.71 & \textbf{0.76} &  -2.02 &  -2.54 & 0.73 &  -4.68 &  -5.20 & 0.75 &  -5.68 &  -7.20 & -0.38 &  \textbf{4.97} &  \textbf{4.93} &  0.06 \\
Age            &  3.78 & -4.85 &  1.39 &  3.10 & -12.67 &  4.31 &  3.25 & -40.96 &  5.15 &  4.35 & -34.96 &  \textbf{5.95} &  \textbf{9.72} &  \textbf{0.30} &  3.07 \\
Ethnicity      &  0.29 & -0.37 & -2.74 &  4.09 &  -3.65 & -1.62 &  -3.90 & -20.27 & -8.60 &  -2.74 & -13.27 & -6.24 &  \textbf{8.23} &  \textbf{3.38} &  \textbf{0.28} \\
Specialty      & -0.06 & -10.32 & -0.35 &  2.42 &  -7.70 & -0.99 &  0.06 & -26.21 & -1.13 &  1.16 & -11.37 & -0.63 &  \textbf{5.00} &  \textbf{0.03} &  \textbf{0.36} \\
PHQ10$>0$      & 0.00 & -0.98 &  -1.39 & -3.67 &  -4.60 &  1.32 &  -3.88 &  -5.76 &  -2.66 &  -2.88 &  -3.76 &  -1.25 &  \textbf{2.04} &  \textbf{0.04} &  \textbf{2.83} \\
\rowcolor{gray!15}\textbf{Mean}  
& 0.80 & -3.45 & -0.47
& 0.78 &  -6.23 &  0.75
& -1.83 & -19.68 &  -1.30
& -1.16 & -14.11 &  -0.51
& \textbf{5.99} &  \textbf{1.74} &  \textbf{1.32} \\

\midrule
\multicolumn{16}{c}{\textbf{Percept-R Dataset}}\\
\midrule
Sex            & -2.11 &  -2.91 &  \textbf{0.93} &  -2.23 &  -3.11 & 0.80 &  0.74 &   0.44 & 0.44 &  3.41 &   2.85 &  0.79 &  \textbf{5.12} &  \textbf{4.70} &  0.59 \\
Age(months)    & -1.90 & -11.83 & 6.75&  -2.35 & -11.87 &   5.71 &  0.66 &  -9.82 &  6.99 &  4.23 &  -8.19 &  \textbf{10.58} &  \textbf{9.60} &  \textbf{0.02} &  3.35 \\
Race           & 0.19 & -18.02 & 2.83 &  3.22 & -18.19 &   5.07 &  2.92 & -15.34 &  \textbf{6.62} &  6.26 & -13.88 &  5.78 &  \textbf{7.43} &  \textbf{0.00} &  5.77 \\
Ethnicity      & -5.03 &  -6.19 & -1.64 & -5.86 & -7.70 &  -2.60 & -1.10 &  -1.28 &  0.26 &  0.04 &  -0.16 & -0.27 &  \textbf{2.84} &  \textbf{1.96} &  \textbf{1.25} \\
\rowcolor{gray!15}\textbf{Mean}  
& -2.21 & -9.74 & 2.22
& -1.81 & -10.22 & 2.25
&  0.81 & -6.50 &  3.58
&  3.49 & -4.85 &  \textbf{4.22}
& \textbf{6.25} & \textbf{1.67} & 2.74 \\

\bottomrule
\end{tabular}}
\arrayrulecolor{black}
\end{table*}
\newpage
\end{document}